\documentclass[cernpreprint, atlasdraft=false, UKenglish, texmf, orcidlogo]{atlasdoc}
\usepackage{atlaspackage}
\usepackage{multirow}
\usepackage{float}
\usepackage{atlasbiblatex}

\usepackage[process=true]{atlasphysics}
\graphicspath{{logos/}{figures/}}

\usepackage{ANA-JETM-2020-03-PAPER-defs}

\AtlasTitle{The performance of missing transverse momentum reconstruction and its significance with the ATLAS detector using 140 fb$^{-1}$ of $\sqrt{s}=13$ TeV $pp$ collisions}

\AtlasAbstract{%
This paper presents the reconstruction of missing transverse momentum ($p_{\text{T}}^{\text{miss}}$) in proton-proton collisions, at a center-of-mass energy of 13 TeV.
This is a challenging task involving many detector inputs, combining fully calibrated electrons, muons, photons, hadronically decaying $\tau$-leptons, hadronic jets, and soft activity from remaining tracks.
Possible double counting of momentum is avoided by applying a signal ambiguity resolution procedure which rejects detector inputs that have already been used.
Several $p_{\text{T}}^{\text{miss}}$ `working points' are defined with varying stringency of selections, the tightest improving the resolution at high pile-up by up to 39\% compared to the loosest.
The $p_{\text{T}}^{\text{miss}}$ performance is evaluated using data and Monte Carlo simulation, with an emphasis on understanding the impact of pile-up, primarily using events consistent with leptonic $Z$ decays.
The studies use $140~\text{fb}^{-1}$ of data, collected by the ATLAS experiment at the Large Hadron Collider between 2015 and 2018.
The results demonstrate that $p_{\text{T}}^{\text{miss}}$ reconstruction, and its associated significance, are well understood and reliably modelled by simulation.
Finally, the systematic uncertainties on the soft $p_{\text{T}}^{\text{miss}}$ component are calculated.
After various improvements the scale and resolution uncertainties are reduced by up to $76\%$ and $51\%$, respectively, compared to the previous calculation at a lower luminosity.
}

\AtlasRefCode{JETM-2020-03}

\PreprintIdNumber{CERN-EP-2024-023}

\AtlasJournal{EPJC}
\AtlasJournalRef{Eur. Phys. J. C 85 (2025) 606}
\AtlasDOI{10.1140/epjc/s10052-025-14062-8}

\hypersetup{pdftitle={ATLAS document},pdfauthor={The ATLAS Collaboration}}

\begin{document}

\maketitle

\tableofcontents
\newpage
\section{Introduction}
\label{sec:intro}

Missing transverse momentum (\ptmiss, also referred to as \met\ or MET) is a crucial observable for the ATLAS experiment at the Large Hadron Collider (LHC).
It is an experimental proxy for the transverse momentum carried by undetected particles produced in proton--proton (\(pp\)) collisions recorded by the ATLAS detector~\cite{PERF-2007-01}.
As such, the \ptmiss\ is the magnitude of the 2-dimensional momentum vector, \ptmissvec, defined transverse to the proton beam direction.
The \ptmiss\ in a given collision event is constructed, using the principle of momentum conservation, from the reconstructed hard objects\footnote{`Hard objects' here refer to the outputs of reconstruction algorithms applied to detector signals, which are candidates to be electrons, muons, jets, hadronically-decaying taus, and photons.} and recorded tracks in the final state.
A non-zero value of `real' \ptmiss\ can indicate not just the production of Standard Model (SM) neutrinos, but potentially the production of certain beyond-SM particles like dark matter, which are stable on detector scales and would escape ATLAS undetected.
Reconstructing \ptmiss\ is a challenging pursuit, since all detector subsystems are involved, and a highly unambiguous representation of all of the hard objects formed in the hard scatter interaction of interest is required --- including calorimeter, tracker and muon spectrometer signals.
This representation is obscured by detector resolution and acceptance limitations, object mis-measurement, calibration errors, and signal remnants from additional \(pp\) interactions occurring in the same --- or neighbouring --- LHC bunch crossings relative to the triggered hard-scatter event (pile-up).
All of these effects cause `fake' \ptmiss, which ATLAS aims to minimise.

To date, ATLAS's approaches to \ptmiss\ reconstruction have prioritised minimising the impact of pile-up.
These were designed based on the data recorded between 2010 and 2012 (\runone)~\cite{PERF-2011-07,PERF-2014-04}, and substantially re-developed using data collected in 2015 (the first year of \runtwo), as described in Ref.~\cite{PERF-2016-07}.
These approaches provide a basis for the \ptmiss\ reconstruction utilised for the full 2015--2018 dataset (\runtwo), described in this paper alongside evaluations of its performance and systematic uncertainties.
In comparison to \runone, there are two major improvements in \ptmiss\ reconstruction: first, the move from using calorimeter to tracker information to form the soft component of the \ptmiss\, as default increases pile-up resilience.
Second, the change to a dynamic approach to \ptmiss\ reconstruction --- such that it is calculated based on the choice of reconstructed and calibrated hard objects considered in any given analysis --- leads to more consistency within an analysis and \ptmiss\ reconstruction to exploit any improvements to hard object calibrations.
The second development is discussed in more detail in Ref.~\cite{Balunas:2023fpu}.
Furthermore, improvements since early \runtwo~\cite{PERF-2016-07} come from the introduction of the particle flow jet algorithm~\cite{PERF-2015-09}, which combines calorimeter and tracking information, and the development of multiple \ptmiss\ working points, which place varying requirements on jets used to build the \ptmiss\ to reduce pile-up contamination, and are each better-suited to different event topologies.
Moving from the loosest to tightest working point improves the \ptmiss\ resolution by 14--39\% for \Zmm\ MC simulated events with average interactions per bunch crossing exceeding 30.
This is countered by a degradation in \ptmiss\ response up to 15\%, and an increase in \ptmiss\ bias from 7--35\%, in \Zmm\ MC simulated events when changing from the loosest to tightest working point.
The modelling and performance of \ptmiss\ is studied in event topologies that permit a focus on the impacts of pile-up, fake \ptmiss\ and the new developments related to jets.
The larger dataset allows for more consideration of the dependence of systematic uncertainties in the scale and resolution of the soft component of the \ptmiss\ on the component of \ptmiss\ built from hard objects.
The uncertainty values in \Zee\ events reduce throughout the kinematic range considered, in comparison to preliminary results in Ref.~\cite{ATLAS-CONF-2018-023} when using particle flow, with these improvements.
Scale uncertainties are reduced by up to $76\%$ and resolution uncertainties are reduced by up to $51\%$.
Finally, a sophisticated \ptmiss\ significance variable was also developed using an object-based approach which significantly improves discrimination between events with real and fake \ptmiss.
This variable has been widely used in ATLAS searches, for example Refs.~\cite{EXOT-2018-29,SUSY-2018-17}.

This paper is organised as follows.
A brief overview of the ATLAS detector is provided in Section~\ref{sec:ATLAS}.
The data and Monte Carlo simulation samples used in the paper are detailed in Section~\ref{sec:samples}, followed by an outline of the hard object and event selections used in Sections~\ref{sec:objects} and~\ref{sec:events} respectively.
The reconstruction of \ptmiss, and other kinematic variables associated with it, is described in Section~\ref{sec:reco}.
The results of \ptmiss\ performance studies are presented in Section~\ref{sec:performance}.
In Section~\ref{sec:systs}, the methodology of the \ptmiss\ systematic uncertainties calculation, and the results of their measurement, are detailed.
Finally, the \ptmiss\ significance is introduced --- and its performance studied --- in Section~\ref{sec:metsig}.


%
\section{ATLAS detector}
\label{sec:ATLAS}

The ATLAS experiment~\cite{PERF-2007-01} at the LHC is a multi-purpose particle detector with a forward-backward symmetric cylindrical geometry and a near \(4\pi\) coverage in solid angle.\footnote{ATLAS uses a right-handed coordinate system with its origin at the nominal interaction point (IP) in the centre of the detector and the \(z\)-axis along the beam pipe.
The \(x\)-axis points from the IP to the centre of the LHC ring, and the \(y\)-axis points upwards.
Polar coordinates \((r,\phi)\) are used in the transverse plane, \(\phi\) being the azimuthal angle around the \(z\)-axis.
The pseudorapidity is defined in terms of the polar angle \(\theta\) as \(\eta = -\ln \tan(\theta/2)\) and is equal to the rapidity \(y = \frac{1}{2} \ln \left( \frac{E + p_z c}{E - p_z c} \right)\) in the relativistic limit.
Angular distance is measured in units of \(\Delta R \equiv \sqrt{(\Delta y)^{2} + (\Delta\phi)^{2}}\).}
It consists of an inner tracking detector (ID) surrounded by a thin superconducting solenoid providing a \SI{2}{\tesla} axial magnetic field, electromagnetic and hadron calorimeters, and a muon spectrometer (MS).
The ID covers the pseudorapidity range \(|\eta| < 2.5\).
It consists of silicon pixel, silicon microstrip, and transition radiation tracking detectors.
Lead/liquid-argon (LAr) sampling calorimeters provide electromagnetic (EM) energy measurements with high granularity.
A hadron (steel/scintillator-tile) calorimeter covers the central pseudorapidity range \(|\eta| < 1.7\).
The end-cap and forward regions are instrumented with LAr calorimeters for both EM and hadronic energy measurements up to \(|\eta| = 4.9\).
The MS surrounds the calorimeters and is based on three large air-core toroidal superconducting magnets with eight coils each.
The field integral of the toroids ranges between \num{2.0} and \SI{6.0}{\tesla\metre} across most of the detector.
The MS includes a system of precision tracking chambers and fast detectors for triggering.
A two-level trigger system is used to select events.
The first-level trigger is implemented in hardware and uses a subset of the detector information to reduce the accepted rate to at most nearly \SI{100}{\kHz}.
This is followed by a software-based trigger that reduces the accepted event rate to \SI{1}{\kHz} on average depending on the data-taking conditions.
An extensive software suite~\cite{ATL-SOFT-PUB-2021-001} is used in data simulation, in the reconstruction and analysis of real and simulated data, in detector operations, and in the trigger and data acquisition systems of the experiment.


%
\section{Data and simulation samples}
\label{sec:samples}

The proton--proton collisions analysed in this paper were collected between 2015 and 2018, at a centre-of-mass energy of \(\sqrt s=13~\TeV\) and a 25~ns inter-bunch spacing.
They correspond to an integrated luminosity of \(140~\ifb\), with an uncertainty of 0.83\%~\cite{DAPR-2021-01} obtained using the LUCID-2 detector \cite{LUCID2} for the primary luminosity measurements, complemented by measurements using the inner detector and calorimeters.

In any given data-taking period, the unprescaled single-lepton triggers with the lowest \(p_\text{T}\), ID or isolation thresholds were used~\cite{TRIG-2018-05,TRIG-2018-01,TRIG-2016-01}.
These thresholds ranged from 20~\GeV\ to 140~\GeV, with the lowest trigger threshold for electrons (muons) at \(\pT=\SI{24}{\GeV}\) (\(\pT=\SI{20}{\GeV}\)).
The offline lepton selection was kept more stringent than the trigger-level requirement to ensure that trigger efficiencies are constant.

Simulated events are used to model the SM processes considered in this paper.
The Monte Carlo (MC) simulated events were processed through a full simulation of the ATLAS detector~\cite{SOFT-2010-01} based on \GEANT~\cite{Agostinelli:2002hh}.
All samples used are listed in Table \ref{tab:sim:mc} along with the relevant parton distribution function (PDF) sets used for the matrix element (ME) and parton shower (PS), the configuration of underlying-event and hadronisation parameters (tune), and the cross-section order in \(\alpha_\mathrm{s}\) (and \(\alpha_\mathrm{EW}\) if corrections are used) used to normalise the event yields for these samples.
Further information on the ATLAS simulations of \ttbar, single-top-quark (\(Wt\)), multiboson and vector-boson plus jets processes can be found in the relevant public notes~\cite{ATL-PHYS-PUB-2018-009,ATL-PHYS-PUB-2016-004,ATL-PHYS-PUB-2017-005,ATL-PHYS-PUB-2017-006}.

The effect of pile-up in the same and neighbouring bunch crossings was modelled by overlaying the simulated hard-scattering event with inelastic proton--proton events generated with \PYTHIA{8.186}~\cite{Sjostrand:2007gs} using the \NNPDF{2.3lo} set of parton distribution functions (PDF)~\cite{Ball:2012cx} and the A3 set of tuned parameters~\cite{ATL-PHYS-PUB-2016-017}.
The MC samples were reweighted so that the distribution of the average number of interactions per bunch crossing reproduces the observed distribution in the data.

\begin{table}[!h]
\tiny
\centering
\caption{\label{tab:sim:mc}Simulated SM event samples with the corresponding matrix element and parton shower generators, cross-section order in \(\alpha_\mathrm{s}\) (and \(\alpha_\mathrm{EW}\) if corrections are used) used to normalise the event yield, underlying-event tune and the generator PDF sets used. \Zll\ \SHERPA{2.2.1} is used for the derivation of systematic uncertainties only.}
\begin{tabular}{lp{0.17\linewidth}p{0.08\linewidth}p{0.16\linewidth}p{0.07\linewidth}p{0.1\linewidth}p{0.1\linewidth}}
\hline
Physics process			& Generator (ME) 									& Parton shower					&  Normalisation						& Tune						& PDF (ME)				& PDF (PS)\\
\midrule
\(t\bar{t}\)			& \POWHEGBOX{v2}\cite{Re:2010bp,Nason:2004rx,Frixione:2007vw,Alioli:2010xd}		& \PYTHIA{8.230}~\cite{Sjostrand:2014zea}	& NNLO+NNLL~\cite{Czakon:2011xx}				& A14~\cite{ATL-PHYS-PUB-2014-021}		& \NNPDF{3.0nlo}~\cite{Ball:2014uwa}	& \NNPDF{2.3lo}~\cite{Ball:2012cx}\\
Single top (\(Wt\))		& \POWHEGBOX{v2}									& \PYTHIA{8.230}				& NLO~\cite{Aliev:2010zk,Kant:2014oha}				& A14						& \NNPDF{3.0nlo}			& \NNPDF{2.3lo}\\
\Zll\ (\SHERPA)			& \SHERPA{2.2.11}\cite{Bothmann:2019yzt,Gleisberg:2008fv}				& \SHERPA{2.2.11}				& NNLO~\cite{Gavin:2010az}					& \SHERPA default~\cite{ATL-PHYS-PUB-2016-003}	& \NNPDF{3.0nnlo}~\cite{Ball:2014uwa}	& \NNPDF{3.0nnlo}~\cite{Ball:2014uwa}\\
\Zll\ (\POWHEG)			& \POWHEGBOX{v1}~\cite{Nason:2004rx,Frixione:2007vw,Alioli:2010xd,Alioli:2008gx}	& \PYTHIA{8.186}~\cite{Sjostrand:2007gs}	& NLO~\cite{ATL-PHYS-PUB-2017-005,Melia:2011tj,Nason:2013ydw}	& \AZNLO~\cite{STDM-2012-23}			& \CT{10nlo}~\cite{Lai:2010vv}		& \CTEQ{6L1}~\cite{Pumplin:2002vw}\\
\Zll\ (\MADGRAPH)		& \MGNLO{2.2.2}~\cite{Alwall:2014hca}							& \PYTHIA{8.186}				& NNLO~\cite{Anastasiou:2003ds}					& A14						& \NNPDF{3.0nlo}			& \NNPDF{2.3lo}\\
\(WW,WZ,ZZ\)			& \POWHEGBOX{v2}~\cite{Nason:2004rx,Frixione:2007vw,Alioli:2010xd}			& \PYTHIA{8.186}				& NLO								& \AZNLO\					& \CT{10nlo}				& \CTEQ{6L1}\\
\Wln\ (\SHERPA)			& \SHERPA{2.2.1}\cite{Bothmann:2019yzt,Gleisberg:2008fv}				& \SHERPA{2.2.1}				& NNLO~\cite{Gavin:2010az}					& \SHERPA default~\cite{ATL-PHYS-PUB-2016-003}	& \NNPDF{3.0nnlo}~\cite{Ball:2014uwa}	& \NNPDF{3.0nnlo}~\cite{Ball:2014uwa}\\
\(ttV\)				& \MGNLO{2.3.3}					  					& \PYTHIA{8.210}				& NLO								& A14						& \NNPDF{3.0nlo}			& \NNPDF{2.3lo}\\
\(WW,WZ,ZZ\)			& \POWHEGBOX{v2}~\cite{Nason:2004rx,Frixione:2007vw,Alioli:2010xd}			& \PYTHIA{8.186}				& NLO								& \AZNLO\					& \CT{10nlo}				& \CTEQ{6L1}\\
\hline
\end{tabular}
\end{table}


%
\section{Object selection}
\label{sec:objects}

This section describes the hard object selection for building \ptmiss\ for the performance studies in this paper.
It is emphasised that other ATLAS papers may use different selection requirements to define the hard objects used to reconstruct \ptmiss, which is made possible by the sophisticated software model described in Ref.~\cite{Balunas:2023fpu}.
Photons and hadronically decaying \(\tau\)-leptons (\tauhad) can be included in the \ptmiss\ calculation, as described in Section~\ref{sec:reco}.
However, since this paper focuses on topologies where they aren't featured (to instead focus on the impact of jets, pile-up and fake \ptmiss), they aren't included here.

ID hits are used to reconstruct tracks originating from a particular collision vertex~\cite{PERF-2015-08}.
Both the tracks themselves and the vertices they are associated with must satisfy basic quality requirements to be accepted, detailed in Ref.~\cite{PERF-2015-08}.
Tracks are required to have \(\pt>\SI{400}{\MeV}\).
Vertices are constructed from at least two tracks that satisfy requirements on the transverse impact parameter \(|d_0|<1.5\)\,mm, and for the longitudinal impact parameter \(|z_0\sin{\theta}|<1.5\)\,mm, relative to the candidate vertex.
A requirement is also placed on the number of hits in the ID.
Amongst the primary vertices in a given event, that with the largest sum of \(\pt^2\) of tracks associated with it is defined as the hard-scatter vertex.
Typically, each event has many reconstructed primary vertices (\npv), and so \npv\ can be used as a measure of the amount of pile-up coming from other collisions in the same bunch crossing (in-time pile-up).
In comparison, the average number of interactions per bunch crossing (\(\mu\)) --- averaged over data in a time interval with assumed constant experimental conditions --- relates more to the out-of-time pile-up coming from collisions in neighbouring bunch crossings.

Electrons are reconstructed using calibrated EM calorimeter clusters of energy depositions which are matched to an ID track.
A likelihood-based identification algorithm is built using both the calorimeter and tracking information, as described in Ref.~\cite{EGAM-2018-01}; electrons are required to satisfy the \workingPoint{Tight} Working Point defined therein.
In addition, electrons must have \(\pt>\SI{25}{\GeV}\) and \(|\eta|<1.37 \text{ or } 1.52<|\eta|<2.47\).
To ensure consistency with the hard-scatter vertex, their impact parameters must satisfy \(|d_0|<5.0\)\,mm and \(|z_0\sin{\theta}|<0.5\)\,mm.
Finally, contributions from semileptonic hadron decays and jets misidentified as electrons are minimised by applying \pt-dependent isolation requirements: the \workingPoint{Tight} Working Point is used, as defined in Ref.~\cite{EGAM-2018-01}.

Muon reconstruction combines ID tracks with muon spectrometer (MS) tracks, and requires that muons possess \(\pt>\SI{25}{\GeV}\) and \(|\eta|<2.5\).
The number of hits in the ID and MS sub-detectors --- along with the significance of the charge-to-momentum ratio --- are used to create the muon identification algorithm~\cite{MUON-2018-03}.
Muons must satisfy the \workingPoint{Medium} identification Working Point defined in Ref.~\cite{MUON-2018-03}.
To suppress muons originating from secondary vertices, the muons' transverse impact parameters must satisfy \(|d_0|<3.0\)\,mm and \(|z_0\sin{\theta}|<0.5\)\,mm.
As with electrons, isolation requirements are applied to reduce contributions from semileptonic hadron decays and misidentified jets.
These are defined in Ref.~\cite{MUON-2018-03}, considering for this paper the \workingPoint{Tight\_VarRad} isolation working point.

The default reconstruction algorithm supported for jets in ATLAS is Particle Flow (PFlow)~\cite{PERF-2015-09}.
This combines information from both the calorimeters and ID to provide improved performance compared with reconstructing jets solely from calorimeter information.
A second --- calorimeter-based --- algorithm, EMTopo~\cite{PERF-2014-07} , was previously the default algorithm, and is still used in a few cases such as long-lived particle searches where PFlow's track use is suboptimal.
More details of EMTopo jets, and the modelling and performance that result from using them to build \ptmiss, are given in Appendix~\ref{sec:app_EMTopo}.

Particle Flow jets~\cite{PERF-2015-09} combine ID and calorimeter measurements to reconstruct the energy flow of the event to improve jet energy resolution at low \pt.
Three-dimensional topological clusters (topo-clusters) of calorimeter energy deposits are used.
Tracks are used to calculate an estimate for the momentum in cases when the tracker resolution is better than the calorimeter resolution, avoiding use of calorimeter energy deposits stemming from charged pile-up.
The algorithm produces two kinds of jet constituent objects from the topo-clusters and tracks: charged particle flow objects which each derive primarily from one ID track associated with the hard-scatter vertex, and neutral particle flow objects each derived from a topo-cluster.
The anti-\(k_t\) algorithm~\cite{Cacciari:2008gp} is used with a radius parameter of \(R=0.4\), taking the charged and neutral particle flow objects as inputs.
The algorithm also improves the jet reconstruction efficiency and increases the accuracy of the jet direction in the \((\eta,\phi)\) plane.

Requirements of \(\pt>\SI{20}{\GeV}\) and \(|\eta|<4.5\) are made on the calibrated PFlow jets.
After reconstruction and calibration, PFlow jets with \(\pt<\SI{60}{\GeV}\) and \(|\eta|<2.4\) are filtered further using the Jet Vertex Tagger (JVT) algorithm to select those originating from the hard-scatter, as detailed in Ref.~\cite{PERF-2014-03}.
This tagger is designed to remove pile-up jets in favour of hard-scatter primary vertex jets, with a 96\% efficiency of correctly identifying hard-scatter jets for the requirements chosen here.
The JVT algorithm uses a likelihood discriminant based on observables derived from the tracks matched to each jet, to produce a JVT score ranging from 0 (pile-up-like), to 1 (hard-scatter-like).
These consider, for example, the fraction of \pt\ carried by tracks matched to a given jet that come from the hard-scatter vertex.
PFlow jets are associated to the hard-scatter interaction by requiring a JVT score greater than 0.5.
Jets outside this \pt\ and \(\eta\) range are considered for analysis without extra requirements.

For event selection purposes, a \(b\)-tagging algorithm is applied to jets with \(\pt>\SI{20}{\GeV}\) and \(|\eta|<2.5\), to identify those likely to have originated from a \(b\)-quark.
The DL1 algorithm described in Ref.~\cite{FTAG-2018-01} is used, with a \(77\%\) efficiency working point.

Finally, Section~\ref{sec:reco:workingpoints} will introduce a set of \ptmiss\ working points.
Each places a different selection on the jets entering the \ptmiss\ calculation, which should be considered in addition to the selections described here.


%
\section{Event selection}
\label{sec:events}

Several event topologies are considered in this paper.
For most studies, a \Zee/\Zmm\ selection is used, but \ttbar\ and \Wmunu\ selections are also considered to inspect events with more hadronic activity and real \ptmiss.
Events are removed if they contain at least one jet failing to meet the \workingPoint{BadLoose} criteria defined in Ref.~\cite{ATLAS-CONF-2015-029}.
For all topologies, events require one lepton to match the fired single-lepton trigger, and said lepton is then required to have \(\pt>30~\)\GeV\ to ensure trigger efficiencies have plateaued.
A summary of the event selections described below is also given in Table~\ref{tab:eventSelection}.

The \Zee\ and \Zmm\ topologies are ideal to study fake \ptmiss, since the dominant Drell-Yan process contains no real sources of \ptmiss\ and they have a high production cross-section.
For the \Zee\ (\Zmm) selection, the event must contain exactly two oppositely-charged electrons (muons) and zero muons (electrons) passing the object selection criteria in Section~\ref{sec:objects}.
The invariant mass of the two leptons in the event (\(\mll\)) must be consistent with a decay from a \(Z\) boson by requiring \(|\mll - m_{Z}|<15~\)\GeV.
For the systematic uncertainty calculation this is loosened to \(|\mll - m_{Z}|<20~\)\GeV\ to reduce statistical uncertainties.

To select \ttbar\ events, a semileptonic \ttbar\ decay (one top quark decays hadronically and the other to a muon, neutrino and \(b\)-quark) is targeted to ensure there is real \ptmiss\ in the final state in addition to substantial hadronic activity.
To reduce backgrounds where jets are falsely reconstructed as electrons, which are hard to model with MC simulation, exactly one muon is required and zero electrons.
Events are required to have at least one \(b\)-tagged jet, and at least four jets overall.

\Wmunu\ events are selected by requiring exactly one muon and zero electrons.
The transverse mass\footnote{Transverse mass is defined as \(\mt = \sqrt{ 2 \ptmiss p_{\text{T}}^{\mu}(1-\cos{\phi})} \), with \(\phi\) as the angle between the \ptmiss\ and the muon, and taking the muon to be massless.} of the muon and \ptmiss, which bounds the mass of the decaying \(W\) boson, is required to be at least 40~\GeV.

\begin{table}[h]
\centering
\small
\caption{Kinematic requirements defining the \Zmm, \Zee, \ttbar\ and \Wmunu\ event selections.}
\begin{tabular}{c|c|c|c}
\hline
variable & \Zmm(\Zee) & \ttbar\ & \Wmunu\ \\
\hline
electron multiplicity & 0 (2) & 0 & 0 \\
muon multiplicity & 2 (0) & 1 & 1 \\
triggering lepton \pt\ \([\GeV]\) & \(>30\) & \(>30\) & \(>30\) \\
second lepton \pt\ \([\GeV]\) & \(>20\) & - & - \\
\(|\mll-m_{Z}|\) \([\GeV]\) & performance: \(<15\), systematics: \(<20\) & - & - \\
\mt\ \([\GeV]\) & - & - & \(>40\) \\
jet multiplicity & - & \(\geq4\) & - \\
\(b\)-tagged jet multiplicity & - & \(\geq1\) & - \\
\hline
\end{tabular}
\label{tab:eventSelection}
\end{table}


%
\section{\ptmiss\ reconstruction}
\label{sec:reco}

Missing transverse momentum reconstruction in ATLAS consists of two aspects.
The first, \pthard, comprises \emph{hard-event} signals in the form of reconstructed and calibrated `hard objects': electrons, photons, \(\tau\)-leptons, muons and jets.
The second aspect (\ptsoft) comes from \emph{soft-event} signals, and currently consists of reconstructed charged-particle tracks that are associated with the hard-scatter vertex but not associated with a hard object.

The procedures implemented by ATLAS to transform the set of detector signals for each event into each type of reconstructed hard object are independent.
This implies that the same detector signals could be used multiple times in an event, for example the same calorimeter deposit could be used to reconstruct both an electron and a jet.
When reconstructing \ptmiss, this can cause double counting of contributions to an event's transverse momentum, leading to an artificial momentum imbalance and fake \ptmiss.
This is resolved by the explicit \emph{signal ambiguity resolution} in the object-based \ptmiss\ reconstruction introduced originally in Refs.~\cite{PERF-2011-07}, \cite{PERF-2014-04} and \cite{PERF-2016-07} and described in Section~\ref{sec:reco:ambiguity}.
Ultimately, the \ptmiss\ is built from a set of mutually exclusive detector signals.

\subsection{\ptmiss\ introduction}
\label{sec:reco:basics}

The reconstruction of missing transverse momentum builds a set of observables from the 2-dimensional transverse momentum vectors (\(\buvec{p}_{\text{T}} = (p_{x},p_{y})\)) of the various event constituents.
The missing transverse momentum vector \(\ptmissvec = (\pxmiss,\pymiss)\) is the first of these observables, and is given by:
\begin{equation}\label{eq:reco:underbraceptmiss}
\ptmissvec = - \Bigg( \underbrace{ \sum_{\substack{\text{selected}\\\text{electrons}}} \ptvecX{e} + \sum_{\substack{\text{accepted}\\\text{photons}}} \ptvecX{\gamma} + \sum_{\substack{\text{accepted}\\\tau - \text{leptons}}} \ptvecX{\tau} + \sum_{\substack{\text{selected}\\\mu}} \ptvecX{\mu} + \sum_{\substack{\text{accepted}\\\text{jets}}} \ptvecX{\text{jet}} }_{\text{hard term}} + \underbrace{\sum_{\substack{\text{unused}\\\text{tracks}}} \ptvecX{\text{track}}}_{\text{soft term}} \Bigg).
\end{equation}

The second is the scalar sum of all transverse momenta (\(\pt = |\buvec{p}_\text{T}|\)) of the \ptmiss\ reconstruction constituent objects, which is given by
\begin{equation}\label{eq:reco:underbrace}
\sumpt = \underbrace{ \sum_{\substack{\text{selected}\\\text{electrons}}} \ptX{e} + \sum_{\substack{\text{accepted}\\\text{photons}}} \ptX{\gamma} + \sum_{\substack{\text{accepted}\\\tau - \text{leptons}}} \ptX{\tau} + \sum_{\substack{\text{selected}\\\mu}} \ptX{\mu} + \sum_{\substack{\text{accepted}\\\text{jets}}} \ptX{\text{jet}} }_{\text{hard term}} + \underbrace{\sum_{\substack{\text{unused}\\\text{tracks}}} \ptX{\text{track}}}_{\text{soft term}}.
\end{equation}
This quantity is useful to calculate in addition to \ptmissvec.
It presents an overall scale for evaluating the hardness of a hard-scatter event in the transverse plane, thus providing a measure of the event activity in physics analyses and \ptmiss\ reconstruction performance studies.

In both the \ptmissvec\ and \sumpt\ definitions, the selected hard objects are chosen by the user, and allow the interpretation of each event to be consistent in a given analysis.
The object selections used specifically for the performance studies in this paper were described in Section~\ref{sec:objects}.
Each reconstructed particle and jet has its own dedicated calibration translating detector signals into a fully corrected four-momentum.
Therefore, for example, rejecting certain electrons in a given analysis can change both the \ptmissvec\ and \sumpt, if the corresponding calorimeter signal is included and is calibrated as a jet or a significant part of a jet.
This also means that systematic uncertainties for the different particles can be consistently propagated into \ptmissvec.
In Eqs.~(\ref{eq:reco:underbraceptmiss}) and~(\ref{eq:reco:underbrace}), the term \emph{selected}, only applicable to electrons and muons, means that the choice of reconstructed particles is given purely by a set of analysis-chosen criteria.
On the other hand, \emph{accepted} implies that the initially selected set of particles has been potentially modified by the signal ambiguity resolution procedure (described in Section~\ref{sec:reco:ambiguity}) or added requirements placed on jets in a given \ptmiss\ `working point' (see Section~\ref{sec:reco:workingpoints}).

The phrase `unused tracks' in Eqs.~(\ref{eq:reco:underbraceptmiss}) and (\ref{eq:reco:underbrace}) refers to ID tracks associated with the hard-scatter vertex but not with any hard object added to the \ptmissvec\ sum.
These are used to calculate the soft-event signal, \ptsoft, as discussed in more detail in Section~\ref{sec:reco:soft}.
As seen in the formulae, observables are also constructed individually for each `term' of \ptmissvec\ coming from each object type.

As part of the signal ambiguity resolution procedure, an ordered sequence is defined for prioritising adding contributions to the \ptmissvec\ sum, following the order of terms in Eq.~(\ref{eq:reco:underbraceptmiss}).
This is explained in detail in Section~\ref{sec:reco:ambiguity}.

Other observables reconstructed from \pxymiss\ include:
\begin{equation*}\label{eq:reco:ptmiss}
\ptmiss = |\ptmissvec| = \sqrt{(\pxmiss)^2 + (\pymiss)^2}\hspace{2mm}, \text{and}
\end{equation*}
\begin{equation*}\label{eq:reco:phimiss}
\phimiss = \begin{cases}
\tan^{-1}{(\pymiss/\pxmiss)} & \text{if}~\pxmiss>0 \\
\tan^{-1}{(\pymiss/\pxmiss)} + \pi & \text{if}~\pxmiss<0~\text{and}~\pymiss\geq0 \\
\tan^{-1}{(\pymiss/\pxmiss)} - \pi & \text{if}~\pxmiss<0~\text{and}~\pymiss<0 \\
\text{indeterminate} & \text{if}~\pxmiss=0~\text{and}~\pymiss=0 \\
\frac{y\pi}{|y|2} & \text{if}~\pxmiss=0~\text{and}~\pymiss\neq0.
\end{cases}
\end{equation*}
The magnitude of the \ptmissvec\ vector gives the amount of missing transverse momentum; its direction in the transverse plane, or azimuthal angle, is given by \phimiss.

Finally, the truth (generator level) \ptmiss\ in MC simulations, \ptmisstrue (magnitude of the 2-dimensional vector \ptmisstruevec), is often used in performance studies.
This is defined by the vector sum of transverse momenta of stable, invisible particles produced in the final state at generator (hadron) level.

\subsection{Object association}
\label{sec:reco:association}

The \ptmissvec\ reconstruction sum and the signal ambiguity resolution procedure rely on knowing which hard objects each track, topo-cluster and particle-flow object in an event are associated with, in order to determine where there is overlap that must be resolved.
Full details of this initial object association procedure, and the sophisticated software used to implement it, are detailed in Ref.~\cite{Balunas:2023fpu}.
Specifics of the ATLAS \runtwo\ workflow to initialise object associations for \ptmiss\ reconstruction before applying the signal ambiguity resolution procedure are given here.

The \ptmiss\ reconstruction algorithm considers the same original ID tracks to be associated with a muon as the muon reconstruction algorithm~\cite{MUON-2018-03} (with the track momentum taken from the combination of the ID and MS tracks).
Topo-clusters, or neutral particle-flow objects, are only considered to be associated with a muon if it is likely they are a result of the muon's calorimeter energy-loss.
A ``muon cluster'' is defined from the calorimeter cells crossed by the muon track, and if the total energy this shares with a given topo-cluster exceeds 20\%, then the topo-cluster is deemed to be associated with the muon.
ID tracks associated with electrons and photons during their reconstruction~\cite{EGAM-2018-01} are again considered associated for \ptmiss\ reconstruction.

Clusters used in electron and photon reconstruction are not the same as the topo-clusters used for jet reconstruction.
However, they are derived from them, and thus can be matched to them\footnote{A topo-cluster associated with a jet is also associated with a given electron if its matched electron cluster is associated with said electron.}.
For topo-clusters within \(\Delta R<0.1\) of an electron/photon cluster, the subset of \(N\) topo-clusters best matching the electron/photon cluster energy are chosen, in order to avoid spurious matches.
This `best-matching' procedure is ordered in decreasing \pt topo-clusters, and considers topo-cluster \(i\) energy \(E_{\text{topo},i}\) and electron/photon cluster energy \(E_{\text{e}/\gamma}\).
If \(E_{\text{topo},i}<1.5\times E_{\text{e}/\gamma}\), and if \(|\sum_{i=1}^{n}E_{\text{topo},i}-E_{\text{e}/\gamma}|<|\sum_{i=1}^{n-1}E_{\text{topo},i}-E_{\text{e}/\gamma}|\) for the \(n\leq N\) topo-clusters so far considered, then topo-cluster \(i\) is associated with the electron/photon.
If no topo-clusters have \(E_{\text{topo},i}<1.5\times E_{\text{e}/\gamma}\) then only the topo-cluster with energy closest to the electron/photon is associated with it.

Neutral PFlow objects are associated with electrons and photons using the same procedure as topo-clusters.
Charged PFlow objects are constructed from an ID track and inherit their associations from this track.

Hadronically-decaying \(\tau\)-leptons are associated with topo-clusters and tracks when they are reconstructed (more detail can be found in Ref.~\cite{ATLAS-CONF-2017-029}).
If using particle flow for \ptmiss, the calibration of topo-clusters may be different to the \(\tau\)-lepton reconstruction and so are considered associated with the \(\tau\)-lepton if they are within \(\Delta R<0.2\) of the \(\tau\)-lepton's seeding-jet axis.

\subsection{Signal ambiguity resolution}
\label{sec:reco:ambiguity}

The previous section defined which tracks, topo-clusters and particle flow objects are initially associated with which hard objects.
This section explains how that information is used to decide which objects to add to the \ptmissvec\ sum in cases where a hard object shares one of these detector signals with another (they overlap).

Electrons enter \ptmiss\ reconstruction as the highest priority object, so are never modified from the analysis selection criteria.
If lower-priority particles (\(\gamma\) then \tauhad) share an ID track, topo-cluster or particle flow object with a higher-priority object that has already entered \ptmissvec\ reconstruction, they are fully rejected from their term in the \ptmissvec.
In this case, their tracks can be used in the \ptsoft.

Muons experience energy loss in the calorimeters, but only non-isolated muons overlap with other objects, most probably jets or \(\tau\)-leptons.
In this case the muon calorimeter energy deposit cannot be separated from the overlapping jets with the required precision, and a more sophisticated treatment of when to reject a jet is needed.
This is discussed in Section~\ref{sec:reco:muonjetOR}.
As indicated by the `selected' notation in Eq.~(\ref{eq:reco:underbraceptmiss}), muons (like electrons) are never modified from the analysis selection criteria.

Jets can also be rejected if they overlap with other accepted higher-priority particles.
In the case of partial or marginal overlap between jets and electrons or photons, signal losses are minimised by applying a more refined overlap removal strategy, as described in Section~\ref{sec:reco:egammajetOR}.

\subsubsection{Muon overlap with jets}
\label{sec:reco:muonjetOR}

There are several scenarios leading to the signal overlap of reconstructed muons and jets.
If a muon overlaps with a pile-up-originating jet, it can lead to the jet being falsely considered as a hard-scatter jet.
This is because the muon's ID track represents a significant amount of hard-scatter vertex \pt, thus increasing the JVT value and making a pile-up jet more likely to satisfy any JVT requirements.
In this case the pile-up jet \pt\ contributes to \ptmiss, where its stochastic nature degrades the \ptmiss\ response and resolution\footnote{Here response is defined as the deviation of the observed \ptmiss\ from its expected value. Resolution of \ptmiss\ considers the root-mean-squared (RMS) width of both the \pxmiss\ and \pymiss.}.

Muon energy deposited in the calorimeter (\Eloss) can also be reconstructed as a hard-scatter primary vertex jet, which will be found in close proximity to the muon's associated ID track.
Because the muon \Eloss\ is already corrected for in the muon \pt\ reconstruction, inclusion of such a jet to the \ptmiss\ reconstruction double-counts it.
Rejection of pile-up jets and muon \Eloss\ jets is achieved through consideration of the four selection criteria.
First, the muon's track is `ghost'-associated with the jet using the anti-\(k_t\) algorithm.
Second, \(\pt^{\mu\text{-ID}}/\pt^{\text{jet-ID}}\) is required to be larger than 0.8: the transverse momentum of the muon's track (\(\pt^{\mu\text{-ID}}\)) represents a significant fraction of the sum of transverse momenta of all hard-scatter primary vertex ID track associated with the jet (\(\pt^{\text{jet-ID}}\)).
Third, the transverse momentum of the jet (\(\pt^{\text{jet}}\)) is less than twice the \(\pt^{\mu\text{-ID}}\).
Finally, the total number of hard-scatter primary vertex tracks associated with the jet (\(N_{\text{track}}^{\text{PV}}\)) is less than five.
If a jet with an overlapping muon meets all of these criteria, it is considered to be either from pile-up or a catastrophic muon \Eloss\ and is rejected from \ptmiss\ reconstruction.

Final state radiation (FSR) can also affect muon contributions to \ptmiss.
Muons can radiate photons at small angles, typically too close to the muon ID track for the photon to be reconstructed.
The mismatch between calorimeter energy and ID track momentum also prevents the FSR photon being reconstructed as an electron.
Instead, the FSR's calorimeter signal is reconstructed as a jet with an associated muon ID track.
The FSR photon's transverse momentum is not recovered in muon reconstruction, hence jets representing this photon must be included in the \ptmiss\ reconstruction.
These jets are characterised by the following selections, which typically indicate photons.
\begin{itemize}
\item The muon's associated ID track is `ghost'-associated with the jet using the anti-\(k_T\) algorithm;
\item \(N_{\text{track}}^{\text{PV}}<3\) --- the jet has a small number of tracks from the hard-scatter primary vertex;
\item \(f_{\text{EM}}=E_{\text{jet}}^{\text{EM}}/E_{\text{jet}}>0.9\) --- the jet energy \(E_{\text{jet}}\) is primarily deposited in the EM calorimeter, as expected for photons;
\item \(\pt^{\text{jet PS}}>\SI{2.5}{\GeV}\) --- an early starting point for the shower is selected by requiring a large transverse momentum contribution of the jet in the presampler (PS) calorimeter;
\item \(w_{\text{jet}}<0.1\) --- the jet width \(w_{\text{jet}}\) is comparable to a dense electromagnetic shower, where jet width is defined as:
\begin{equation*}
w_{\text{jet}} = \frac{\sum_{i}\Delta R_{i}p_{\text{T},i}}{\sum_{i}p_{\text{T},i}},
\end{equation*}
the angular distance of topo-cluster \(i\) from the jet axis is \(\Delta R = \sqrt{(\Delta \eta_{i})^{2}+(\Delta \phi_{i})^{2}}\) and \(p_{\text{T},i}\) is the cluster's transverse momentum;
\item \(\pt^{\text{jet-ID}}/\pt^{\mu \text{-ID}}>0.8\) --- the transverse momentum of hard-scatter primary vertex tracks associated with the jet is close to the muon ID track transverse momentum.
\end{itemize}

If a jet meets all of the above criteria, it is deemed to be an FSR photon and is included in the \ptmiss\ reconstruction in the jet term.
The energy scale of the jet is set to the EM scale to reflect its interpretation as a photon, and further scaled both to remove the fraction of the \pt\ overlapping with the reconstructed muon and the muon energy loss in the calorimeter.

Muons and jets can also overlap if a muon is produced in the decay of a heavy-flavour hadron, inside a jet.
In this case, both the muon and jet should be kept for the \ptmiss\ calculation, but any double-counting removed.
If a jet and muon overlap and, after the checks detailed above, the jet is not deemed to be an FSR photon or rejected from the calculation, it is kept in the \ptmiss\ jet term.
Similar to the case above, its momentum is scaled, both to remove the fraction of the \pt\ overlapping with the reconstructed muon and the muon energy loss in the calorimeter.
The overlapping muon is added to the \ptmiss\ reconstruction without any adjustment.

\subsubsection{Electron/Photon overlap with jets}
\label{sec:reco:egammajetOR}

In the case where electrons/photons overlap with a jet, two discriminating variables are used to establish whether the jet should be treated as real and enter the \ptmiss\ calculation.
These use the energy and \pt\ of the jet and electron or photon, calibrated at the EM scale.

The first variable is the ratio \foverlap, the ratio of the electron (or \(\gamma\) or \tauhad) energy \EegtEM to the jet energy \EjetEM:
\begin{equation*}\label{eq:foverlap}
\foverlap = \frac{\EegtEM}{\EjetEM}.
\end{equation*}

The second variable represents the unique \pt\ of the jet, \jetptunique, which is defined thus:
\begin{equation*}\label{eq:jetptunique}
\jetptunique = \pt^{\text{EM,jet}} - \pt^{\text{EM},e(\gamma,\tau)}.
\end{equation*}

In the scenario where a jet shares an ID track with a high momentum electron (\(\pt > \SI{90}{\GeV}\)), and carries a large amount of \pt\ from tracks not associated with other objects (\((\sum_{i=1}^{n}p_{\text{T}}^{\text{track},i} - \sum_{j=1}^{m}p_{\text{T}}^{\text{track},j}) < \SI{10}{\GeV}\) for a jet with \(n\) associated tracks, \(m\) of which are non-overlapping) then it is likely that both the electron and jet are real and should be treated as such in the \ptmiss.
These requirements can be encapsulated in a boolean variable \texttt{KeepJet}, which is always false for jet-photon overlaps since photons have no associated tracks.

To treat the jet as real and include it in the \ptmiss\ jet term along with the (higher priority) electron/photon, \((\foverlap<1.0~\text{or}~\texttt{KeepJet})\) and \(\jetptunique>\SI{20}{\GeV}\) are required.
To avoid any double-counting the jet \pt\ is scaled by (\(1-\foverlap\)) if it is included in the \ptmiss\ jet term.

\subsection{\ptmiss\ working points}
\label{sec:reco:workingpoints}

When reconstructing \ptmiss, the requirements on jets entering the calculation have a large impact on performance.
More stringency leads to a reduction in contamination from pile-up and jet mismeasurement, however it also leads to an increased likelihood of excluding real and well-measured jets.
In different use cases, the optimal stringency can be different; thus ATLAS recommends several working points for analysers to choose from.

The requirements placed on jets for a given \ptmiss\ working point act in addition to those chosen by the analysis.
If jets are rejected from \pthard\ by working point requirements, their tracks are not added to the soft term because the jet is deemed to have originated from pile-up.
Four working points are supported, as illustrated in Table~\ref{tab:MET_WPs} in order of increasing stringency.
fJVT is the forward Jet Vertex Tagger (fJVT), described in Ref.~\cite{PERF-2016-06}, used to remove pile-up jets with \(2.5<|\eta|<4.5\) and \(20<\pt<\SI{50}{\GeV}\).
The fJVT uses the angular kinematics of other jets in the event to associate forward jets --- which lack tracking information --- to pile-up vertices by minimising the other vertices' reconstructed \ptmiss.

The main change in jet selection as the working point is tightened is increasing the \pt\ threshold for jets in the forward \(\eta\) region of the detector.
In this region, pile-up jets (which tend to have a lower \pt\ than hard-scatter jets) are more commonly found.
Different JVT selections are also used to remove pile-up jets.
The \workingPoint{Tenacious} working point takes an aggressive approach, using a very tight JVT requirement for low \pt\ jets.

\begin{table}[h]
\centering
\small
\caption{Selections for the \ptmiss\ working points supported for PFlow jets.}
\begin{tabular}{ l| l l l l }
\hline
& \multicolumn{4}{|c}{Selections} \\
& \multicolumn{2}{|c}{\(\pT\ [\GeV]\) for } & & fJVT for jets with \\
& \multicolumn{2}{|c}{jets with:} & JVT for jets with & \(2.5<|\eta|<4.5\) \& \\
Working point	& \(|\eta|<2.4\)	& \(2.4 < |\eta| < 4.5\)	& \(|\eta|<2.4\)	& \(\pt<120~\)\GeV \\
\hline
\workingPoint{Loose}	& \(>20\)	& \(>20\)	& \(>0.5\text{ for } \pt<\SI{60}{\GeV}\)	& - \\
\workingPoint{Tight}	& \(>20\)	& \(>30\)	& \(>0.5\text{ for } \pt<\SI{60}{\GeV}\)	& - \\
\workingPoint{Tighter}	& \(>20\)	& \(>35\)	& \(>0.5\text{ for } \pt<\SI{60}{\GeV}\)	& - \\
\workingPoint{Tenacious}	& \(>20\) & \(>35\)	& \(>0.91\text{ for } 20<\pt<\SI{40}{\GeV}\)	& \(<0.5\) \\
& & & \(>0.59\text{ for } 40<\pT<\SI{60}{\GeV}\) & \\
& & & \(>0.11\text{ for } 60<\pT<\SI{120}{\GeV}\) & \\
\hline
\end{tabular}
\label{tab:MET_WPs}
\end{table}

\subsection{\ptmiss\ soft term}
\label{sec:reco:soft}

The current soft term reconstruction approach exclusively uses hard-scatter vertex ID-tracks, and so only includes the \pt\ from charged soft particles.
However, this choice ensures that the soft term has a high resilience to pile-up contamination.
The inclusion of the soft term into the \ptmiss\ improves the \ptmiss\ resolution and agreement with truth \ptmiss.
It also improves the \ptmiss\ scale, which is defined in Section~\ref{sec:perf} and partly measures how well the \ptmiss\ accounts for the hadronic recoil in an event.
The soft term particularly improves the scale in events with a low multiplicity of hard objects, by capturing components of the event that are not represented by reconstructed and calibrated objects and would thus otherwise be ignored.

Tracks are required to satisfy the requirements described in Section~\ref{sec:objects}.
If tracks are not associated with any hard object in the event, then they are used to built the \ptsoft.
Contributions to \ptsoft\ also come from ID tracks associated with jets that have been rejected by the signal ambiguity resolution procedure, but not ID tracks associated with jets that were rejected by the working point cuts (since these are deemed to originate from pile-up).
ID tracks are also vetoed from inclusion in the \ptsoft\ if any of the following signal-overlap resolution requirements are met:
\(\Delta R (\text{track}, e/\gamma~\text{cluster})<0.05\);
\(\Delta R (\text{track}, \tau-\text{lepton})<0.2\);
the track is associated with a muon or is ghost-associated with contributing jet.

Alternative calorimeter-based soft term definitions have been used in the past~\cite{PERF-2016-07}.
These benefit from the inclusion of neutral soft particles, but are very susceptible to pile-up contamination.
Due to the higher-pile-up conditions of \runtwo, these aren't currently supported, as the track-based soft term was found to provide a better resolution and general agreement with truth.
However, they may be revisited in the future.


%
\section{Modelling and performance of \ptmiss}
\label{sec:performance}

\subsection{\ptmiss\ modelling in MC simulation and data}
\label{sec:modelling}

To assess the modelling of \ptmiss, comparisons between data and MC simulation are made for several variables.
Events must satisfy either a \Zmm\ or \Zee\ selection, as defined in Section~\ref{sec:events}, using objects selected according to Section~\ref{sec:objects}.
By default, PFlow jets are used to build \ptmiss\ using the \workingPoint{Tight} working point.
Unless otherwise stated, the \Zll\ MC events are generated using \SHERPA.

After looking at this default configuration, the modelling is studied when using different \ptmiss\ working points, jet collections, and \Zll\ MC generators in turn.
The uncertainty bands on the SM MC contributions are formed from a quadrature sum of the MC statistical uncertainty, luminosity uncertainty and relevant detector uncertainties.
Detector uncertainties include those on the \ptmiss\ soft term (discussed in Section~\ref{sec:systs}); lepton reconstruction efficiency, energy scale and resolution, and trigger efficiency differences between MC simulation and data~\cite{EGAM-2018-01,MUON-2018-03}; uncertainties in the jet-energy scale and resolution~\cite{JETM-2018-05}; JVT efficiencies~\cite{PERF-2014-03}; and uncertainties in the pile-up profile used for the MC events.
It is emphasised that systematic uncertainties on the MC modelling and cross-sections are not included.

Figure~\ref{fig:termsZmmZee} shows the overall \ptmiss\ distribution, the hard and soft terms, for \Zmm\ and \Zee\ selections.
The plots show a `jet inclusive' selection, where no additional requirements are placed on the jets in the event beyond those described in previous sections.
The \ptmiss\ and \pthard\ distributions show very good agreement between MC simulation and data within uncertainties.
The dominant systematic uncertainties leading to the bump seen around \SI{100}{\GeV} comes from the jet energy scale and resolution.
In Figure~\ref{fig:termsZmm:soft}, the soft term shows a slight excess in data in the tails, expected to be covered if modelling uncertainties were included on the MC.

\begin{figure}
\centering
\subfloat[]{\label{fig:termsZmm:ptmiss}\includegraphics[width=0.35\textwidth]{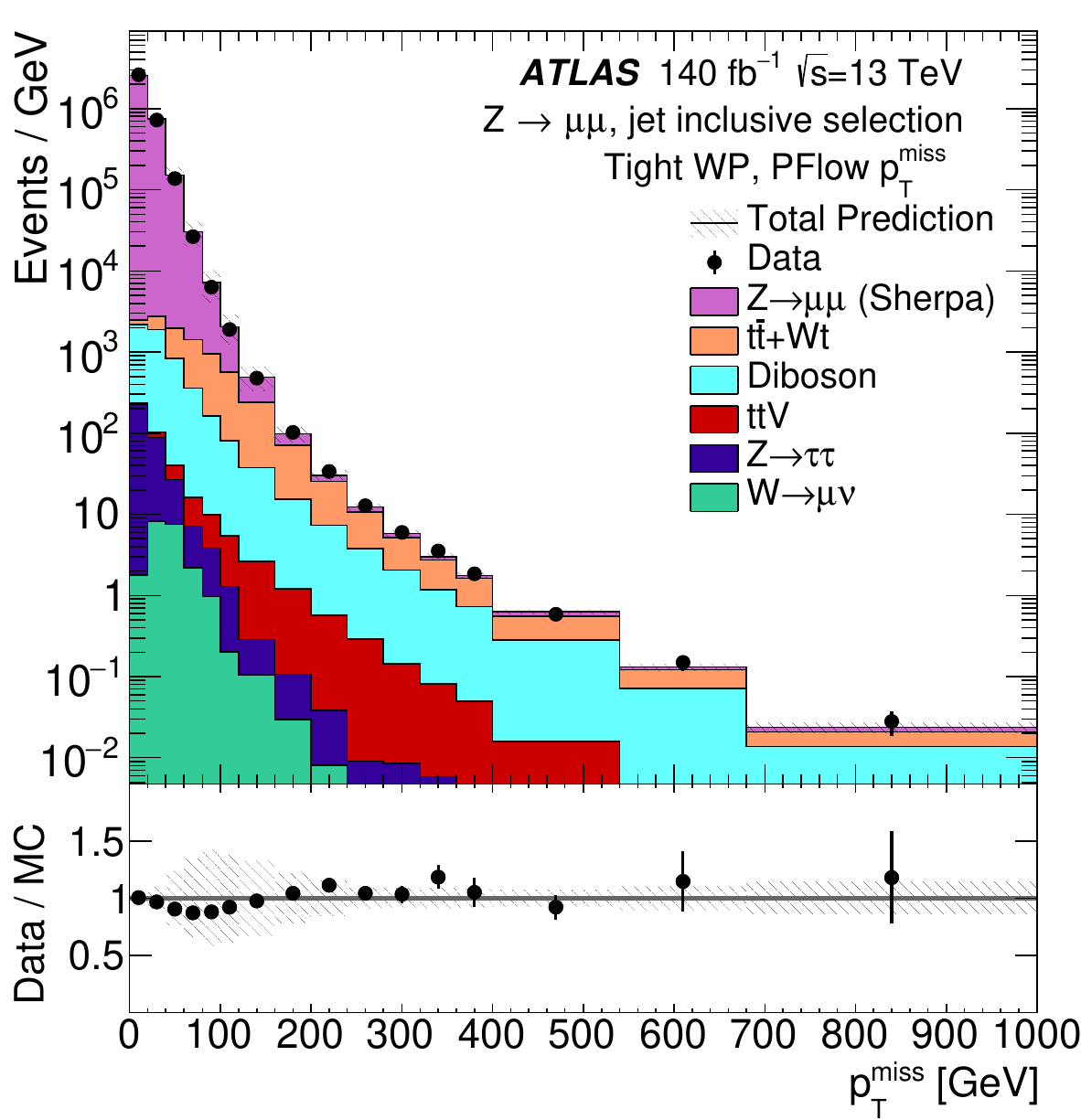}}\qquad
\subfloat[]{\label{fig:termsZee:ptmiss}\includegraphics[width=0.35\textwidth]{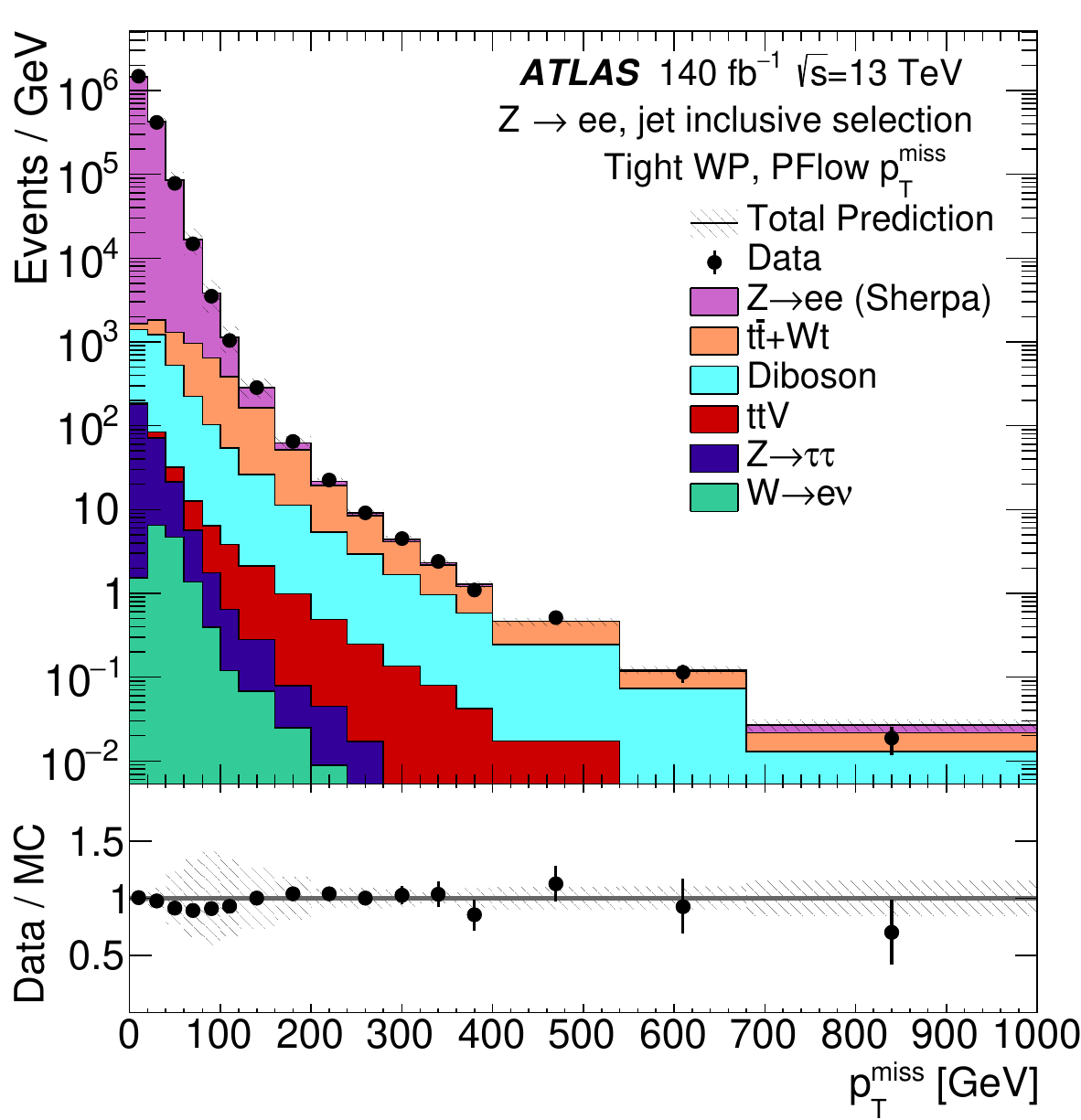}}\\
\subfloat[]{\label{fig:termsZmm:hard}\includegraphics[width=0.35\textwidth]{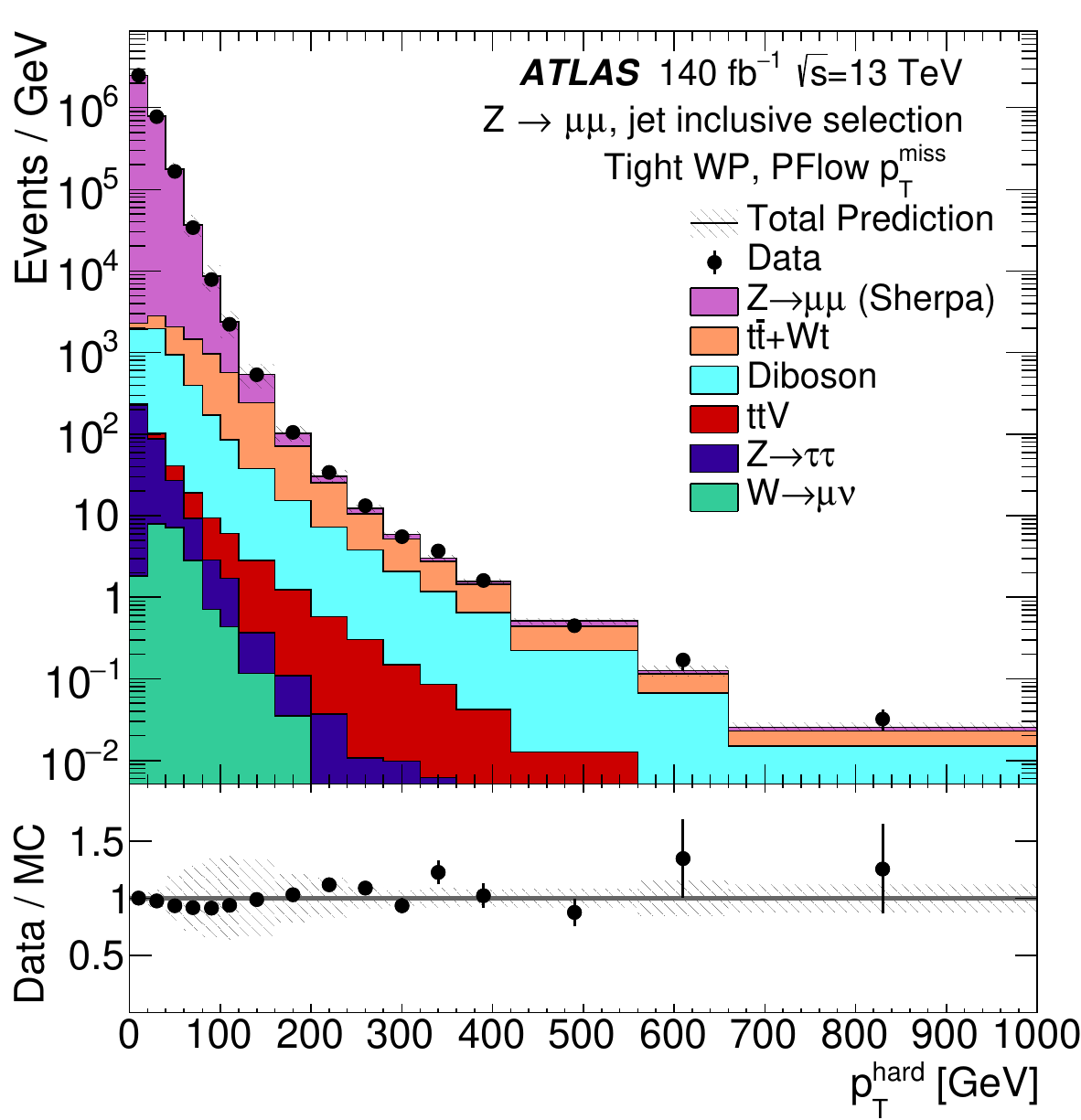}}\qquad
\subfloat[]{\label{fig:termsZee:hard}\includegraphics[width=0.35\textwidth]{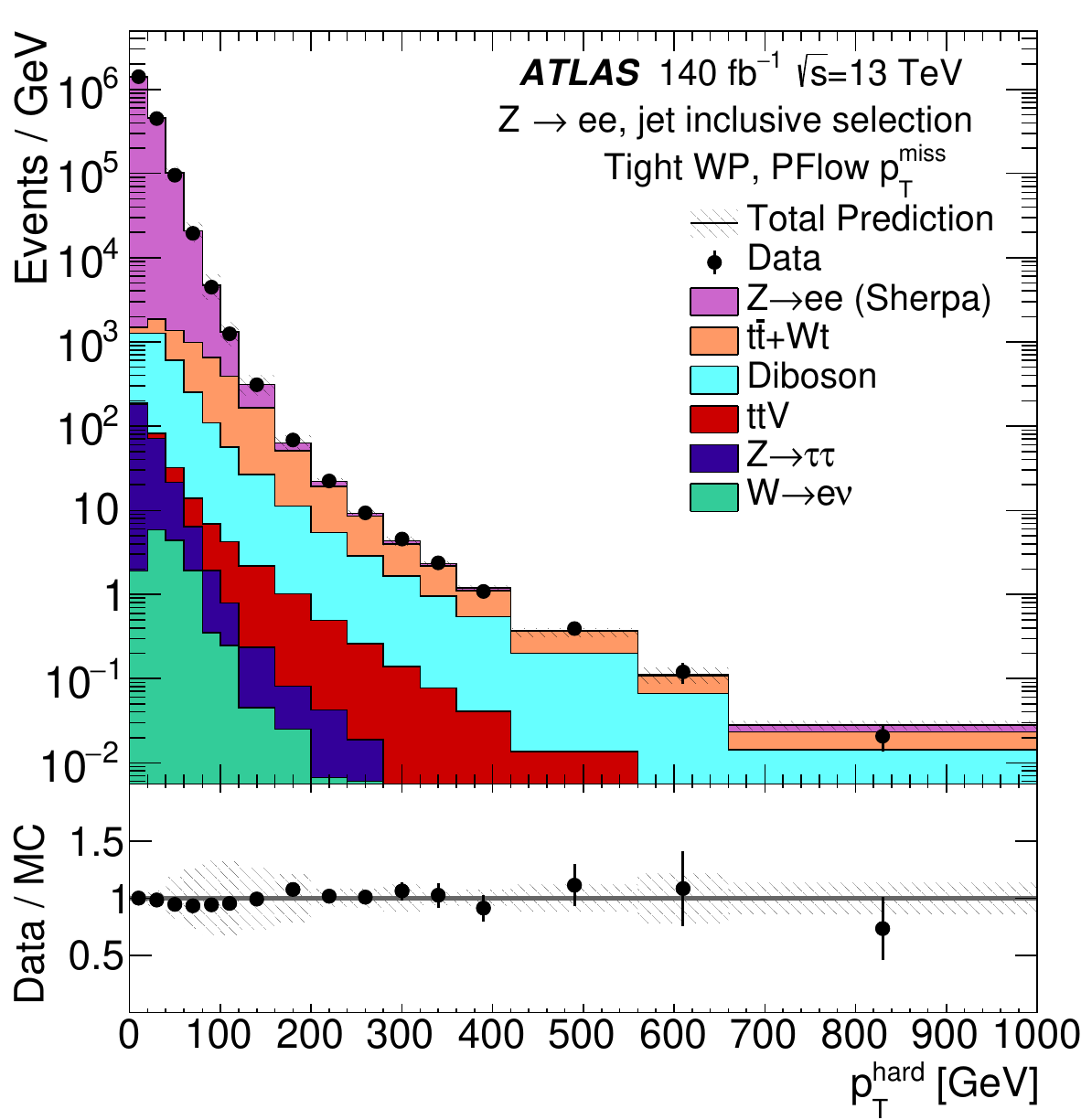}}\\
\subfloat[]{\label{fig:termsZmm:soft}\includegraphics[width=0.35\textwidth]{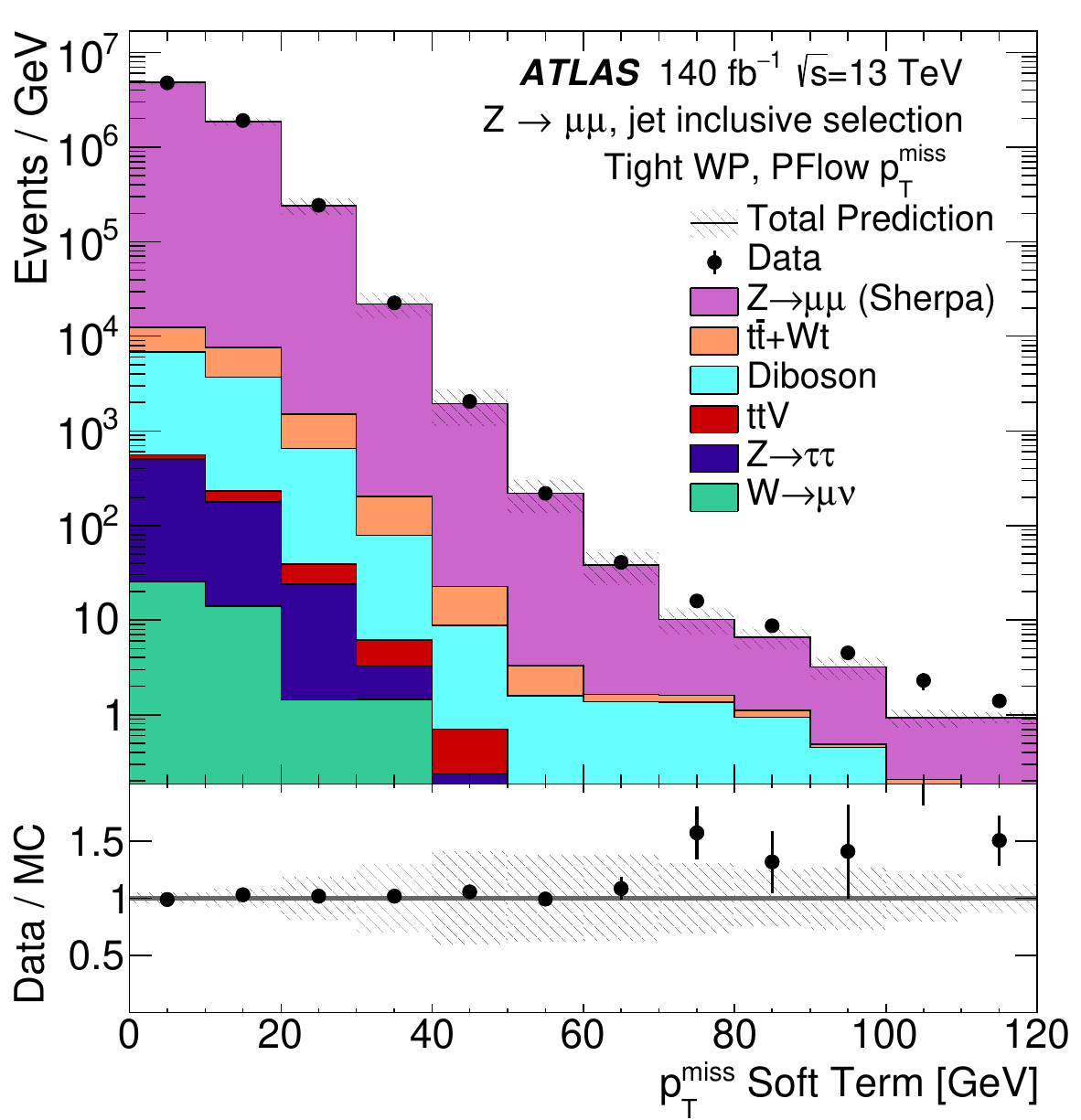}}\qquad
\subfloat[]{\label{fig:termsZee:soft}\includegraphics[width=0.35\textwidth]{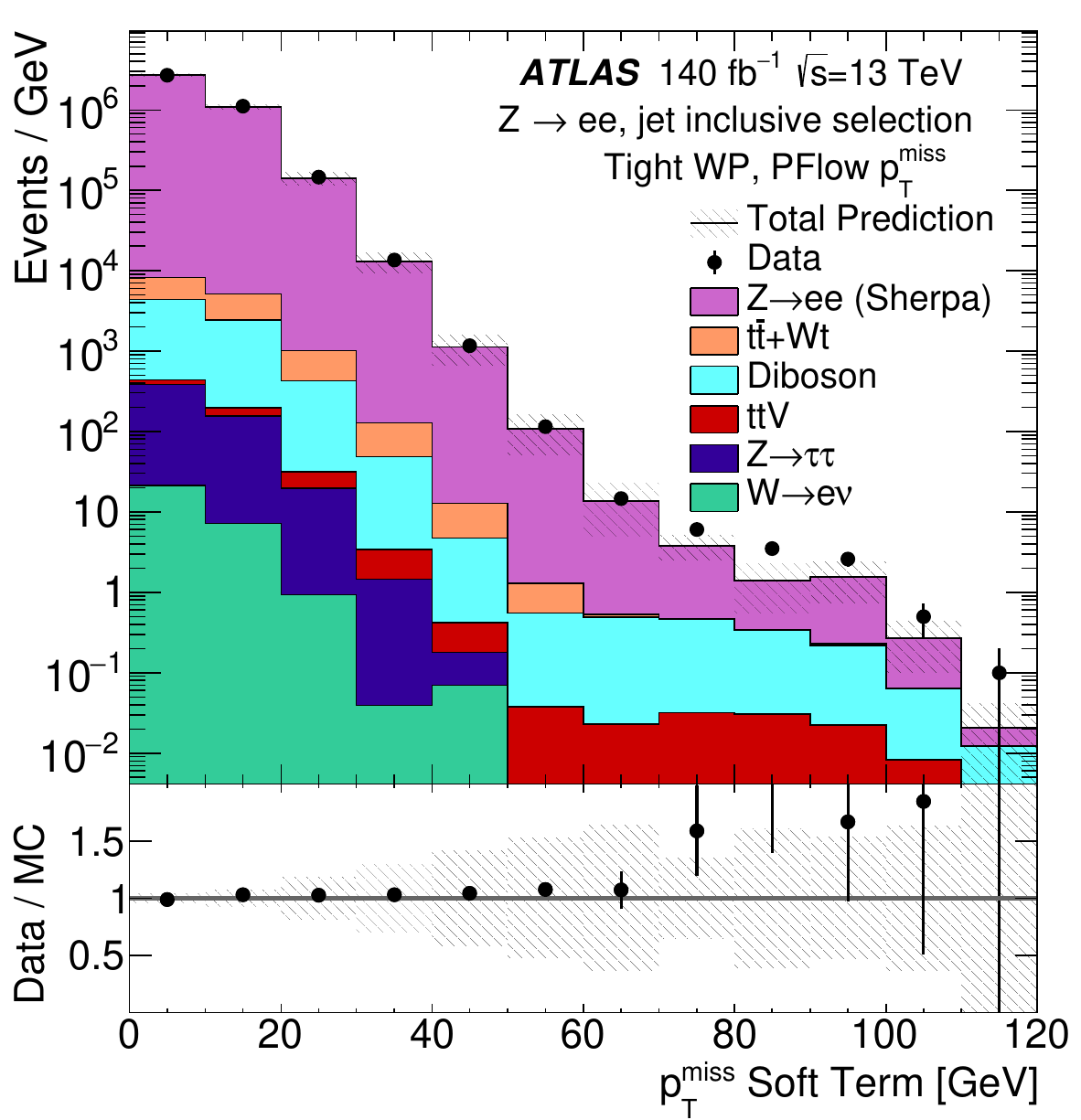}}
\caption{Distributions of \ptmiss\ ((a) and (b)) and its constituent hard ((c) and (d)) and soft ((e) and (f)) terms in MC simulation and data. Events satisfy a \Zmm\ ((a), (c) and (e)) or \Zee\ ((b), (d) and (f)) selection. PFlow jets are used with a jet inclusive selection, and the \workingPoint{Tight} \ptmiss\ working point. \SHERPA\ is used to generate the \Zee/\Zmm\ events. The error band includes MC statistical, luminosity and detector uncertainties.}
\label{fig:termsZmmZee}
\end{figure}

Figure~\ref{fig:wpsZmm} shows the \ptmiss\ distributions for the \workingPoint{Loose}, \workingPoint{Tighter} and \workingPoint{Tenacious} working points.
As the working point is tightened from \workingPoint{Loose} to \workingPoint{Tenacious}, the modelling improves, due primarily to the removal of low \pt\ forward jets.
These have relatively large uncertainties in the jet energy resolution, stemming partly from large pile-up contamination.
The error band decreases with the tightening working point, which is caused by a large reduction in the impact of jet energy resolution uncertainties.

\begin{figure}
\centering
\subfloat[]{\label{fig:wpsZmm:PFlowLoose}\includegraphics[width=0.4\textwidth]{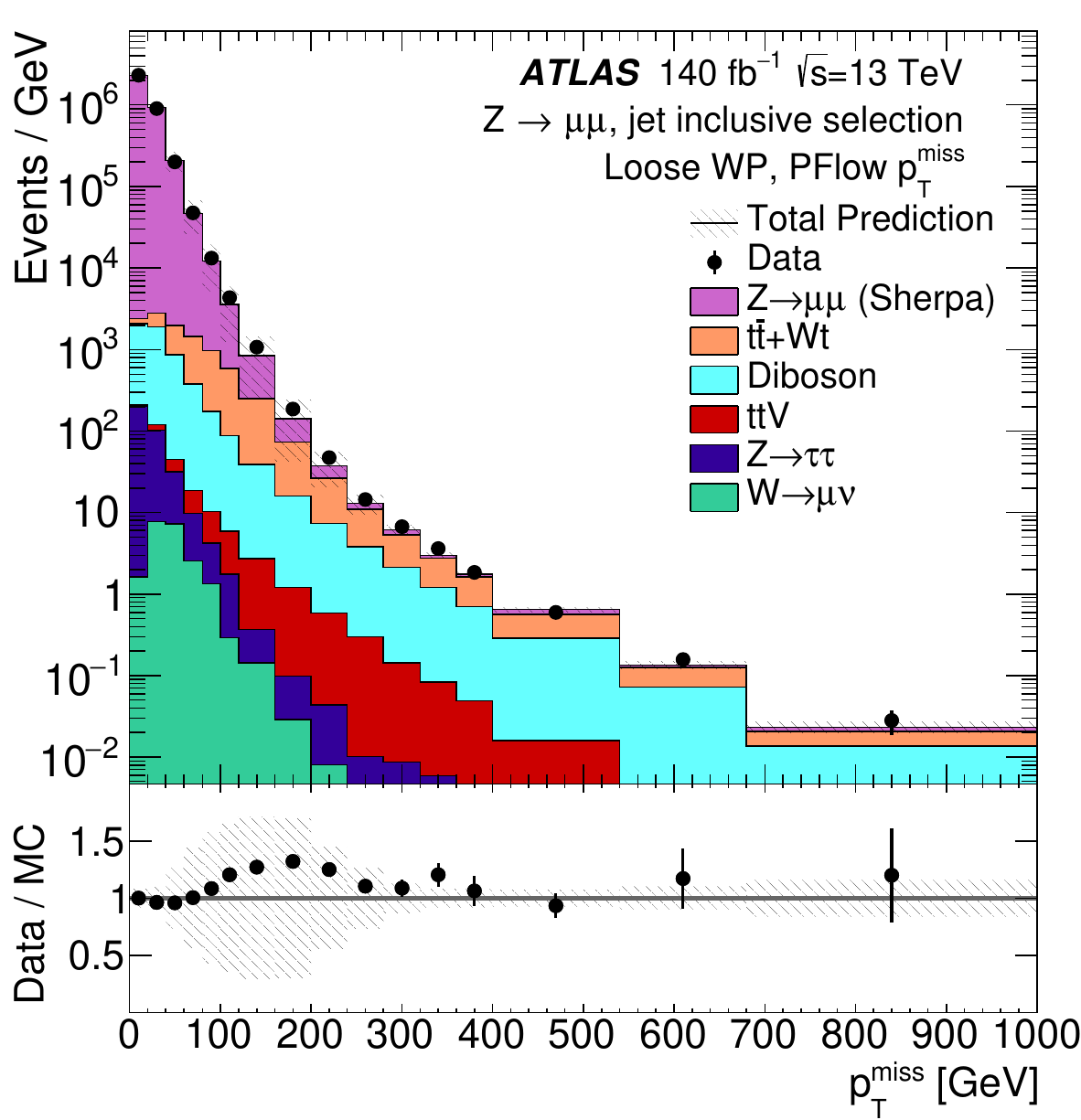}}\qquad
\subfloat[]{\label{fig:wpsZmm:PFlowTighter}\includegraphics[width=0.4\textwidth]{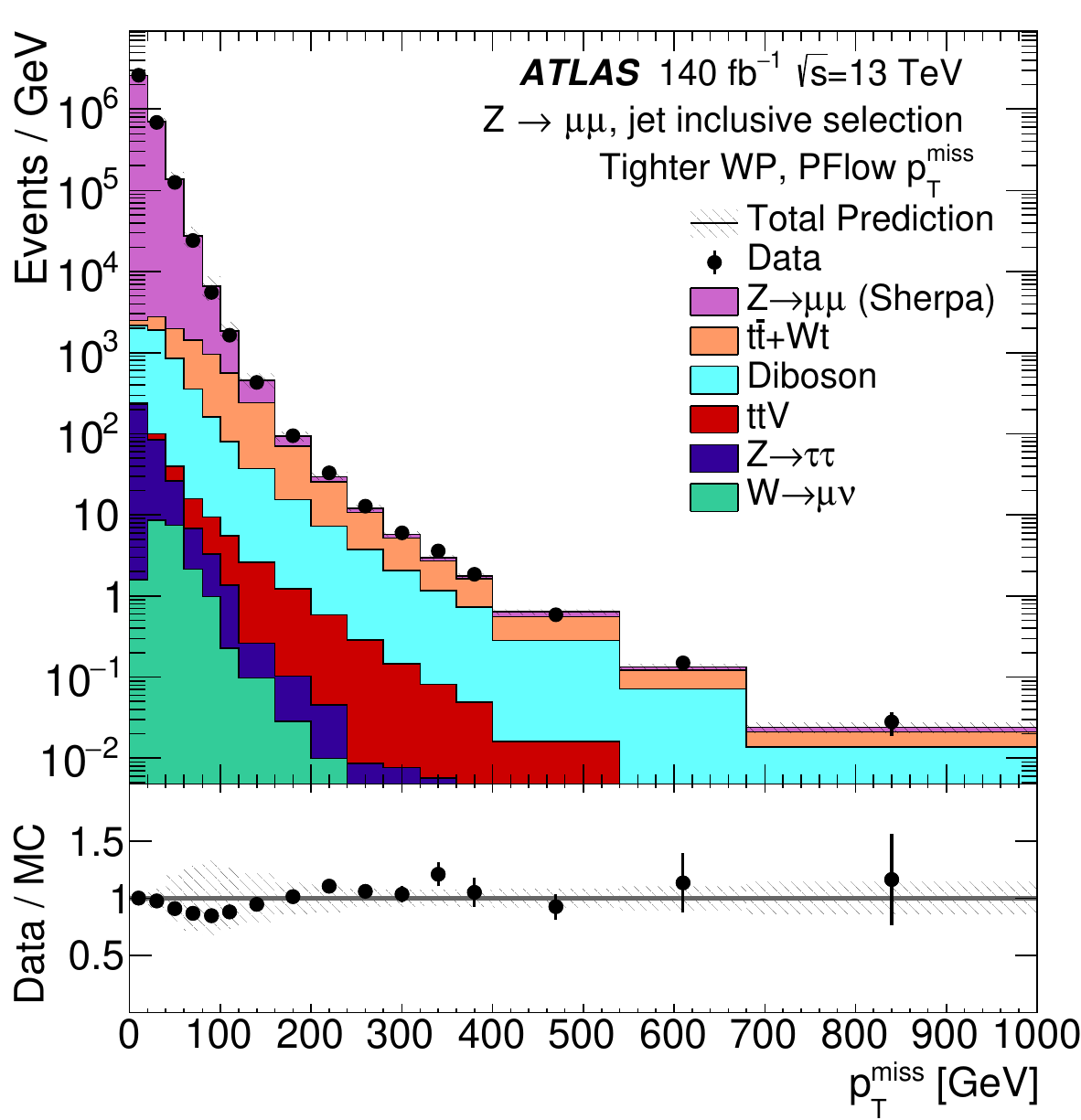}}\\
\subfloat[]{\label{fig:wpsZmm:PFlowTenacious}\includegraphics[width=0.4\textwidth]{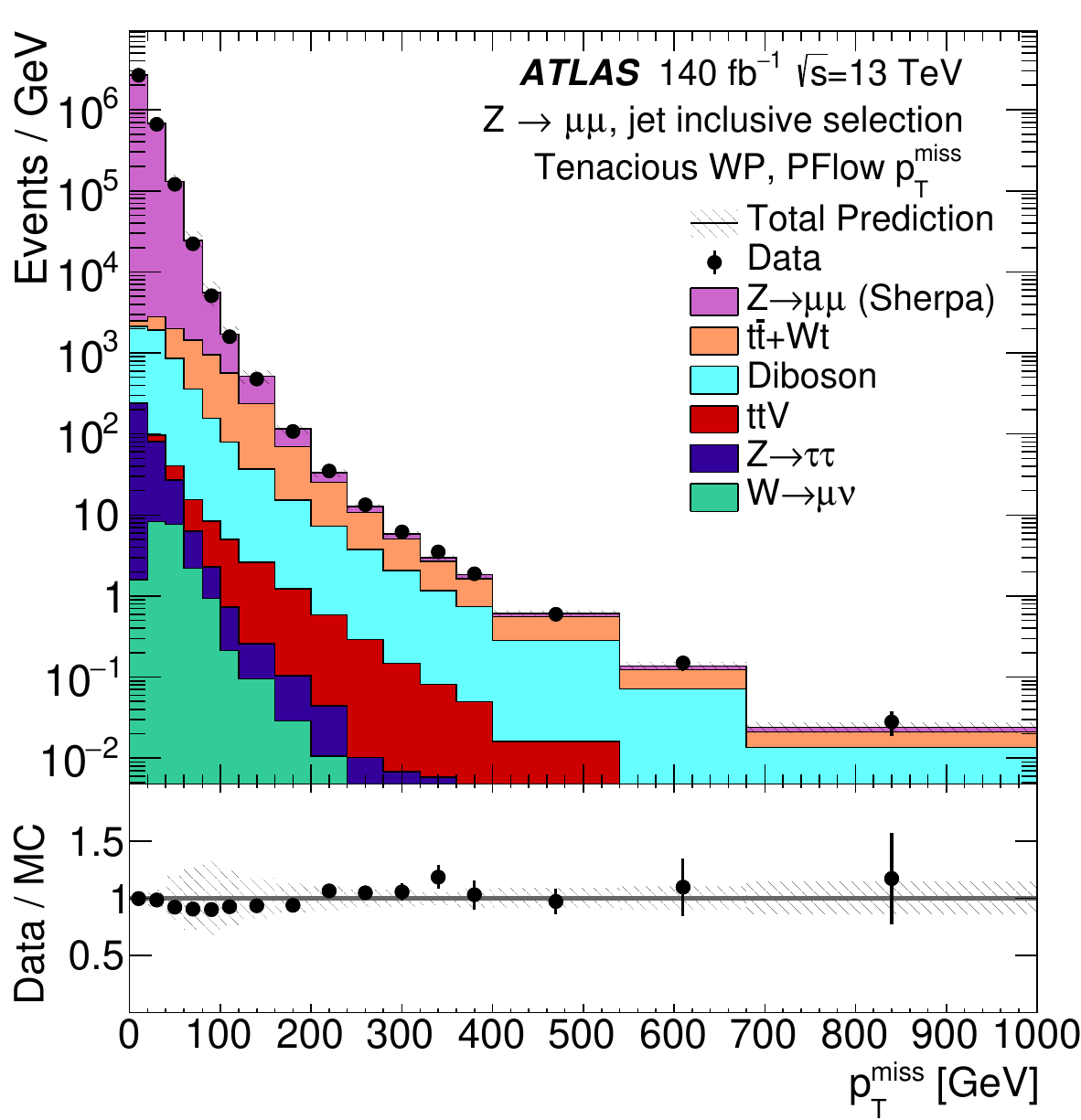}}
\caption{Distributions of \ptmiss\ in MC simulation and data for different working points: \workingPoint{Loose} (a), \workingPoint{Tighter} (b) and \workingPoint{Tenacious} (c). Events satisfy a \Zmm\ selection and \ptmiss\ is built using PFlow jets. \SHERPA\ is used to generate the \Zmm\ events. The error band includes MC statistical, luminosity and detector uncertainties.}
\label{fig:wpsZmm}
\end{figure}

Figure~\ref{fig:genZmm} shows the distributions for \ptmiss\ and the soft term, using \POWHEG+\Pythia\ to produce \Zmm\ events.
\POWHEG+\Pythia\ performs well throughout the whole \ptmiss\ distribution, however when considering \ptsoft, \POWHEG+\Pythia\ models the data worse than \sherpa.
For \POWHEG+\Pythia, extra jets in an event are produced at the parton shower level, where they are less well-modelled, in comparison to \sherpa\ where they are produced at matrix element level.
The tail of the \ptsoft\ distribution will have a significant contribution from events with a high multiplicity of these poorly-modelled soft jets.
Additionally, \POWHEG+\Pythia\ has a different representation of the underlying event, which can be a significant contribution to the soft momenta in the event.

\begin{figure}
\centering
\subfloat[]{\label{fig:genZmm:PowhegMET}\includegraphics[width=0.4\textwidth]{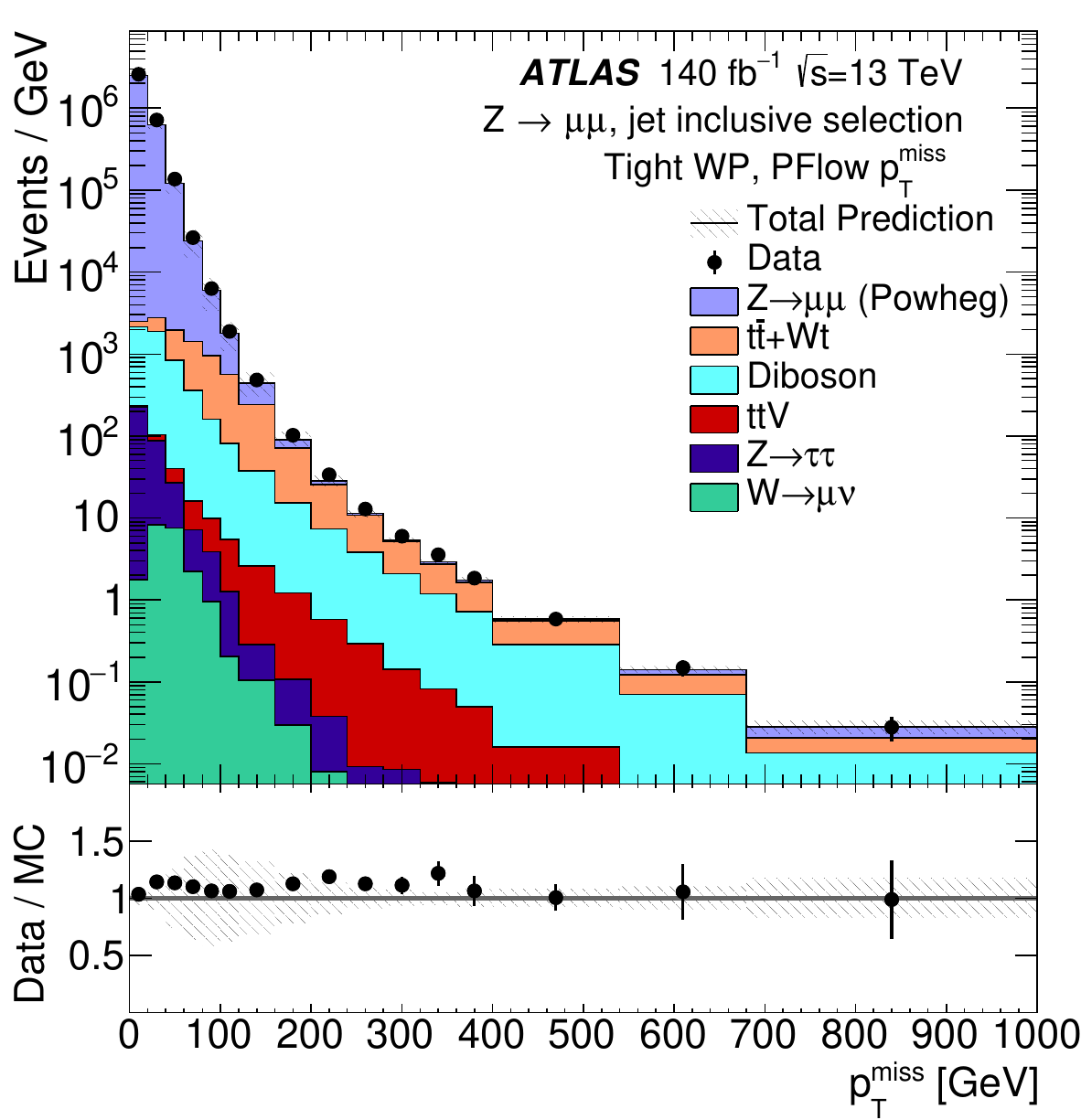}}\qquad
\subfloat[]{\label{fig:genZmm:PowhegSoft}\includegraphics[width=0.4\textwidth]{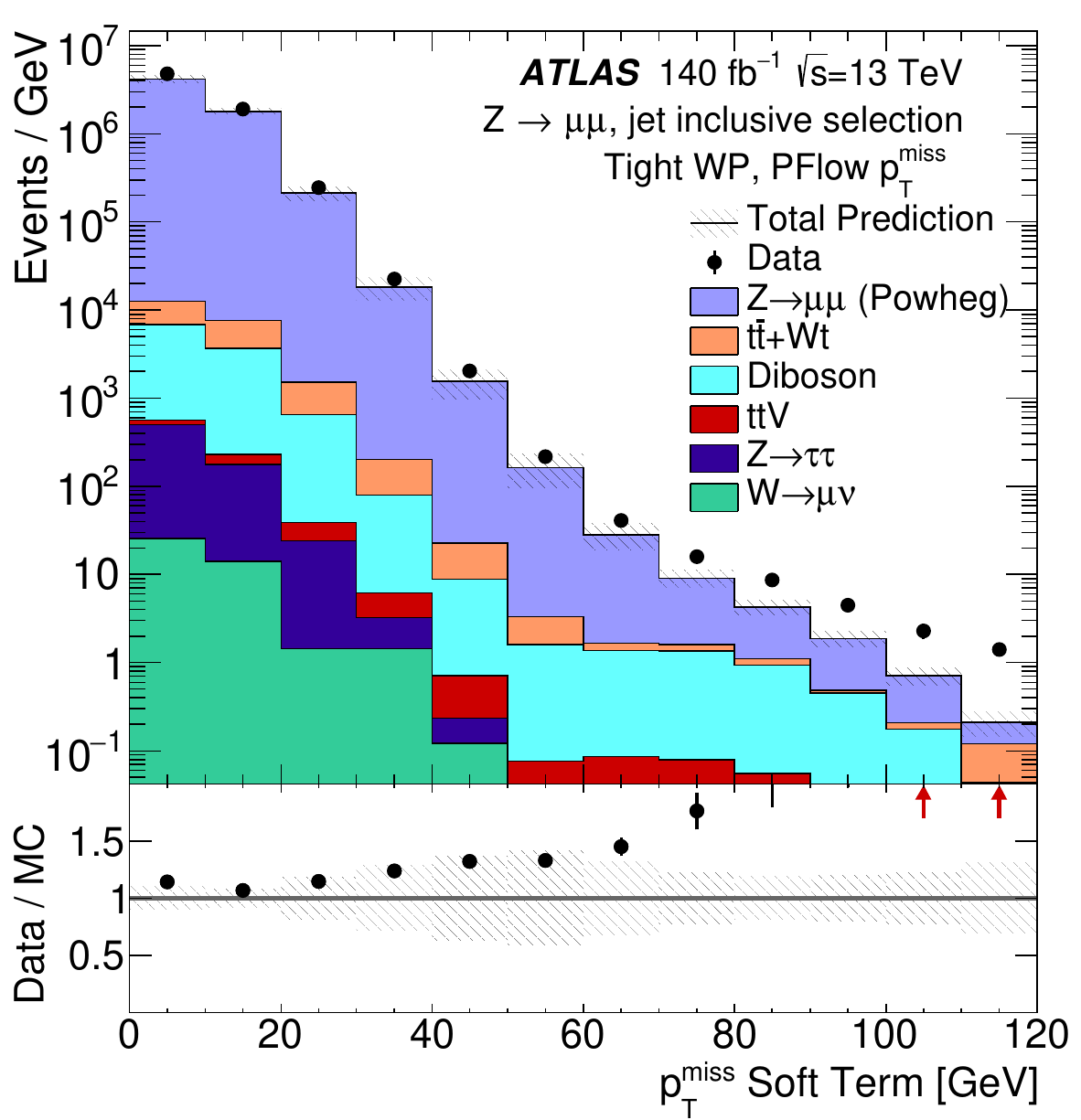}}
\caption{Distributions of \ptmiss\ (a) and the \ptX{\text{soft}} (b) in MC simulation and data using the \POWHEG+\Pythia\ \Zmm\ generator. Events satisfy a jet inclusive \Zmm\ selection. PFlow jets are used with the \workingPoint{Tight} \ptmiss\ working point. The error band includes MC statistical, luminosity and detector uncertainties.}
\label{fig:genZmm}
\end{figure}

\FloatBarrier
\subsection{\ptmiss\ performance}
\label{sec:perf}

An important measure for the quality of \ptmiss\ reconstruction is the resolution.
For \Zll events, the \pxmiss\ and \pymiss\ are approximately Gaussian-distributed about zero, except for events with very large \sumpt\ or noise.
Non-Gaussian tails are expected, so to appropriately represent the distributions, the root-mean-square error (RMS) is used to measure the \pxmiss\ and \pymiss\ resolution.
For MC simulation, the truth \pxmiss\ and \pymiss\ (defined in Section~\ref{sec:objects}) are subtracted.

To understand the impact of pile-up on \ptmiss\ resolution, Figure~\ref{fig:resZmm} shows the \pxmiss\ and \pymiss\ resolutions in \SHERPA\ \Zmm\ MC simulations, binned in the variables introduced in Section~\ref{sec:objects} which parametrise the amount of pile-up present: \npv\ and \(\mu\).
For the jet inclusive selection, the resolution degrades as the amount of pile-up increases, as expected.
The resolution improves dramatically as events containing jets are vetoed, until the pile-up dependence almost entirely disappears.

The intention of the various \ptmiss\ working points is to try to reduce fake \ptmiss\ contamination further.
As can be seen for \SHERPA\ \Zmm\ MC simulated events satisfying the \Zmm\ selection in Figures~\ref{fig:resZmmwps:muPFlow} and \ref{fig:resZmmwps:npvPFlow}, the tighter working points have a reduced pile-up dependence and better resolution, indicating they are indeed less susceptible to fake \ptmiss\ generally, and specifically from pile-up contamination.

\begin{figure}[!h]
\centering
\subfloat[]{\label{fig:resZmmjets:muPFlow}\includegraphics[width=0.4\textwidth]{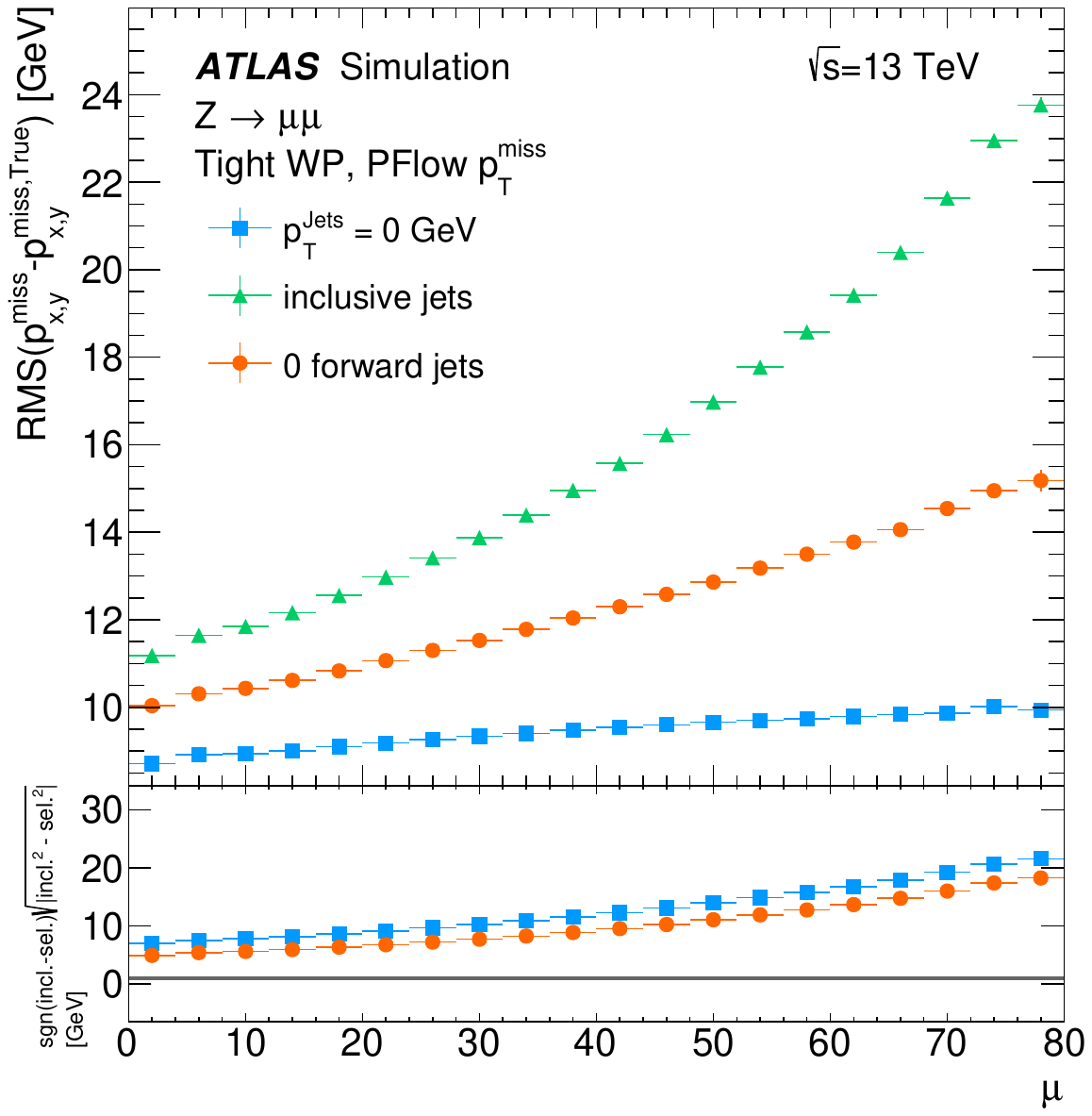}}\qquad
\subfloat[]{\label{fig:resZmmjets:npvPFlow}\includegraphics[width=0.4\textwidth]{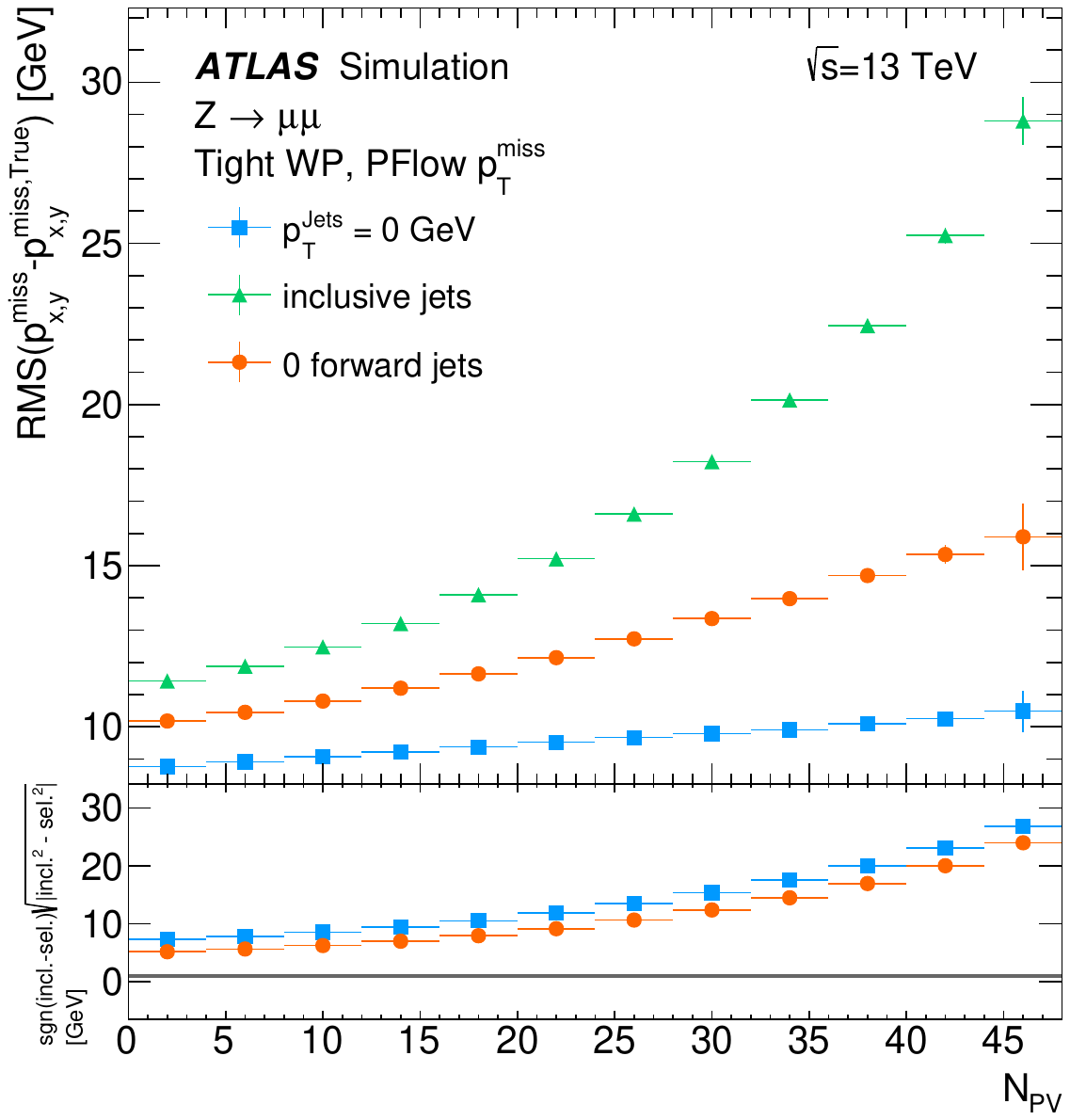}}\\
\subfloat[]{\label{fig:resZmmwps:muPFlow}\includegraphics[width=0.4\textwidth]{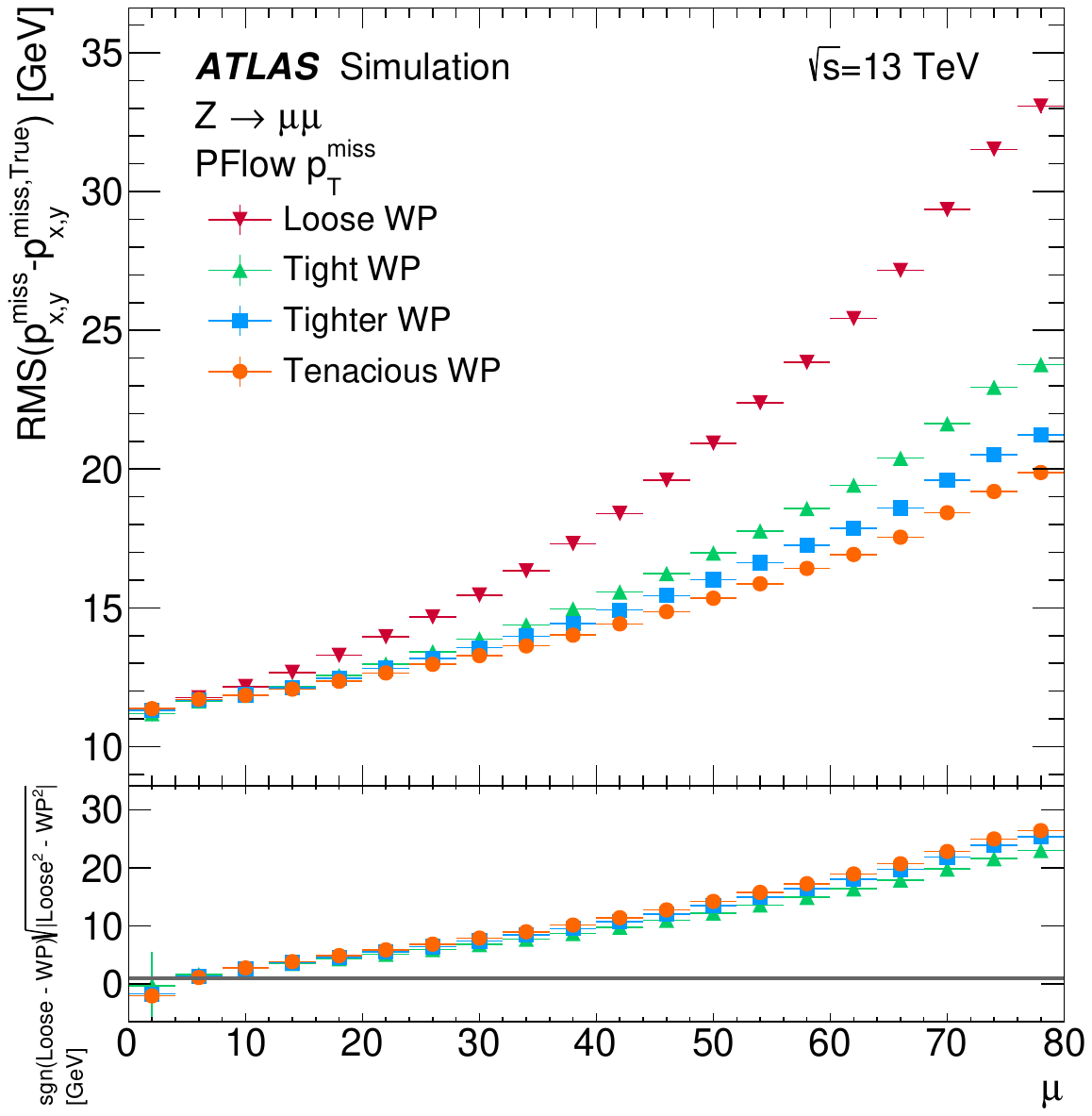}}\qquad
\subfloat[]{\label{fig:resZmmwps:npvPFlow}\includegraphics[width=0.4\textwidth]{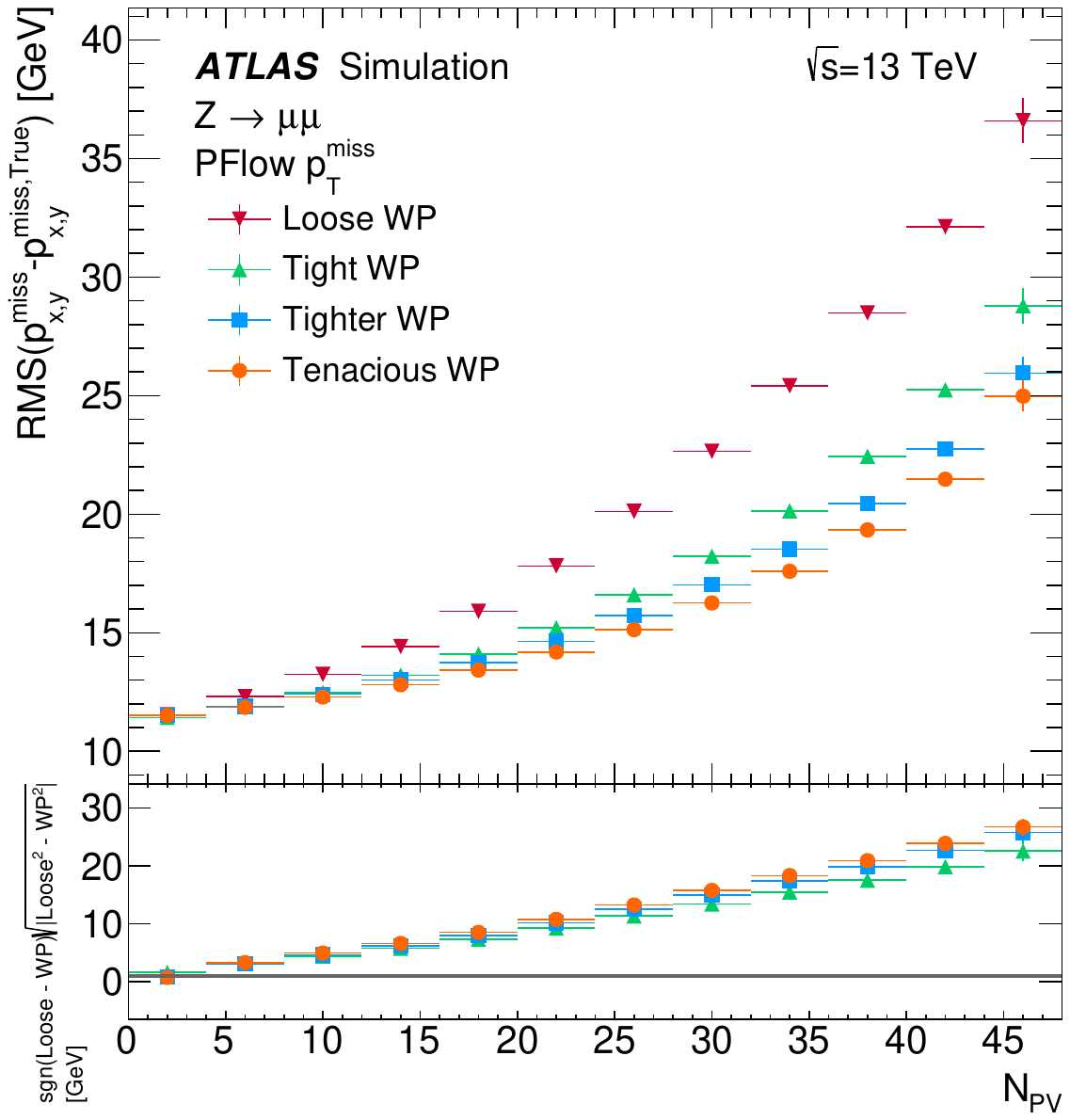}}
\caption{The \pxmiss\ and \pymiss\ resolution for different jet selections (sel.) ((a) and (b)) and different \ptmiss\ working points ((c) and (d)) as a function of \(\mu\) ((a) and (c)) or \npv\ ((b) and (d)). PFlow jets and the \workingPoint{Tight} \ptmiss\ working point are used, on \SHERPA\ \Zmm\ MC simulated events. The error bars include the MC statistical uncertainty. In the \(y\)-axis label of the lower panels, `incl.' refers to the inclusive jet selection, `sel.' to the alternate jet selection under consideration and `WP' to the working point under consideration. `True' refers to MC-generated quantities.}
\label{fig:resZmm}
\end{figure}

In Figure~\ref{fig:resttbarWmunuwps}, the working point resolution dependence is shown for \ttbar\ or \Wmunu\ MC simulations.
For the \ttbar\ processes, the amount of hadronic activity in the hard-scatter process increases substantially relative to \Zmm.
At high pile-up, tighter working points improve the resolution for \ttbar\ and \Wmunu\ topologies by removing more pile-up jets from the jet term.
For the \ttbar\ process at low pile-up, the majority of reconstructed jets in the event come from the hard-scatter, so the tighter working points are more likely to remove jets originating from the hard-scatter, leading to a degradation in the resolution.
The performance for \Wmunu\ is very similar to \Zmm, suggesting that the working point performance is minimally effected by the amount of real \ptmiss\ in the event.
The topology dependence in the choice of `best' working point leads to the support of all of them for analysis use.

\begin{figure}[!h]
\centering
\subfloat[]{\label{fig:resttbar:muPFlow}\includegraphics[width=0.4\textwidth]{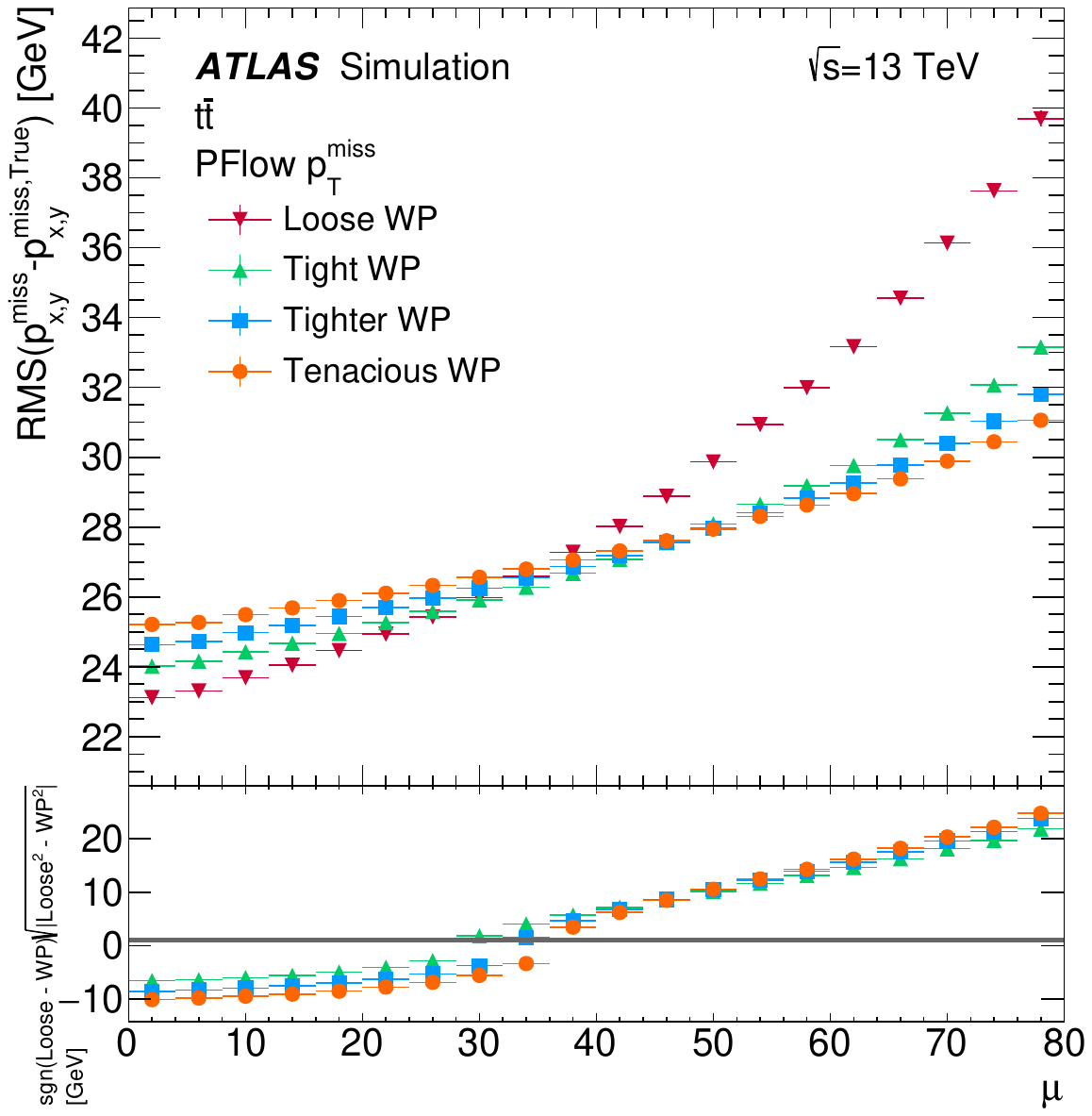}}\qquad
\subfloat[]{\label{fig:resttbar:npvPFlow}\includegraphics[width=0.4\textwidth]{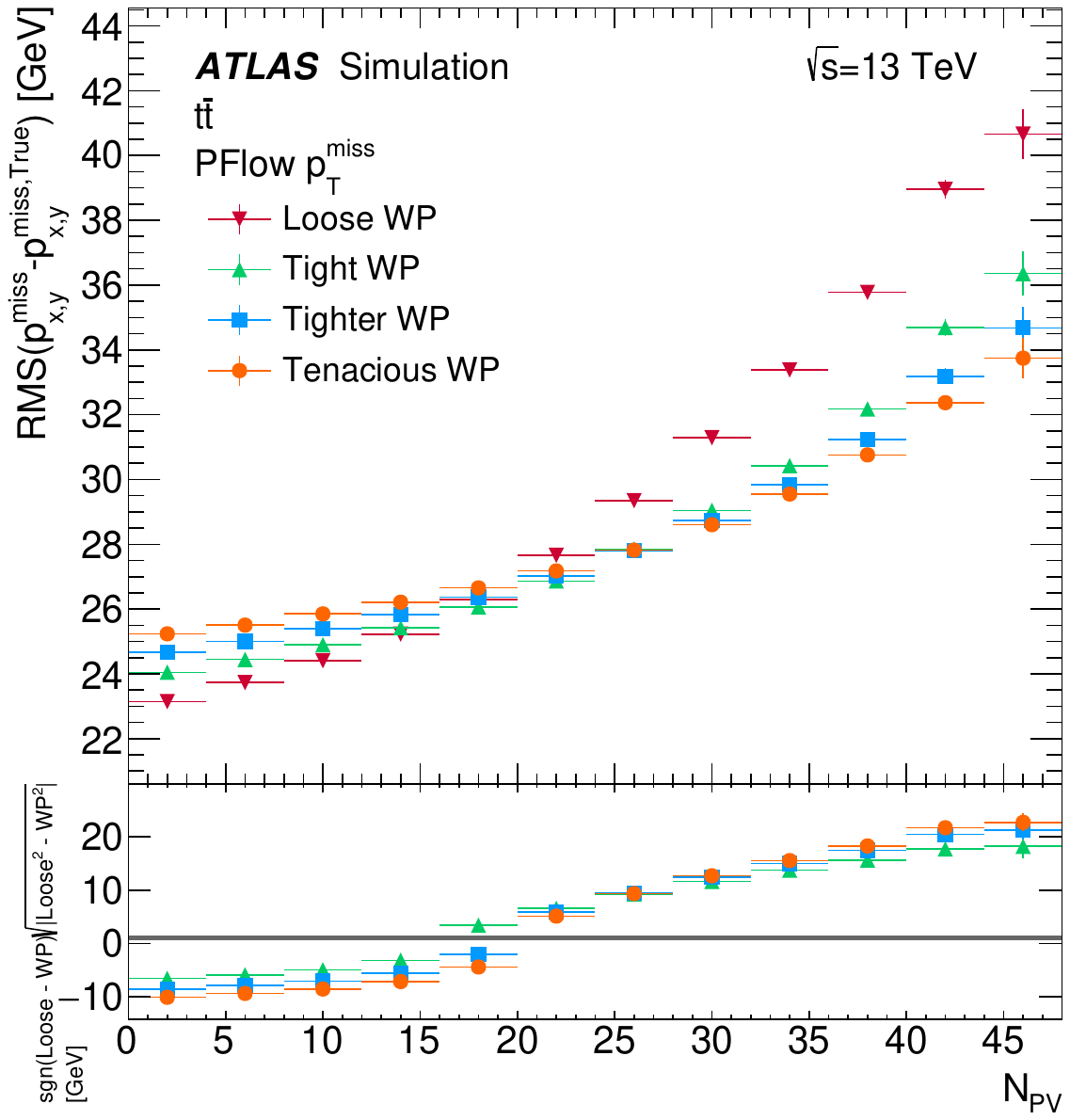}}\\
\subfloat[]{\label{fig:resWmunu:muPFlow}\includegraphics[width=0.4\textwidth]{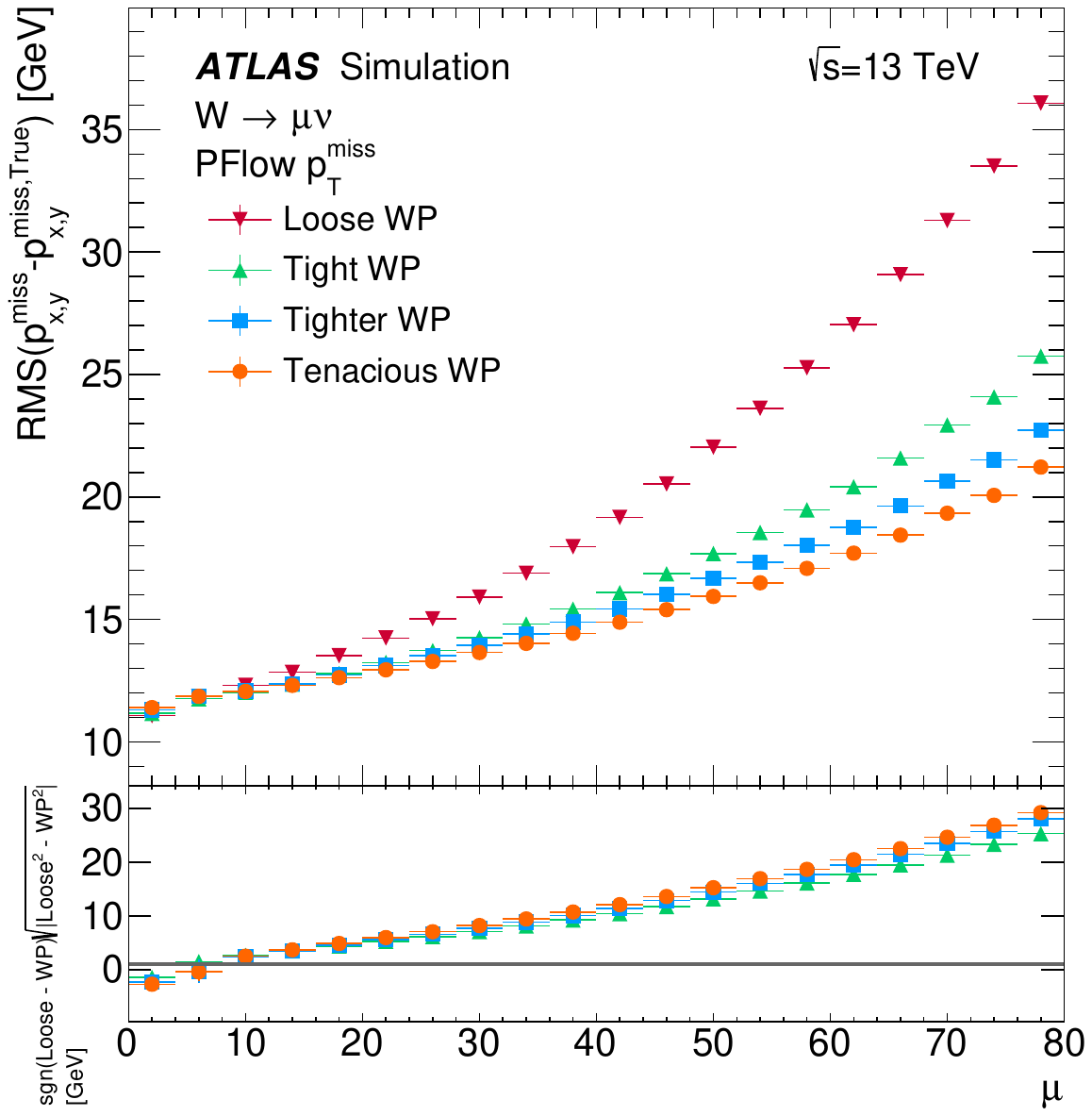}}\qquad
\subfloat[]{\label{fig:resWmunu:npvPFlow}\includegraphics[width=0.4\textwidth]{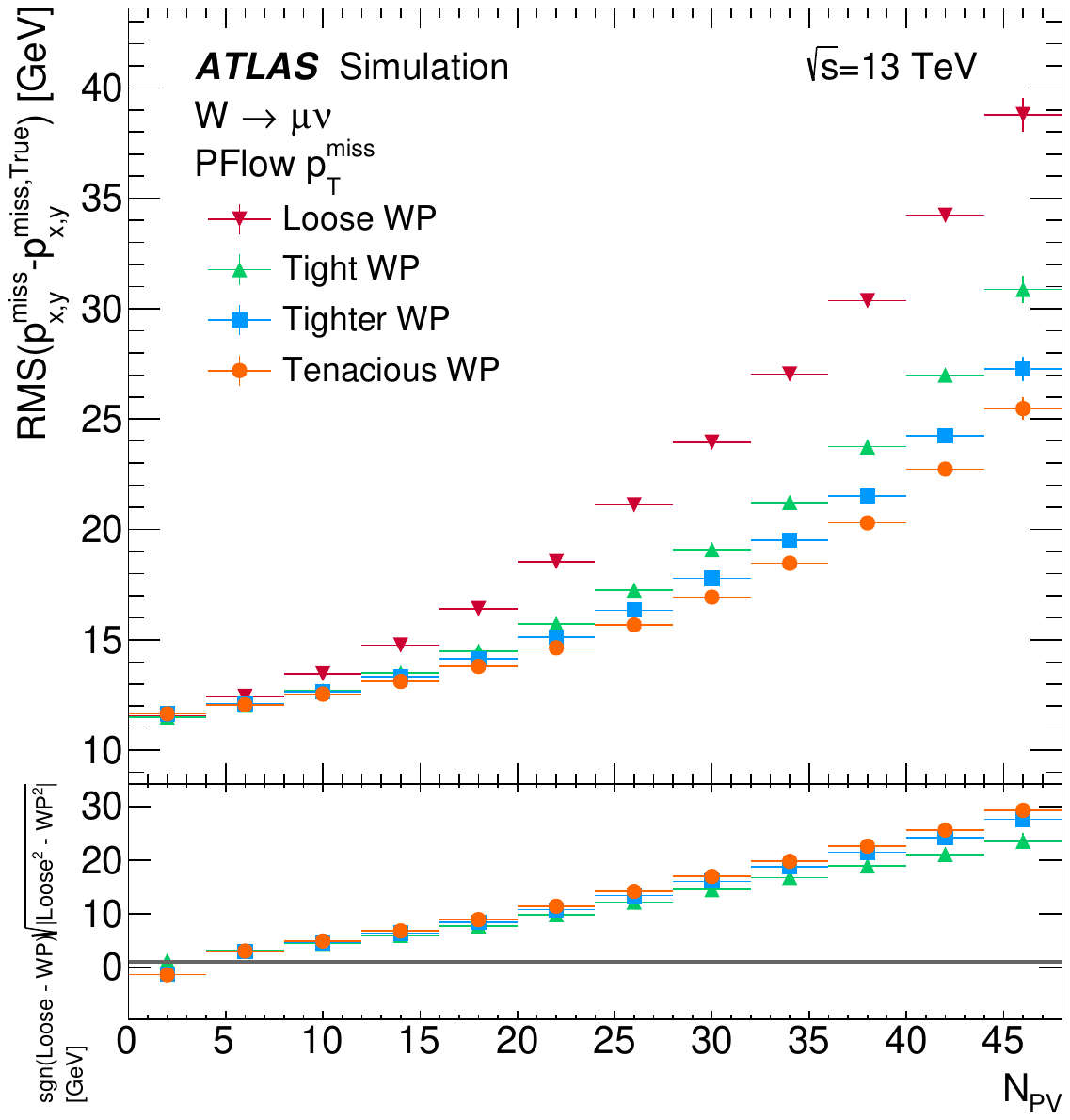}}
\caption{The \pxmiss\ and \pymiss\ resolution for different \ptmiss\ working points as a function of \(\mu\) ((a),(c)) or \npv\ ((b),(d)). MC simulated events are shown: \ttbar\ events are used in (a) and (b), and \Wmunu\ events in (c) and (d). PFlow jets are used. The error bars include the MC statistical uncertainty. In the \(y\)-axis label of the lower panels, `WP' refers to the working point under consideration. `True' refers to MC-generated quantities.}
\label{fig:resttbarWmunuwps}
\end{figure}

To confirm that the \ptmiss\ resolution in MC simulation represents data well, the \pxmiss\ and \pymiss\ resolutions are shown in Figure~\ref{fig:resdataMCZmm} as a function of \(\mu\) and \npv\ in the default \Zmm\ configuration comparing MC simulation (including \Zmm\ as well as the background processes) and data.
In this case the truth values are not subtracted from the MC simulation values.
The resolutions agree within the error band which includes the MC statistical, luminosity and detector uncertainties.

\begin{figure}[!h]
\centering
\subfloat[]{\label{fig:resZmm:mudataMC}\includegraphics[width=0.4\textwidth]{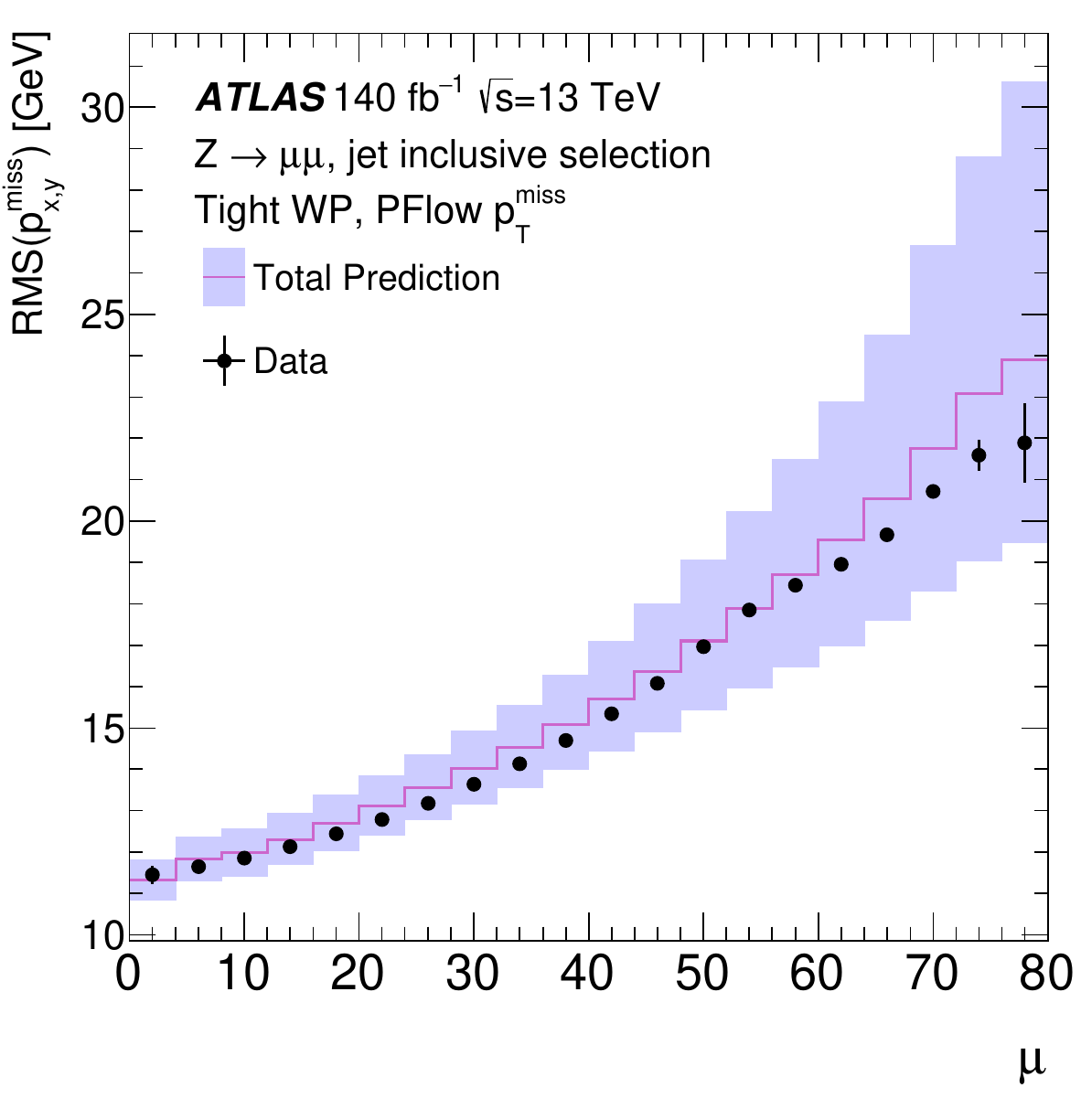}}\qquad
\subfloat[]{\label{fig:resZmm:npvdataMC}\includegraphics[width=0.4\textwidth]{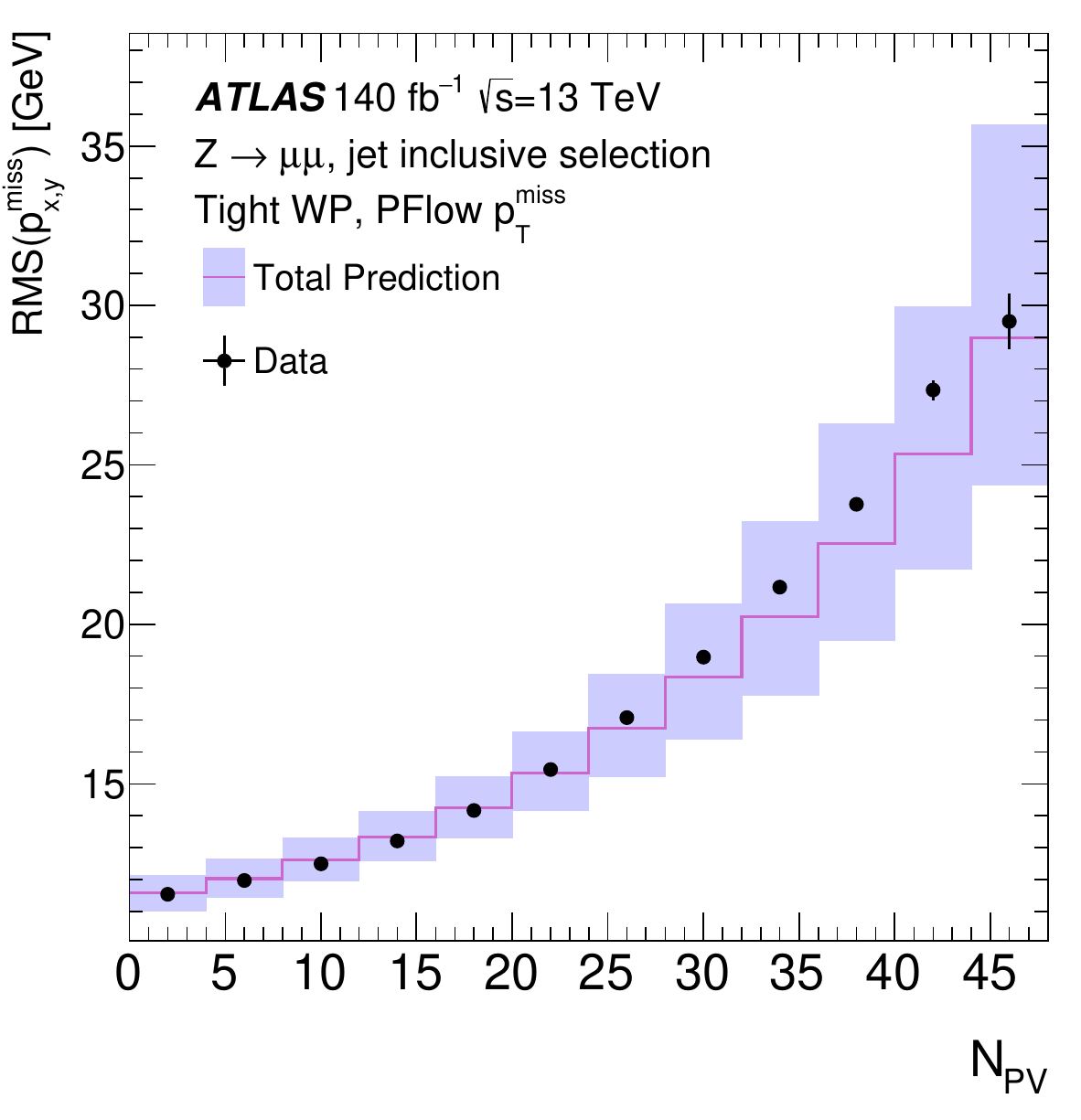}}
\caption{The \pxmiss\ and \pymiss\ resolution in data and simulation with the Tight working point as a function of (a) \(\mu\) or (b) NPV. A \Zmm\ selection is applied with \SHERPA\ used to generate the \Zmm\ events. PFlow jets are used with an inclusive selection. The error band includes MC statistical, luminosity and detector uncertainties.}
\label{fig:resdataMCZmm}
\end{figure}

Comparing the reconstructed and truth \ptmiss\ is a way to assess bias in events with real \ptmiss.
Figure~\ref{fig:linWPs} shows the response (\(\ptmiss/\ptmisstrue\)) in each case as a function of truth \ptmiss, for all four working points, in events satisfying the \Wmunu\ and \ttbar\ selections.
Since the track-based soft term means soft neutral contributions to the event are ignored, it is expected that some bias from truth \ptmiss\ will be seen at low values where the \ptmiss\ is more dominated by the soft term.
For \Wmunu\ events, tightening the working point slightly reduces the bias at low values, due to the removal of pile-up, which contributes to the bias.
For \ttbar\ events, the \workingPoint{Tight} performs slightly better at low values, consistent with the \workingPoint{Loose} working point leaving too much pile-up, and the tighter working points removing some of the hard-scatter jets.

\begin{figure}[!h]
\centering
\subfloat[]{\label{fig:linWmunu}\includegraphics[width=0.4\textwidth]{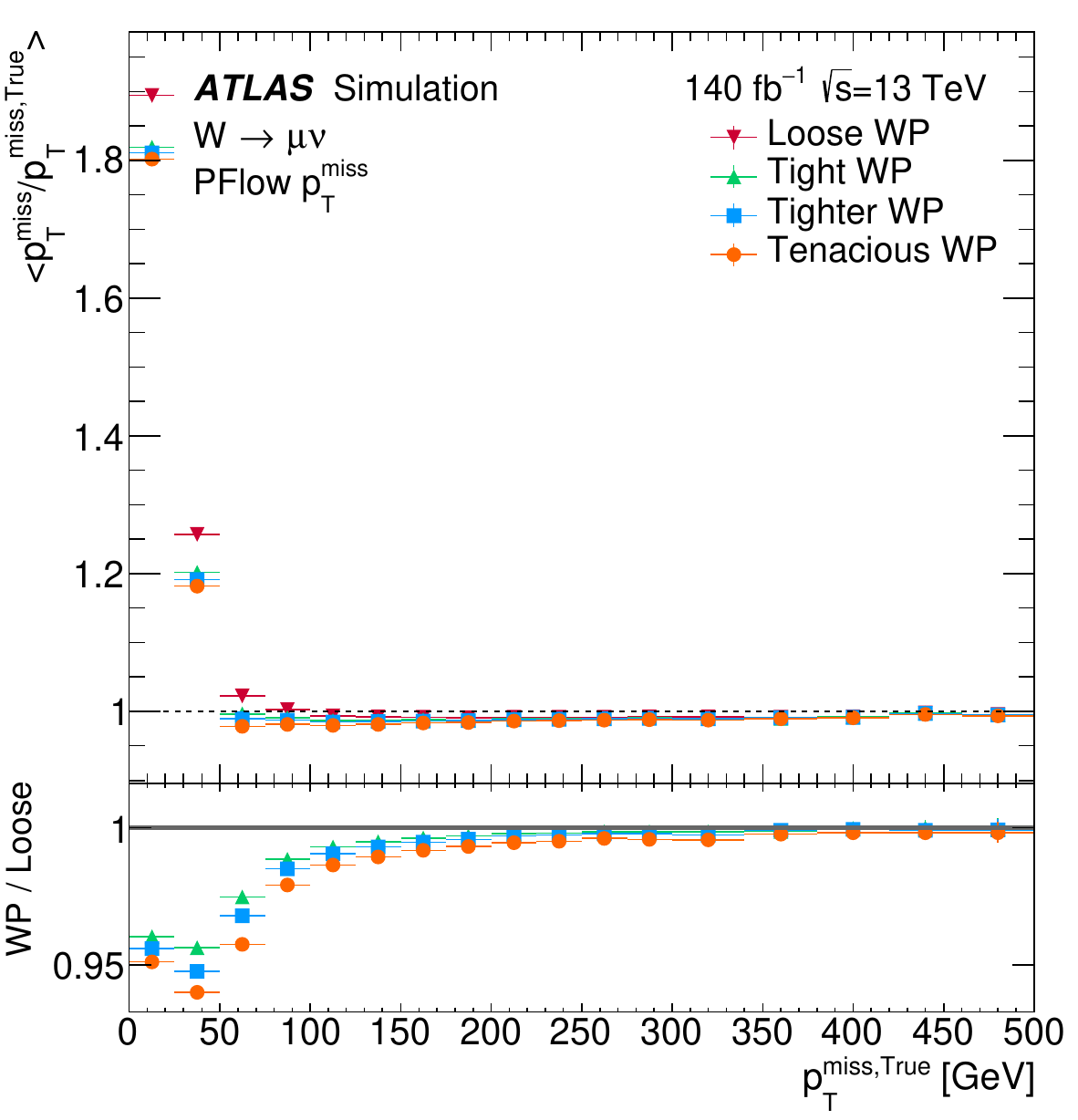}}\qquad
\subfloat[]{\label{fig:linttbar}\includegraphics[width=0.4\textwidth]{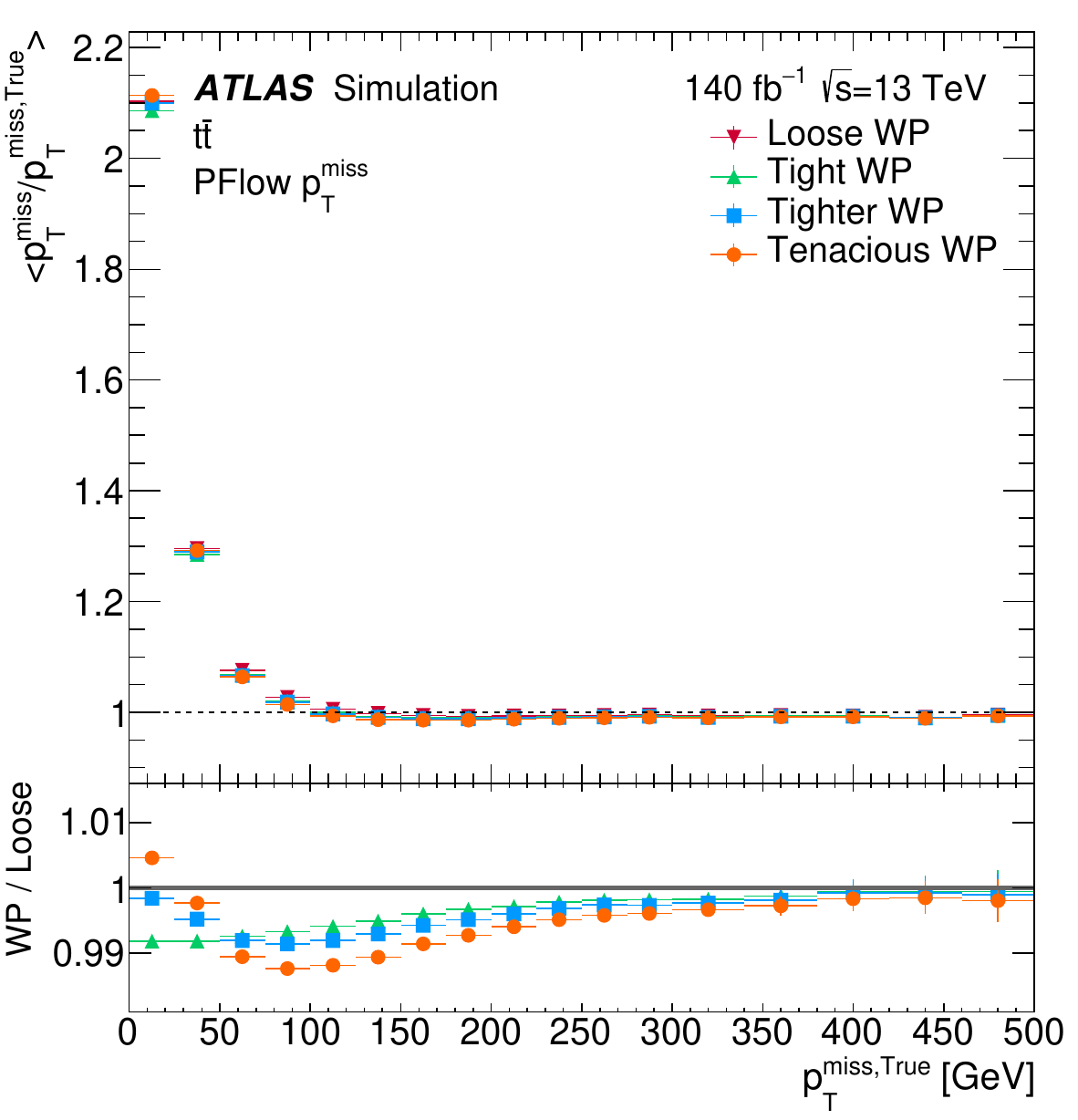}}
\caption{The \ptmiss\ response for different working points as a function of truth (generated) \ptmiss. MC simulated (a) \Wmunu or (b) \ttbar\ events are used. PFlow jets are used. The error bars include the MC statistical uncertainty.}
\label{fig:linWPs}
\end{figure}

In \Zll\ events, where there is no real \ptmiss, the transverse momentum of the \(Z\) (\(p_{\mathrm{T}}^{Z}\)) can be used as a measure of the hardness of the interaction and provides a scale for the evaluation of the \ptmiss\ response.
One can define an axis in the transverse plane from the \ptvec\ of the \(Z\) which is constructed by using the \ptvec\ of each of the leptons,
\begin{equation*}
\mathbf{\hat{A}}_{Z} = \frac{\mathbf{p}_{\mathrm{T}}^{\ell^{+}} + \mathbf{p}_{\mathrm{T}}^{\ell^{-}} }{| \mathbf{p}_{\mathrm{T}}^{\ell^{+}} + \mathbf{p}_{\mathrm{T}}^{\ell^{-}} |} = \frac{\mathbf{p}_{\mathrm{T}}^{Z} }{p_{\mathrm{T}}^{Z}}.
\end{equation*}

With this reference axis the \ptmissvec\ can be projected onto it with,
\begin{equation}
\mathcal{P}^{Z} = \ptmissvec \cdot \mathbf{\hat{A}}_{Z}.
\label{eq:scale}
\end{equation}

This quantity --- the scale \scale\ --- is sensitive to any mis-reconstruction in the \ptmiss\ and provides an excellent way to gauge the performance of the \ptmiss\ reconstruction.
It is particularly sensitive to the impact of the hadronic recoil against the \(Z\) boson.
For a completely balanced reaction, where the \(Z\) boson is produced in perfect balance with the hadronic recoil, the expectation is \(\mathcal{P}^{Z} = 0\).
If \(\mathcal{P}^{Z} < 0\) then there is not enough hadronic recoil to balance the momentum of the \(Z\) and when \(\mathcal{P}^{Z} > 0\) there is too much reconstructed recoil.
The hardness of the interaction (roughly the amount of \pt\ produced in the event) can be assessed by taking the average of the projection, \(\langle \mathcal{P}^{Z} \rangle\), and binning it as a function of \(p_{\mathrm{T}}^{Z}\).

Figure~\ref{fig:scaleZmmPFlow:MCData} shows the average value of \scale in bins of \pTZ\ for data and MC simulation in a \Zmm\ selection.
Overall there is an underestimation of the hadronic balance with the \(Z\) boson, caused by the missing component of neutral soft energy and finite detector acceptance, and an offset between data and prediction that is within uncertainties.
The scale is worst at very low values of \pTZ, where the missing neutral component of the soft term means much of the hadronic recoil is missed.
At higher values the scale improves as jets are reconstructed allowing for better hadronic recoil determination.
At very high values the event selection is dominated by the non-\Zmm\ processes, which causes the scale to increase again.
\FloatBarrier

The impact of the varying working points on the \scale\ is assessed in Figure~\ref{fig:scaleZmmPFlow:WPs}.
Moving from the \workingPoint{Loose} to the \workingPoint{Tenacious} working points, the hadronic balance becomes increasingly underestimated, as the tightening jet selection increases the potential for part of the hadronic recoil to be missed.

\begin{figure}[!h]
\centering
\subfloat[]{\label{fig:scaleZmmPFlow:MCData}\includegraphics[width=0.4\textwidth]{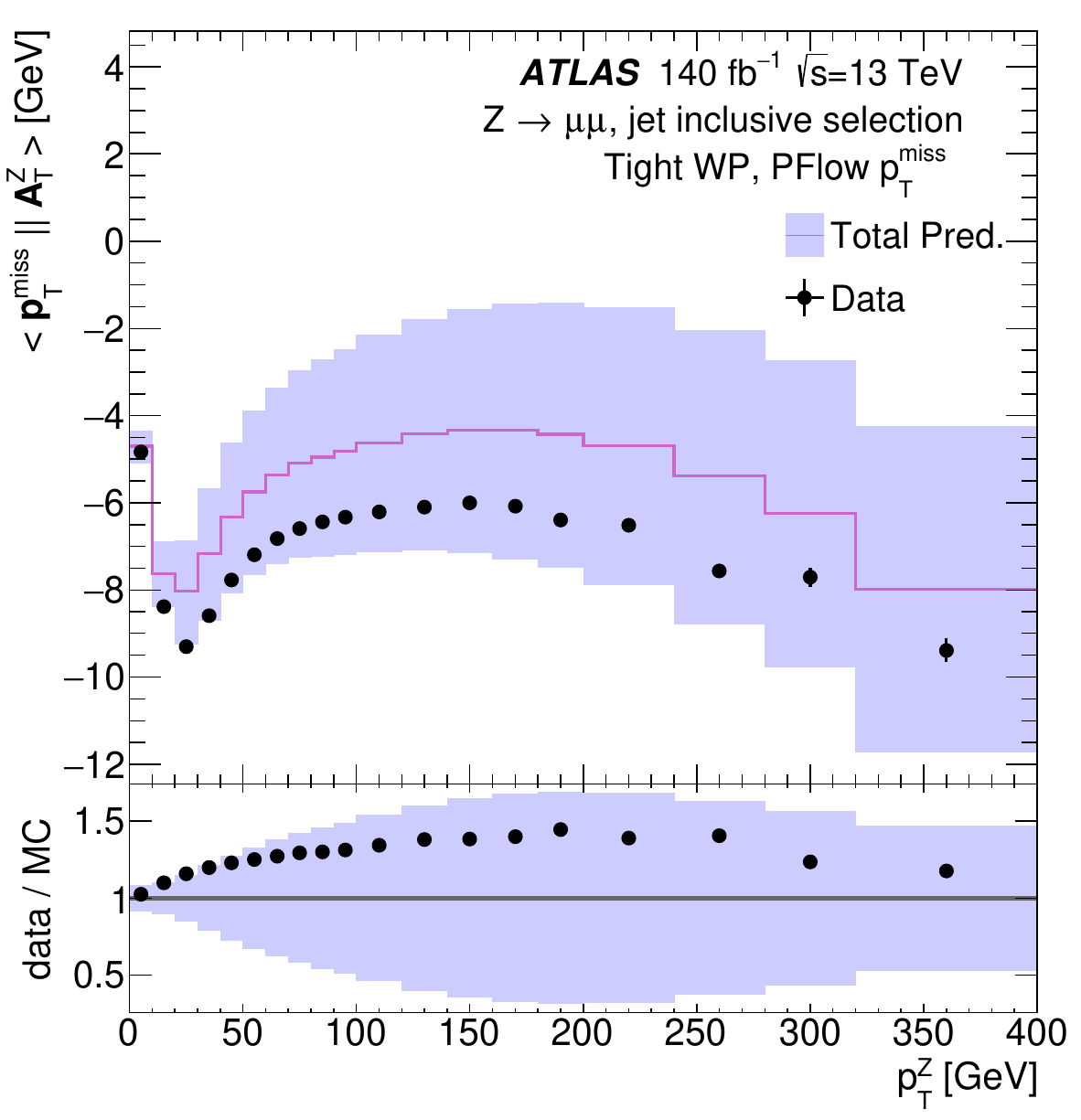}}\qquad
\subfloat[]{\label{fig:scaleZmmPFlow:WPs}\includegraphics[width=0.4\textwidth]{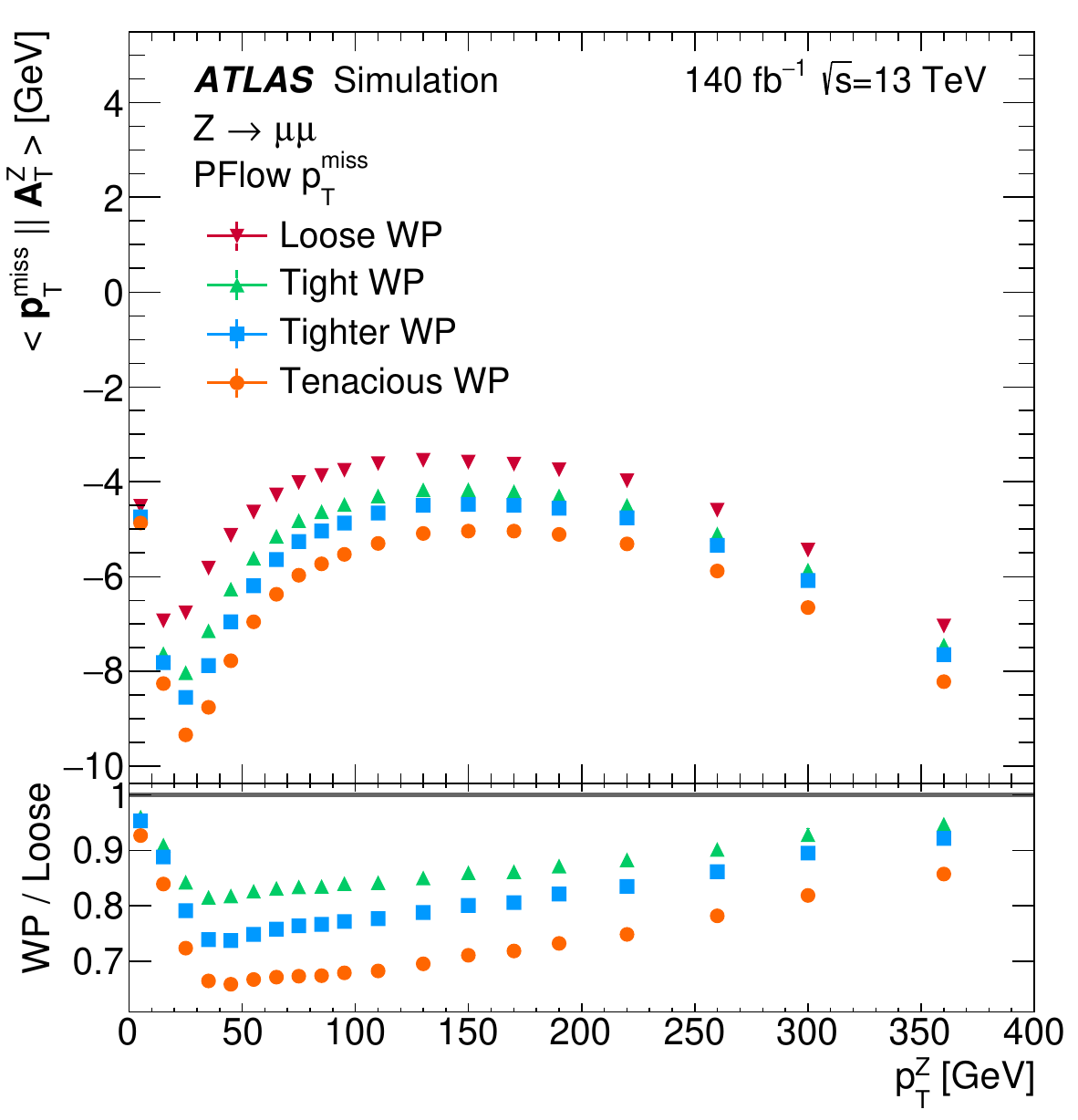}}
\caption{The average \scale\ (projection of \ptmissvec\ on to unit vector in the direction of the \(Z\) boson \(\mathbf{A}_{Z}\)) as a function of the \(Z\) boson's transverse momentum.
Events satisfy a \Zmm\ selection with no requirements placed on the jets, with \SHERPA\ used to generate the \Zmm\ events. In Figure (a) all events passing the event selection are shown, using the \workingPoint{Tight} \ptmiss\ working point, and the error bars include statistical and detector uncertainties. In Figure (b), only \Zmm\ MC simulated events are shown, comparing each \ptmiss\ working point, and the error bars include statistical uncertainties only.}
\label{fig:scaleZmmPFlow}
\end{figure}

Furthermore, the \ptmiss\ response in this system can be defined by:
\begin{equation}
\C = 1 + \frac{\langle\scale\rangle}{\langle p_{\text{T}}^{Z} \rangle}.
\label{eq:C}
\end{equation}
A comparison of \C\ for the different \ptmiss\ working points as a function of \pTZ\ is shown in Figure~\ref{fig:C:C}, for \Zmm\ MC simulated events.
At low values of \pTZ\ the response decreases below one as the missing neutral component of the soft term means more of the hadronic recoil is missed and \ptmiss\ is reconstructed opposing the \(Z\).
In consistency with the behaviour of the \scale, as the working point changes from \workingPoint{Loose} to \workingPoint{Tenacious}, the response decreases further, as more of the hadronic recoil has potential to be missed.

In Figure~\ref{fig:C:RMSScaleOverC}, the RMS of the scale is shown, correcting for the response \C, for \Zmm\ MC simulated events.
At low values of \pTZ\ the RMS worsens for the same reasons that the scale itself worsens, and as the contribution of pileup becomes relatively more important, and for very high \pTZ\ as the absolute resolution of the muons and jets used to reconstruct the \(Z\) and \ptmiss\ degrades.
At these extremes, the \workingPoint{Loose} working point has the worst RMS as it removes the fewest poorly-measured jets or those originating from pileup.
In the medium \pTZ\ regime, the \workingPoint{Tight} working point has the best RMS, providing the best compromise between removing pileup and not removing too much of the hard-scatter hadronic recoil.

Finally, the RMS of the \pxmiss\ and \pymiss\ of \Zmm\ MC simulated events, corrected by \C\ in order to compare the resolutions for each working point at the same energy scale, is shown in Figure~\ref{fig:RMSOverC}.
In this case the behaviour is very similar to the uncorrected RMS, apart from at very low values of pileup, where the more stringent working points no longer perform the best.

\begin{figure}[!h]
\centering
\subfloat[]{\label{fig:C:C}\includegraphics[width=0.4\textwidth]{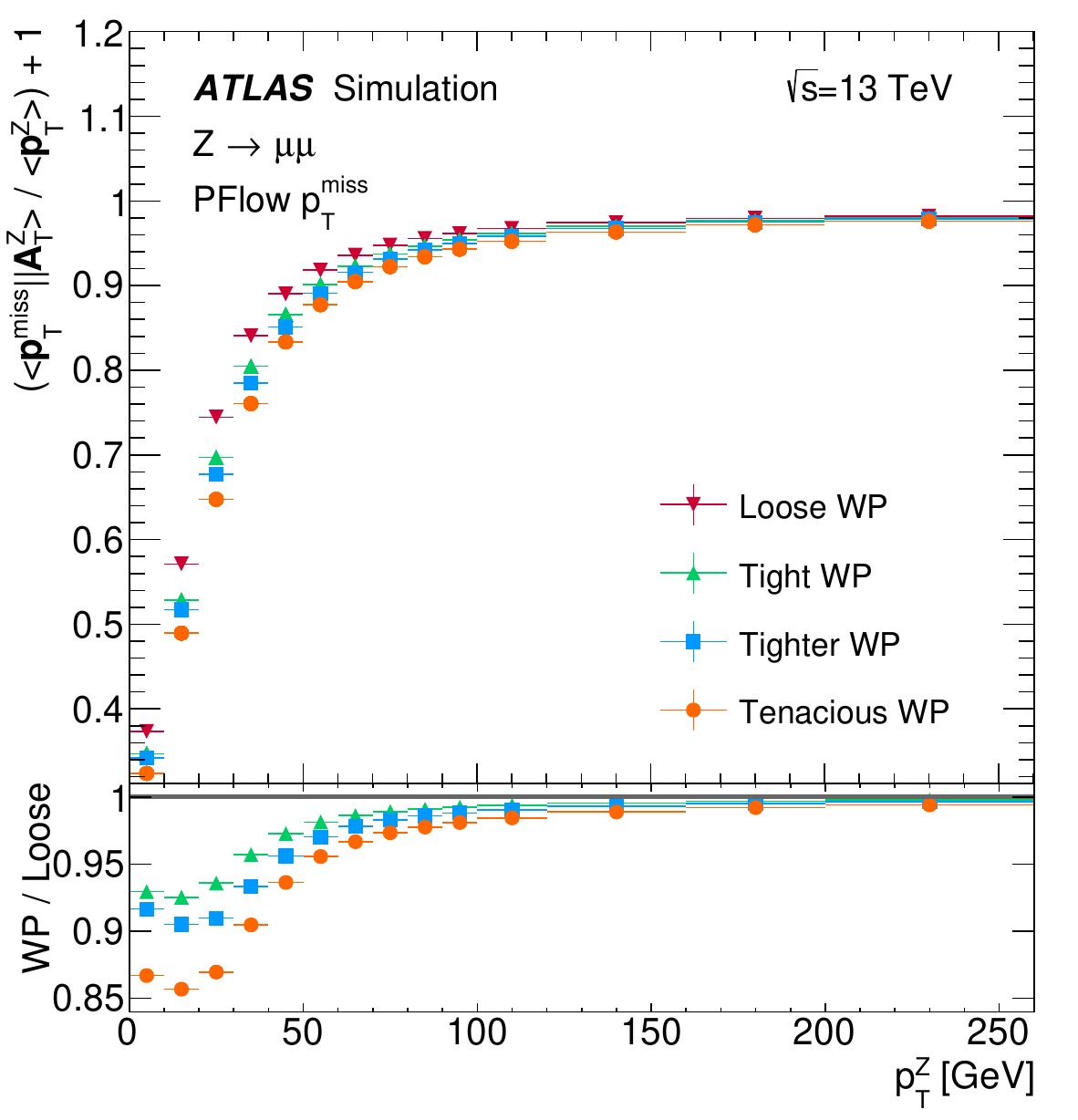}}\qquad
\subfloat[]{\label{fig:C:RMSScaleOverC}\includegraphics[width=0.4\textwidth]{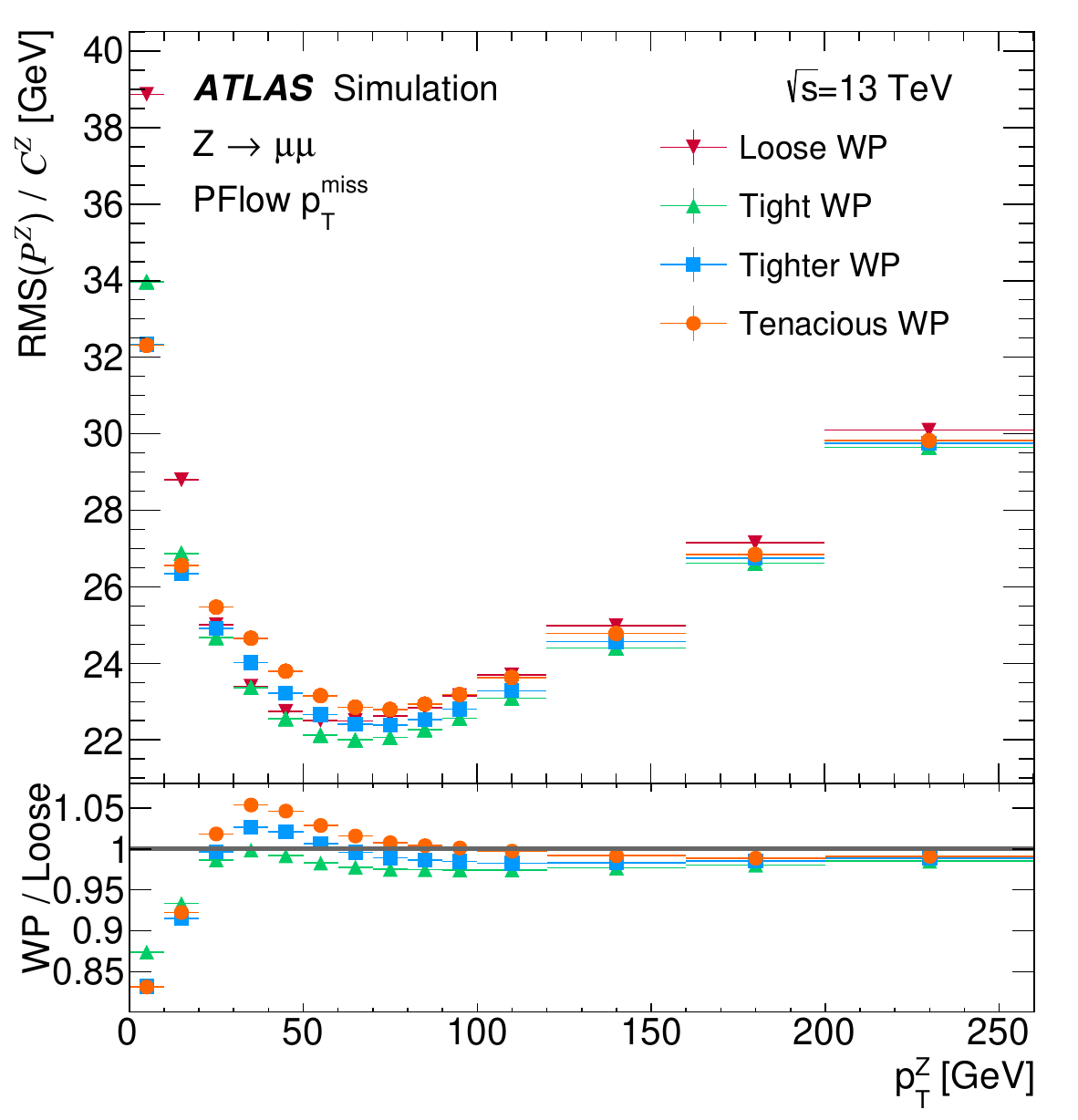}}
\caption{
The (a) average value and (b) RMS of the \scale\ (defined in Eq.~\ref{eq:scale}), corrected by the \(Z\) system \ptmiss\ response \C (defined in Eq.~\ref{eq:C}), for different \ptmiss\ working points as a function of \pTZ.
PFlow jets are used, and \SHERPA\ \Zmm\ MC simulated events.
The error bars include the MC statistical uncertainty. In the lower panel, `WP' refers to the working point under consideration. `True' refers to MC-generated quantities.
}
\label{fig:C}
\end{figure}

\begin{figure}[!h]
\centering
\subfloat[]{\label{fig:RMSOverC:mu}\includegraphics[width=0.4\textwidth]{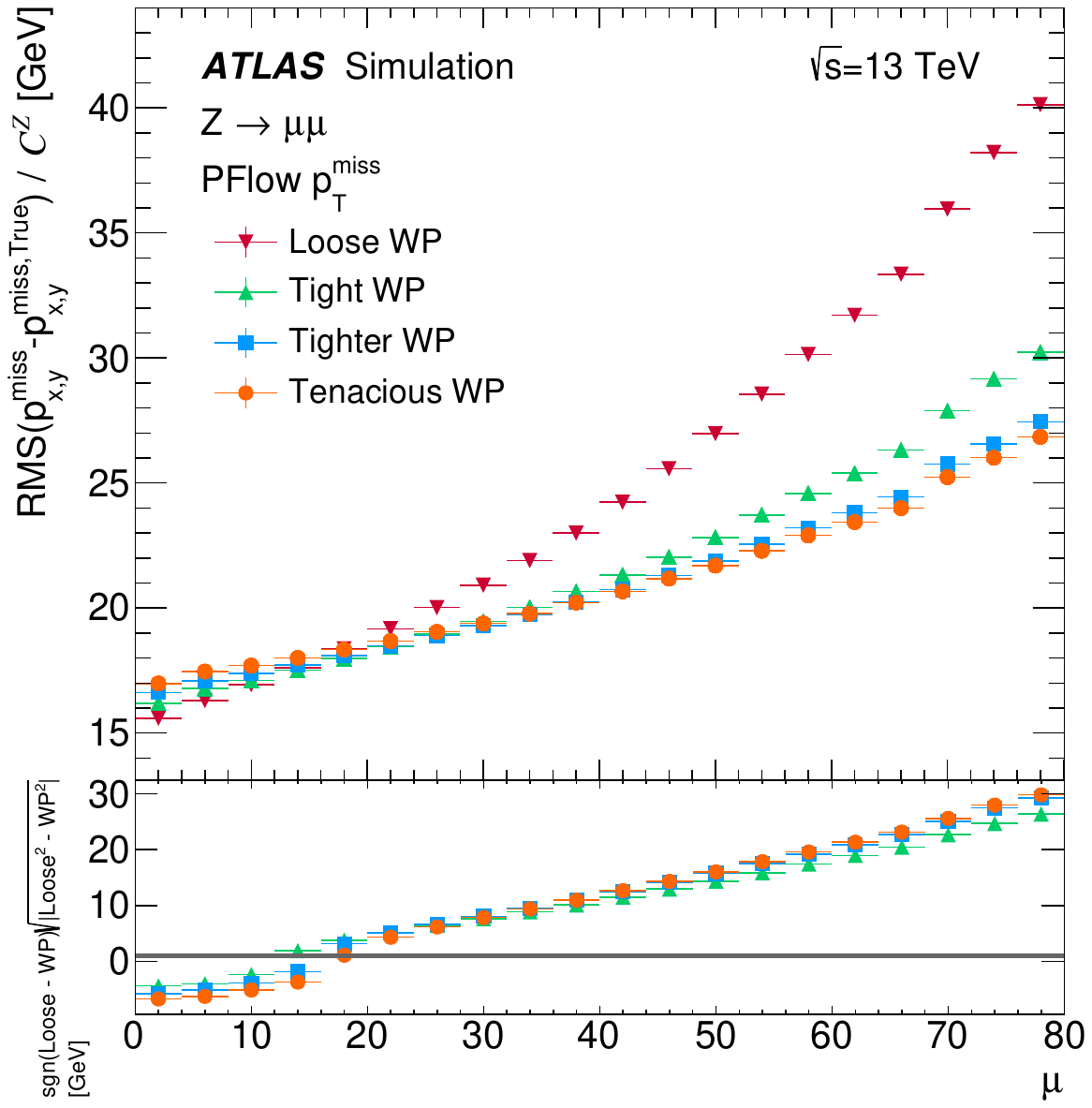}}\qquad
\subfloat[]{\label{fig:RMSOverC:NPV}\includegraphics[width=0.4\textwidth]{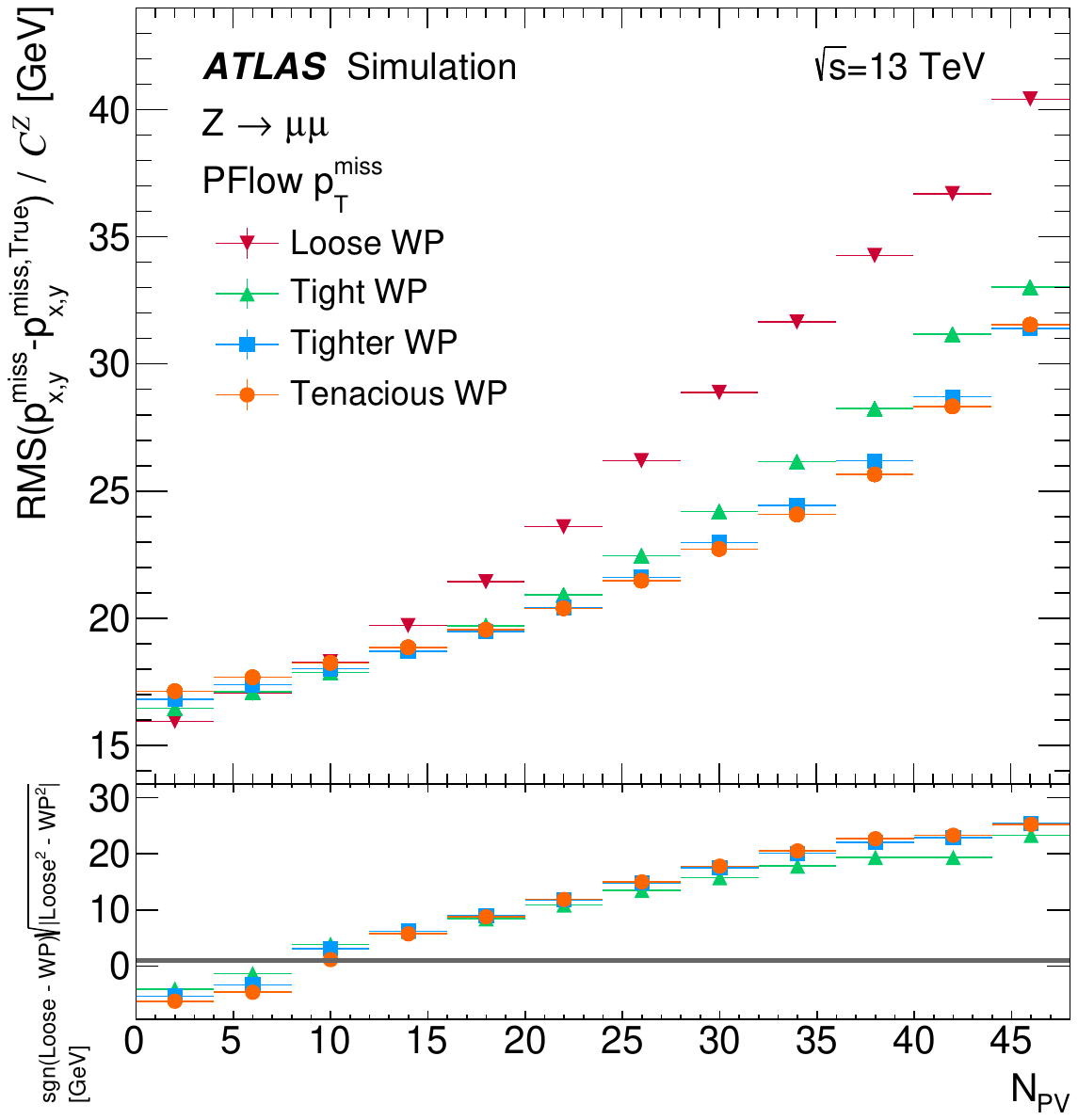}}
\caption{
The \pxmiss\ and \pymiss\ resolution, corrected by the \(Z\) system \ptmiss\ response \C (defined in Eq.~\ref{eq:C}), for different \ptmiss\ working points as a function of (a) \(\mu\) or (b) \npv.
PFlow jets are used, and \SHERPA\ \Zmm\ MC simulated events.
The error bars include the MC statistical uncertainty. In the lower panel, `WP' refers to the working point under consideration. `True' refers to MC-generated quantities.
}
\label{fig:RMSOverC}
\end{figure}

\FloatBarrier


%
\section{Systematic uncertainties}
\label{sec:systs}

Uncertainties on the measurement of \ptmiss\ are calculated for the scale and resolution.
These uncertainties depend on every object entering the \ptmiss\ reconstruction, and thus on both the soft term and the composition of the hard term.
Since the hard term's composition is defined individually for any given analysis, the scale and resolution uncertainty of each of the hard objects must be extracted based on the object definitions used.
This is done for each analysis, using the uncertainty recommendations provided for each object type.
In propagating these uncertainties through \ptmiss\ reconstruction, correlations between systematic uncertainties for the same type of object are taken into account.
However, the systematic uncertainties of each of the different types of object in the hard term are taken to be uncorrelated since independent reconstruction and calibration algorithms are applied to each.
As seen in Section~\ref{sec:modelling}, for topologies dominated by fake \ptmiss\ the dominant uncertainty in the \ptmiss\ distribution can come from the uncertainties in the reconstruction of jets entering the hard term.
For the case of the \ptmiss\ soft term, the scale and resolution uncertainties are calculated as described in the remainder of this section, and these are used for any analysis.
It is expected that the soft term uncertainties only have a significant effect on the overall \ptmiss\ uncertainty when the soft term itself dominates the \ptmiss\ calculation, either because the topology contains few hard objects to contribute to the hard term or because it contains a relatively large amount of soft activity.

\subsection{Methodology}
\label{sec:systs:method}

The uncertainty in the soft term is assumed to be dominated by how well it is modelled by Monte Carlo simulation.
This is best studied in events with no true \ptmiss, where \(\buvec{p}_{\text{T}}^{\text{miss}} = -\buvec{p}_{\text{T}}^{\text{soft}} - \buvec{p}_{\text{T}}^{\text{hard}} = 0\) if the soft term is perfectly reconstructed.
In this case the soft term behaviour can be easily studied by comparing the soft and hard terms.
In practice, fake \ptmiss\ will spoil this balance.
The \Zee\ selection defined in Section~\ref{sec:events} is used for this uncertainty derivation, and it is validated in a \Zmm selection.

The soft term's uncertainty is calculated by quantifying the balance between the hard and soft terms by considering the projection of the soft term onto the hard term.
This leads to three variables used to parametrise the uncertainties, which can be defined with the help of Figure~\ref{fig:systs:projection}.
These are:
\begin{itemize}
\item the parallel scale (\scalepara) --- defined as the mean of the parallel projection of \ptsoftvec\ along \pthardvec, \(\langle \ptsoftpara \rangle\);
\item the parallel resolution (\resopara) --- defined as the root-mean-square of \ptsoftpara;
\item and the perpendicular resolution (\resoperp) --- defined as the root-mean-square of the perpendicular projection of \ptsoftvec\ along \pthardvec, \ptsoftperp.
\end{itemize}
As expected, the perpendicular scale was found to be consistent with zero in both the Monte Carlo and data in Ref.~\cite{PERF-2014-04}, so is not of interest.
\begin{figure}[h]
\centering
\includegraphics[width=0.45\textwidth]{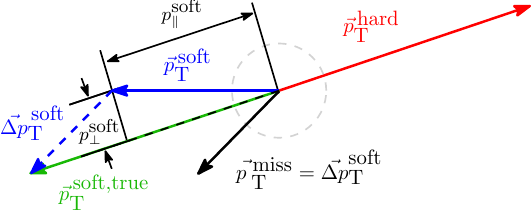}
\caption{\ptsoftvec\ projections along \pthardvec, taken from Ref.~\cite{ATLAS-CONF-2018-023}.}
\label{fig:systs:projection}
\end{figure}

The values of these variables are calculated in different bins of \pthard.
Separate soft term uncertainties are calculated for \ptmiss\ built from EMTopo and PFlow jets, using the \workingPoint{Tight} \ptmiss\ working point, by considering the maximal difference between the data and the different Monte Carlo generators, and taking the maximum of these between both the jet inclusive and 0-jets selections.
The three generators considered are \POWHEG+\Pythia, \MADGRAPH+\Pythia\ and \SHERPA, which are all the standard options available for \Zjets\ processes in ATLAS.

Up to a \pthard\ of 60~\GeV, both the jet inclusive and 0-jets selections are considered.
Due to decreased statistical precision, the 0-jets selection is not used \(\pthard>60~\)\GeV.

To account for contamination of non-\Zee\ events passing the \Zee selection in data, MC simulations of \(VV\) and \ttbar\ processes were included in addition to the various \Zee\ simulations.
At the point in the \pthard\ distributions where these processes start to dominate, the crucial initial assumption of the \ptsoft-\pthard\ balance breaks down.
As was seen in Figure~\ref{fig:termsZmmZee}, this occurs at around 100~\GeV, where the \Zee\ events would require the \(Z\) boson to be increasingly boosted.
As a result, the measurement of the soft term uncertainty stops at \(\pthard=100~\)\GeV, and the value obtained in the final bin up to 100~\GeV is used for any event with a higher \pthard.

\subsection{Uncertainty values}
\label{sec:systs:values}

Figure~\ref{fig:systs:fullPFlow} shows the three variables for the jet inclusive and 0-jets selection using PFlow jets, in the same bins of \pthard\ used for the uncertainty calculation.
The distributions are given for data and the different Monte Carlo generators, with the uncertainty values (labelled as `TST syst. uncert.', short for track soft term systematic uncertainty) illustrated as a shaded band centred on the data.

In comparison to preliminary results presented in Figure 6 of Ref.~\cite{ATLAS-CONF-2018-023}, a reduction in the uncertainty values for scale and resolution is seen throughout the \pthard\ distribution, after the improvements described here.
Scale uncertainties are reduced by up to $76\%$ and resolution uncertainties are reduced by up to $51\%$.
For a representative example considering the \([30,35]~\)\GeV\ bin of \pthard, the parallel scale uncertainty is reduced in comparison to the previous results by \(52\%\) (dropping from 0.97 \GeV\ to 0.47 \GeV), the parallel resolution uncertainty is reduced by \(43\%\) (dropping from 2.59 \GeV\ to 1.47 \GeV), and the perpendicular resolution uncertainty is reduced by \(13\%\) (dropping from 2.29 \GeV\ to 2.00 \GeV).

Below \(\sim20~\)\GeV, the uncertainties are dominated by the 0-jets selections where the \pt\ of the \(Z\)-boson directly balances the soft term.
Above this, the jet inclusive selection starts to dominate, where the soft term consists mainly of diffuse radiation which hasn't formed jets.
The values of the soft term uncertainties calculated for PFlow, are shown in Figure~\ref{fig:systs:summary}.
The PFlow uncertainties are generally smaller than EMTopo (shown in Figure~\ref{fig:systs:summaryEMTopo} in Appendix~\ref{sec:app_EMTopo}), attributed to better rejection of poorly modelled pile-up, which is consistent with the performance seen in the previous section.

The parallel resolution uncertainty, which relates largely to mismeasurement of the jets which recoil the \(Z\) and grow in \pt\ with the \(Z\), increases with \pthard.
The transverse resolution uncertainty relates to other effects and is less dependent on the \pthard.
Thus, the \resoperp\ uncertainty dominates (in terms of absolute uncertainty value) at low values and \resopara\ dominated beyond around \(\pthard=60~\)\GeV.
To validate the uncertainties for the \workingPoint{Tight} working point, they are applied to the three variables calculated for \Zmm\ events, and successfully cover \Zmm\ MC/data discrepancies.
To validate the use of the uncertainties for other working points, they are applied to \Zee\ events where \ptmiss\ is reconstructed using the \workingPoint{Loose}, \workingPoint{Tighter}, or \workingPoint{Tenacious} working points.
Again the uncertainty band successfully covers MC/data differences.

To apply the calculated systematic resolution uncertainties in an ATLAS analysis, the soft term projection is smeared by a Gaussian function with a width corresponding to the resolution uncertainty in the relevant \pthard\ bin.
It is conventional to symmetrise the variation of the soft-term to produce a \(\pm\) error band.
The systematic uncertainty in the scale is applied by either adding or subtracting the scale uncertainty (\scalepara) for the appropriate \pthard\ bin to the value of the parallel component of the soft term, \ptsoftpara.

\begin{figure}[!htb]
\centering
\subfloat[]{\includegraphics[width=0.4\textwidth]{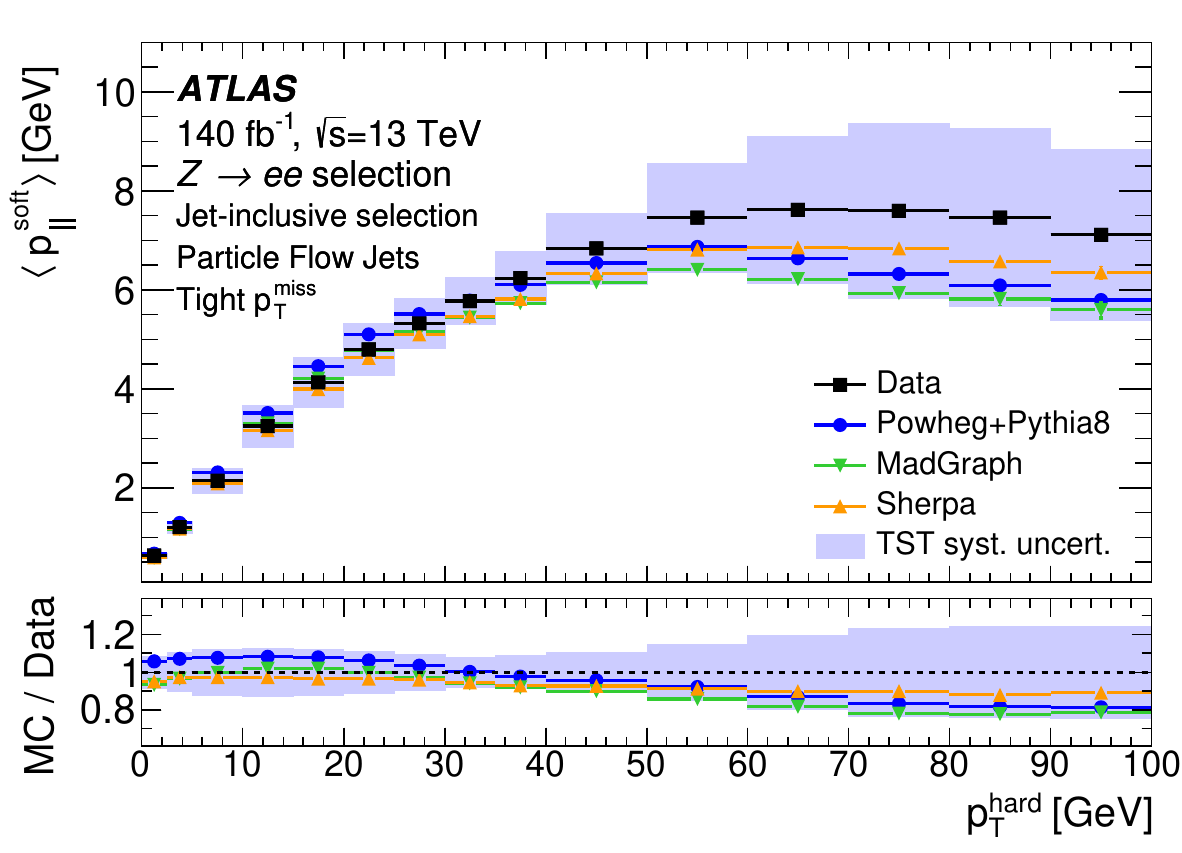}}\qquad
\subfloat[]{\includegraphics[width=0.4\textwidth]{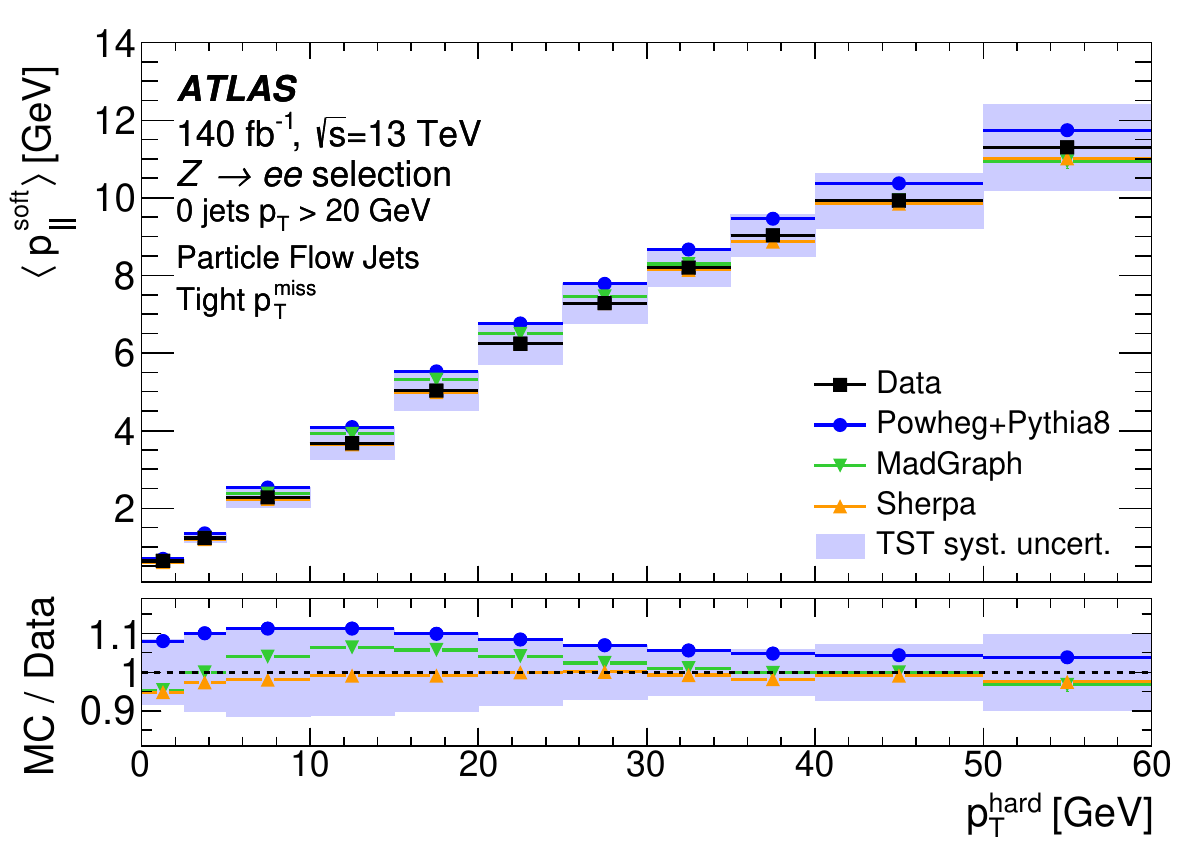}}\\
\subfloat[]{\includegraphics[width=0.4\textwidth]{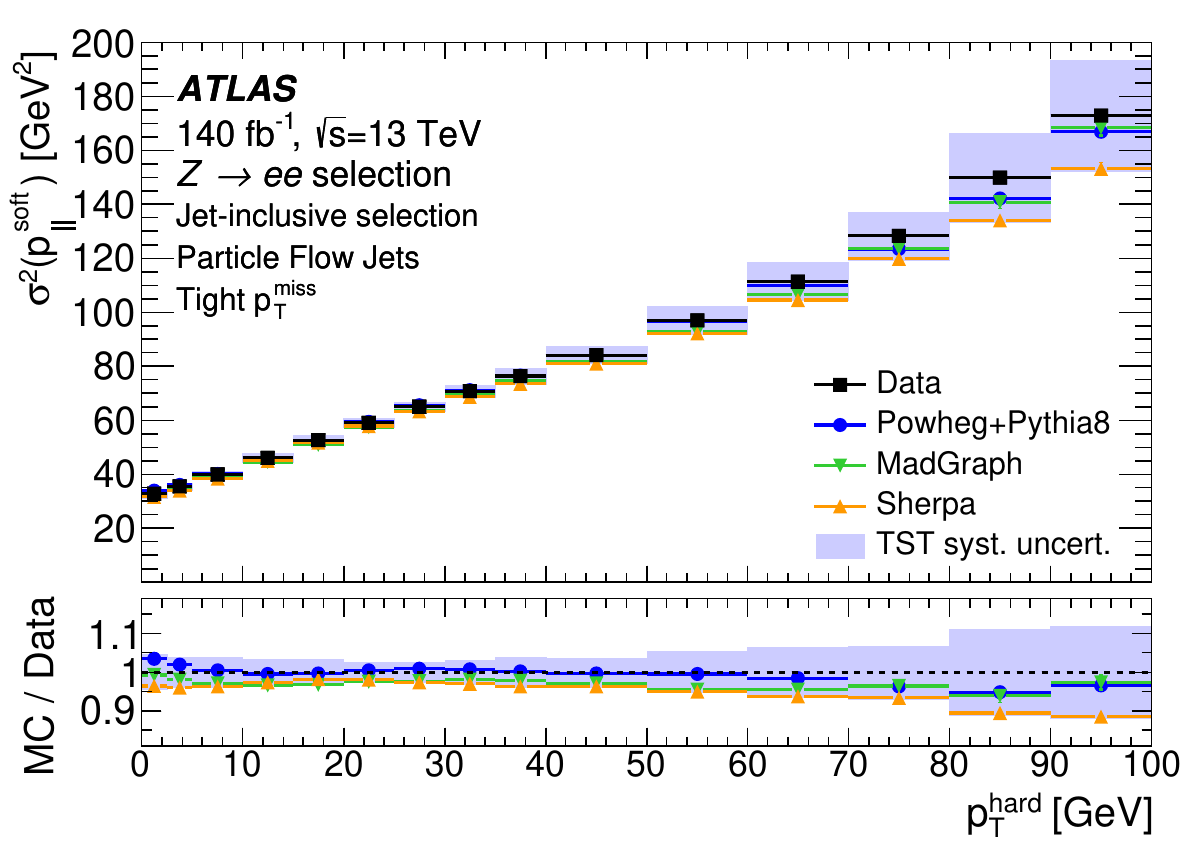}}\qquad
\subfloat[]{\includegraphics[width=0.4\textwidth]{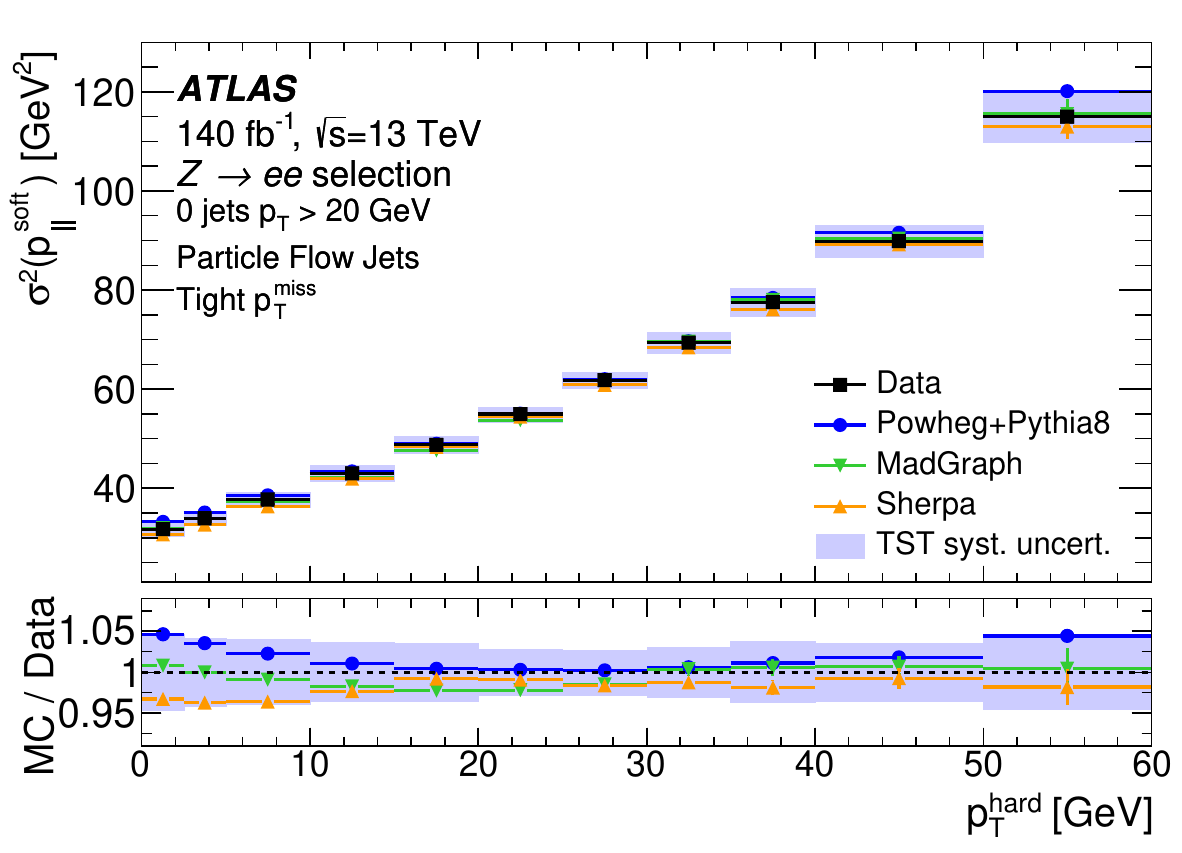}}\\
\subfloat[]{\includegraphics[width=0.4\textwidth]{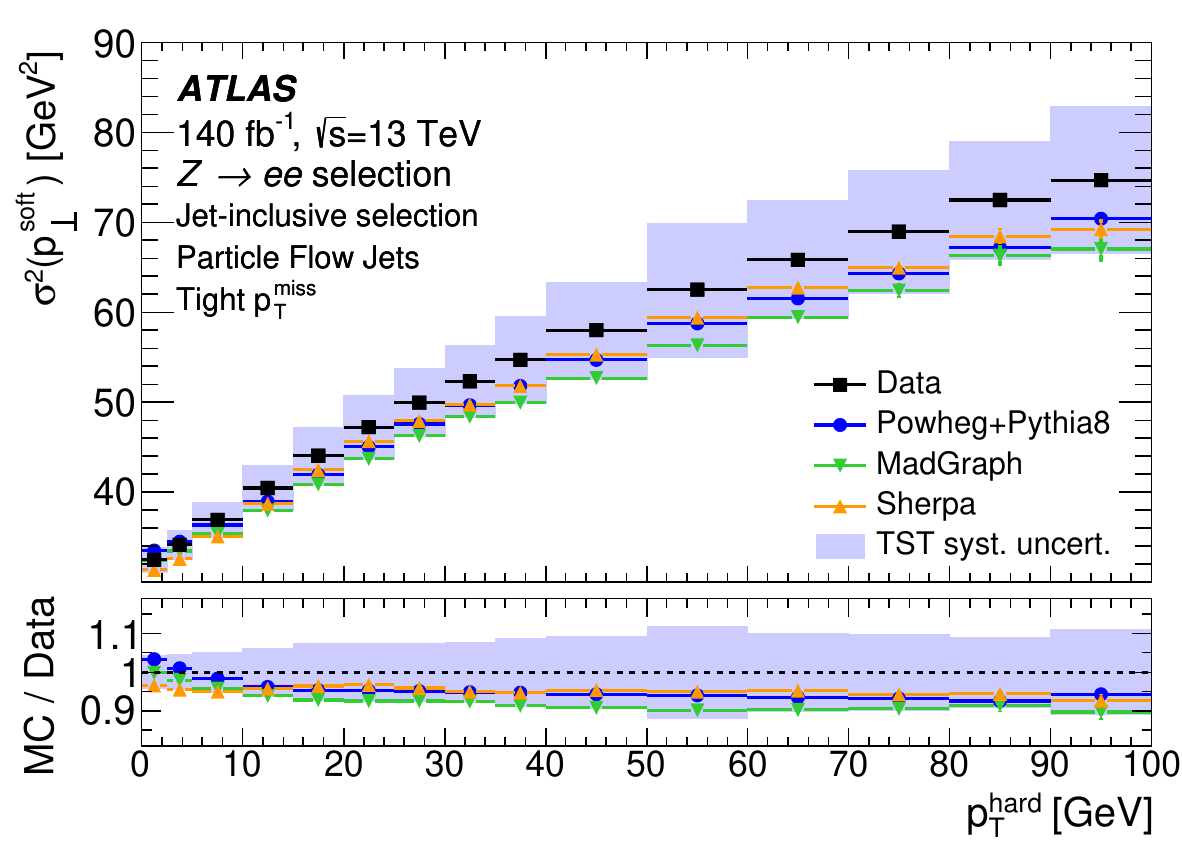}}\qquad
\subfloat[]{\includegraphics[width=0.4\textwidth]{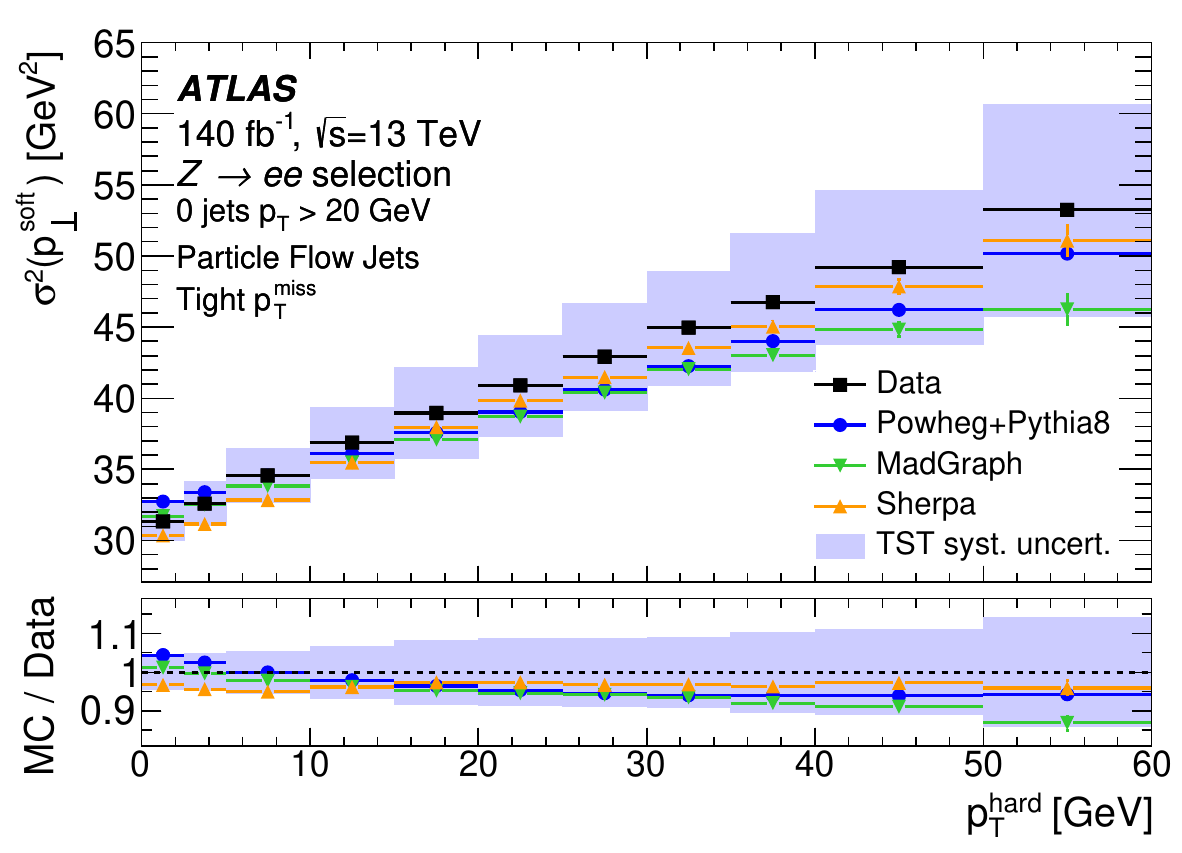}}
\caption{Parallel scale (\scalepara, (a) and (b)), parallel resolution squared (\resopara, (c) and (d)) and transverse resolution squared (\resoperp, (e) and (f)) plots for the \ptsoft\ (TST, track soft term) in bins of \pthard. Full \runtwo\ data and MC simulated samples are shown with a \Zee\ event selection applied using PFlow jets, in the jet inclusive ((a), (c) and (e)) or 0-jets selections ((b), (d) and (f)). Full \runtwo\ uncertainties are shown as a shaded band about the data.}
\label{fig:systs:fullPFlow}
\end{figure}

\begin{figure}[!htb]
\centering
\includegraphics[width=0.45\textwidth]{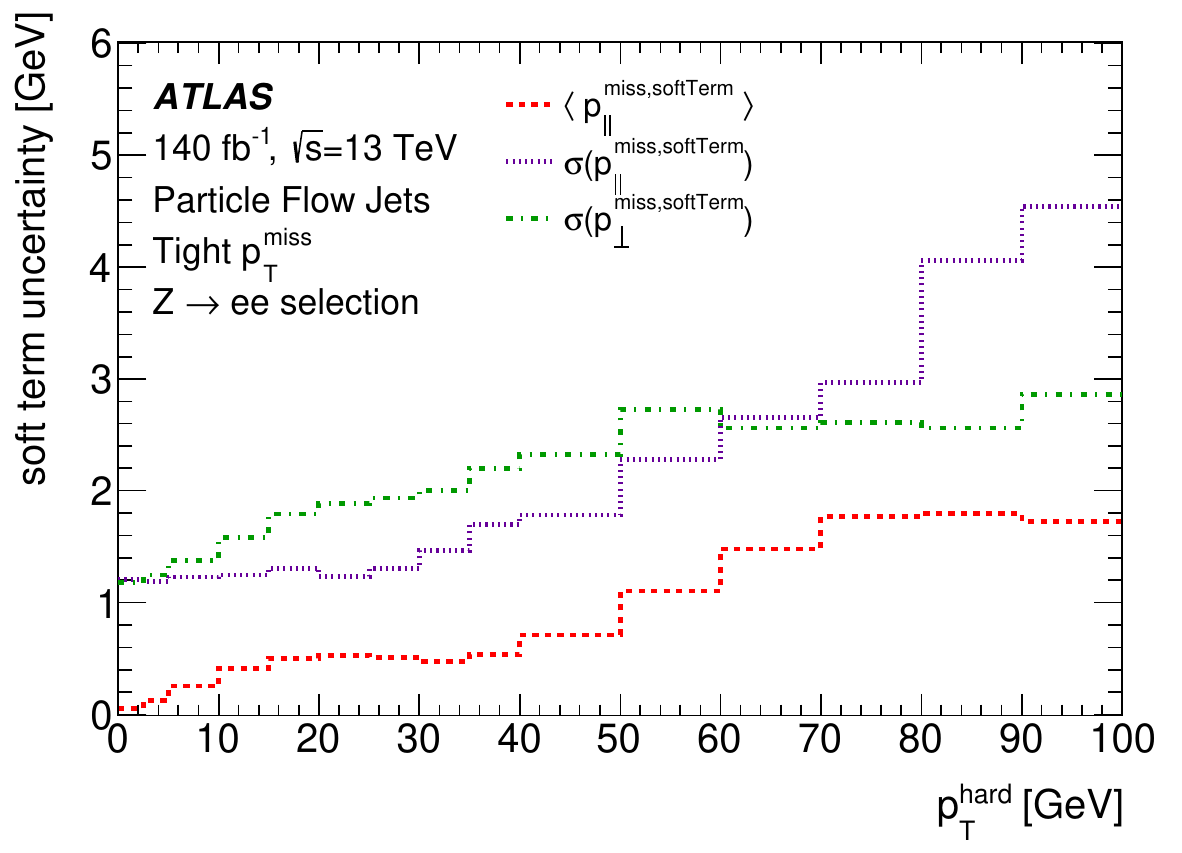}
\caption{A summary of the \ptsoft\ systematic uncertainties. The parallel scale (\scalepara), parallel resolution (\resopara) and transverse resolution (\resoperp) of the \ptsoft\ projection onto the \pthard, binned in \pthard. Full \runtwo\ data and Monte Carlo samples are shown with a \Zee\ event selection applied.}
\label{fig:systs:summary}
\end{figure}
\FloatBarrier


%
\section{\ptmiss\ significance}
\label{sec:metsig}

\subsection{\ptmiss\ significance definitions}

In association with \ptmiss, the concept of a \ptmiss\ `significance' can be defined to quantify the belief that the reconstructed \ptmiss\ is real.
As well as being useful to identify SM processes with neutrinos in the final state, such a variable is extremely useful in searches for new stable particles, where typically a large amount of real \ptmiss\ is expected in the new-physics signal process but not in SM background processes, or in the opposite scenario of searches for new physics processes with no real \ptmiss\ in the final state and SM backgrounds with neutrinos present.
In any case, a \ptmiss\ significance variable can often more effectively discriminate between the signal and backgrounds than \ptmiss\ alone.
Ten examples of searches or  measurements where this has been the case can be found in Refs.~\cite{SUSY-2018-32,SUSY-2018-17,SUSY-2023-01,HDBS-2021-02,HDBS-2019-02,SUSY-2019-12,EXOT-2018-29,EXOT-2020-06,STDM-2019-09,STDM-2017-19}.

ATLAS initially used event-based \ptmiss\ significance approximations.
Subsequently, the \ptmiss\ significance definition has adroitly evolved to follow a similar object-based approach to that used in calculating \ptmiss\ itself.
This new object-based \ptmiss\ significance performs better at discriminating between real and fake \ptmiss.
Both approaches are discussed here.

\subsubsection{Event-based \ptmiss\ significance}
\label{sec:metsig:event}

As a first attempt at quantifying a measure of the `realness' of \ptmiss, a heuristic definition was considered that approximated the resolution of \ptmissvec\ using the square root of the scalar sum of all jet \pT\,
\begin{equation*}
\HT = \sum_{i} p_{\mathrm{T, i}},
\end{equation*}
where the index runs over the jets in an event. The approximation of \ptmiss\ significance (\(\mathcal{S}\)), made possible because \HT\ scales with \ptmissvec\ resolution, is written as
\begin{equation*}\label{eq:METSigHT}
\mathcal{S}_{\HT} = \frac{\ptmiss}{\sqrt{\HT}}.
\end{equation*}

Another approximation for the resolution was based on the sum of all the reconstructed objects in the detector defined in Eq.~(\ref{eq:reco:underbrace}), \(\sqrt{\sumpt}\), and allowed the significance to be written as:
\begin{equation*}\label{eq:METSigSumET}
\mathcal{S}_{\Sigma} = \frac{\ptmiss}{\sqrt{\sumpt}}.
\end{equation*}

These definitions are formed from proxies for the resolution of \ptmissvec\ and so are not true dimensionless significances.
Both \(\sqrt{\HT}\) and \(\sqrt{\sumpt}\) are event-by-event proxies for resolution that scale linearly with \ptmissvec\ resolution under the assumption that only calorimeter signals are used to build \ptmissvec.
This is not the case when one wishes to use the tracker for its improved pile-up rejection and better \pT\ resolution at low momentum for charged particles.

\subsubsection{Object-based \ptmiss\ significance}
\label{sec:metsig:obj}

Section~\ref{sec:reco} introduced the concept of an object-based approach to \ptmissvec, described in Eq.~(\ref{eq:reco:underbraceptmiss}).
An analogous approach using these objects and their detector resolutions can be used to define an improved, object-based, \ptmiss\ significance.
This significance encodes the resolutions of all reconstructed objects\footnote{This considers the \pt\ and \(\eta\) dependence of objects' detector resolution.} and accounts for the correlations between each object in an event.
Appendix~\ref{sec:app_metsig} provides a detailed derivation of this quantity; in this section a more concise overview is presented.

To determine if the observed missing transverse momentum is real or fake in origin, a hypothesis test can be performed.
This compares the hypothesis with no momentum carried by invisible particles (\(\ptmisstruevec = 0\)) to that with there being genuine \pT\ carried by invisible particles (\(\ptmisstruevec \neq 0\)).
The missing transverse momentum significance (\ptmisssig) definition,
\begin{equation}\label{eq:METSigHypoTest}
\mathcal{S}^2 = 2 \mathrm{ln} \left(\frac{\mathrm{max}_{\ptmisstruevec \neq 0} \mathcal{L}\left( \ptmissvec | \ptmisstruevec \right)}{\mathrm{max}_{\ptmisstruevec = 0} \mathcal{L}\left( \ptmissvec| \ptmisstruevec \right)} \right),
\end{equation}
is formed by this test, where \(\mathcal{L}\) is the likelihood (the `true' label refers to MC generated quantities).
This log likelihood ratio, based on the Neyman-Pearson lemma \cite{Neyman:1933wgr}, assumes that each of the likelihoods depends on all the objects measured in an event; their multiplicities, types and kinematic properties.

In addition to the log likelihood ratio, the functional form of \(\mathcal{L}\left( \ptmissvec | \ptmisstruevec \right)\) is required to calculate \ptmisssig. This can be found following a few assumptions.
Firstly, the \pT\ measurement for each object, \pTobjvec, is assumed to be independent of all others (where \(\mathrm{Obj \in \{ e, \, \gamma, \, \tau, \, \mu, \, \mathrm{jet} \}}\)).
For all objects, \pTobjvec\ (given a true value of \pTobjtruevec) is taken to follow a Gaussian probability distribution of the form \(\mathrm{Gaus} \left( \pTobjvec - \pTobjtruevec \right)\).
The probability distribution for each object has a covariance matrix labelled \metsigcovobj, which is the sum of covariances quantifying the resolutions of each object, in \pT\ and \(\phi\), entering the \ptmissvec\ calculation.
Finally, conservation of momentum in the transverse plane means that if the true momentum of each measured particle were to be summed this would balance with the negative signed invisible particle momentum: \(\sum_{\mathrm{Objects}}\pTobjtruevec = - \ptmisstruevec.\)
With these assumptions made, the form of the likelihood is a two dimensional Gaussian distribution.
Entering this into the maximised log likelihood ratio, Eq.~(\ref{eq:METSigHypoTest}), results in the cancellation of any preceding coefficients and leaves:
\begin{equation}\label{eq:METSigChiSquare}
\mathcal{S}^2 = \left( \ptmissvec \right)^{\mathrm{T}} \left( \sum_{\mathrm{Objects}} \metsigcovobj \right)^{-1} \left( \ptmissvec \right).
\end{equation}

This is now a sum of independent standard Gaussian-shaped variables in two dimensions, or more simply a \chiSq\ hypothesis test in two dimensions.
Equation~(\ref{eq:METSigChiSquare}) links \ptmissvec\ to all the object resolutions which are encoded in the covariance matrix summation.

In this format the results of the \chiSq\ test are easily interpreted with a single value that indicates how likely it is that the null hypothesis (\(\pTobjtruevec = 0\)) holds.
Low values of \(\mathcal{S}^2\) indicate that the \ptmissvec\ comes from fake sources like mismeasurement or resolution effects while high values show that it is likely the \ptmissvec\ comes from a real invisible particle leaving the detector without interactions. The covariance matrix for each object is defined with an axis along the measured transverse momentum vector of the object under consideration, \pTobjvec.

After some matrix calculation covered in detail in Appendix~\ref{sec:app_metsig}, one obtains the final definition of \ptmisssig:
\begin{equation}\label{eq:METSigDefroot}
\ptmisssig = \frac{\ptmiss}{\sqrt{\sigma_{\mathrm{L}}^{2} \left( 1 - \rho_{\mathrm{LT}}^2\right)}}.
\end{equation}
Here \(\sigma_{\mathrm{L}}\) defines the resolution longitudinally to \ptmissvec\ and \(\rho_{\mathrm{LT}}\) is the correlation between the transverse and longitudinal resolutions relative to \ptmissvec, calculated from the covariance matrix.
This dimensionless variable contains the measured quantity in the numerator, along with a measure of its variance in the denominator.
Code to implement the object-based \ptmiss\ significance externally to the ATLAS Collaboration, including the object resolution values used, can be found in the \workingPoint{SimpleAnalysis} Framework~\cite{ATL-PHYS-PUB-2022-017}.

\subsection{\ptmiss\ significance modelling and performance}
\label{sec:sigperf}

Figure~\ref{fig:oldmetsigZmm:METOverSumpT} shows the original calorimeter dependent significance proxy, \(\mathcal{S}_{\Sigma}\), for events that satisfy a \Zmm\ selection.
The \workingPoint{Tight} and PFlow jets are used to build the \ptmiss, and the jet inclusive selection is applied.
\SHERPA\ is used to generate the \Zmm\ MC simulation events.
The low values are dominated by events with an expected truth \ptmiss\ of zero, which have some fake \ptmiss.
The high valued tails are more dominated by events from other processes that have a high energy neutrino produced and satisfy the \Zmm\ selection in data.
Figure~\ref{fig:oldmetsigZmm:METSigHT} shows a different event-based significance estimate, \(\mathcal{S}_{\HT}\), which indicates a larger estimate of events which are likely to have real \ptmiss\ in them.

The object-based missing transverse momentum significance derived in Section~\ref{sec:metsig} is presented in Figure~\ref{fig:newmetsigZmm:obj}.
The \ptmisssig\ distribution for the \Zmm\ events moves closer to the expected value of zero, whilst the other processes move to higher values.
It shows good agreement between data and MC in the bulk where \Zmm\ events dominate and the MC simulations used in this paper are expected to be more representative of the data.
The behaviour here is closer to that of Figure~\ref{fig:oldmetsigZmm:METOverSumpT} than Figure~\ref{fig:oldmetsigZmm:METSigHT} and reinforces the statement that \(\sqrt{\sum \pT}\) is a good proxy for the resolution of \ptmiss.

One can also investigate how the resolution terms in the denominator impact the agreement between data and prediction by defining a directional \ptmiss\ significance (\(\mathcal{S}_{\mathrm{dir}}\)) that only has the longitudinal resolution in Eq.~(\ref{eq:METSigDefroot}) and so remove any input from \(\sigma_{\mathrm{T}}\).
This is shown in Figure~\ref{fig:newmetsigZmm:dir}, which looks very similar to Figure~\ref{fig:newmetsigZmm:obj} suggesting a small impact in this \Zmm\ event topology.

\begin{figure}
\centering
\subfloat[]{\label{fig:oldmetsigZmm:METOverSumpT}\includegraphics[width=0.4\textwidth]{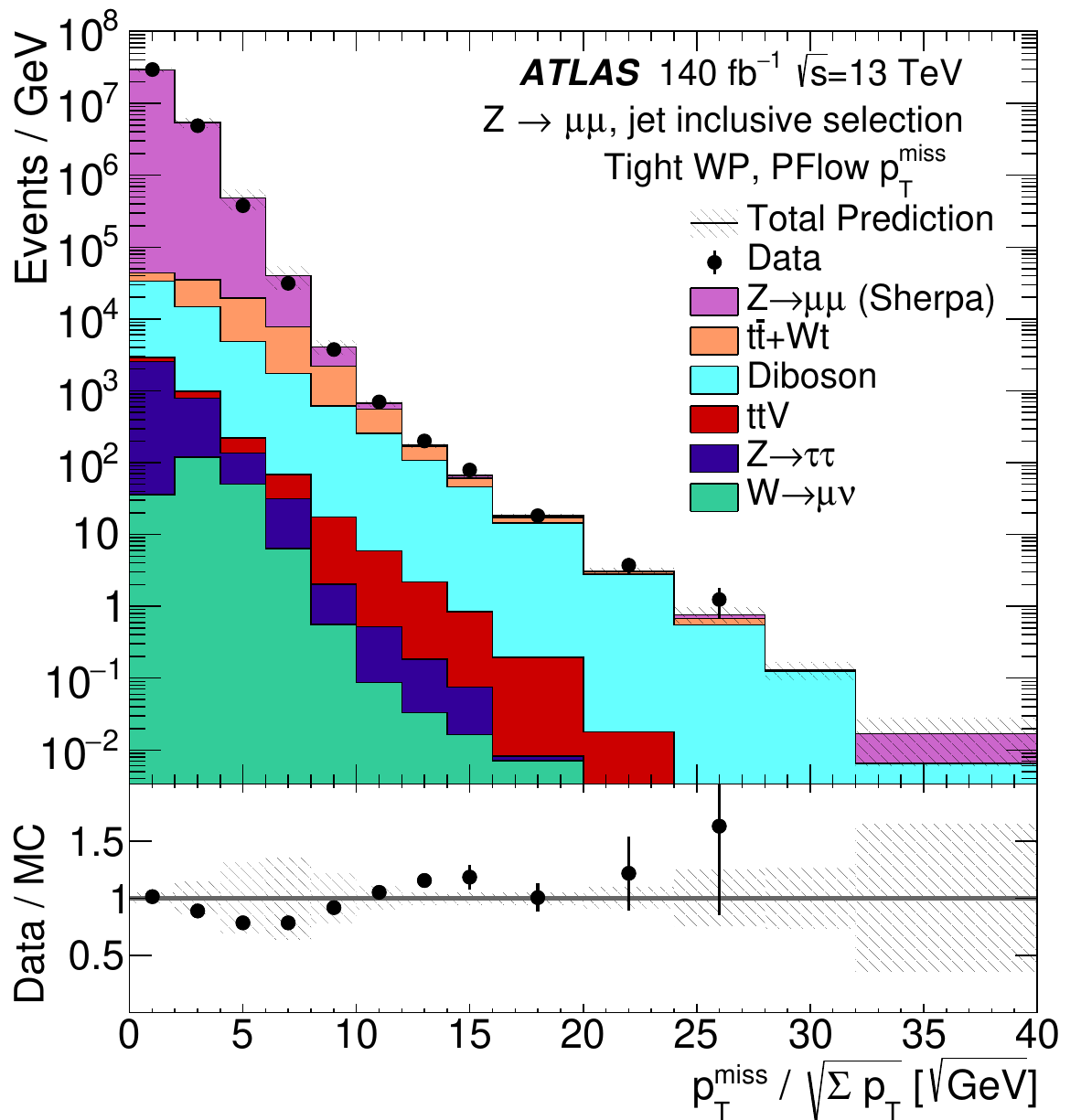}}
\subfloat[]{\label{fig:oldmetsigZmm:METSigHT}\includegraphics[width=0.4\textwidth]{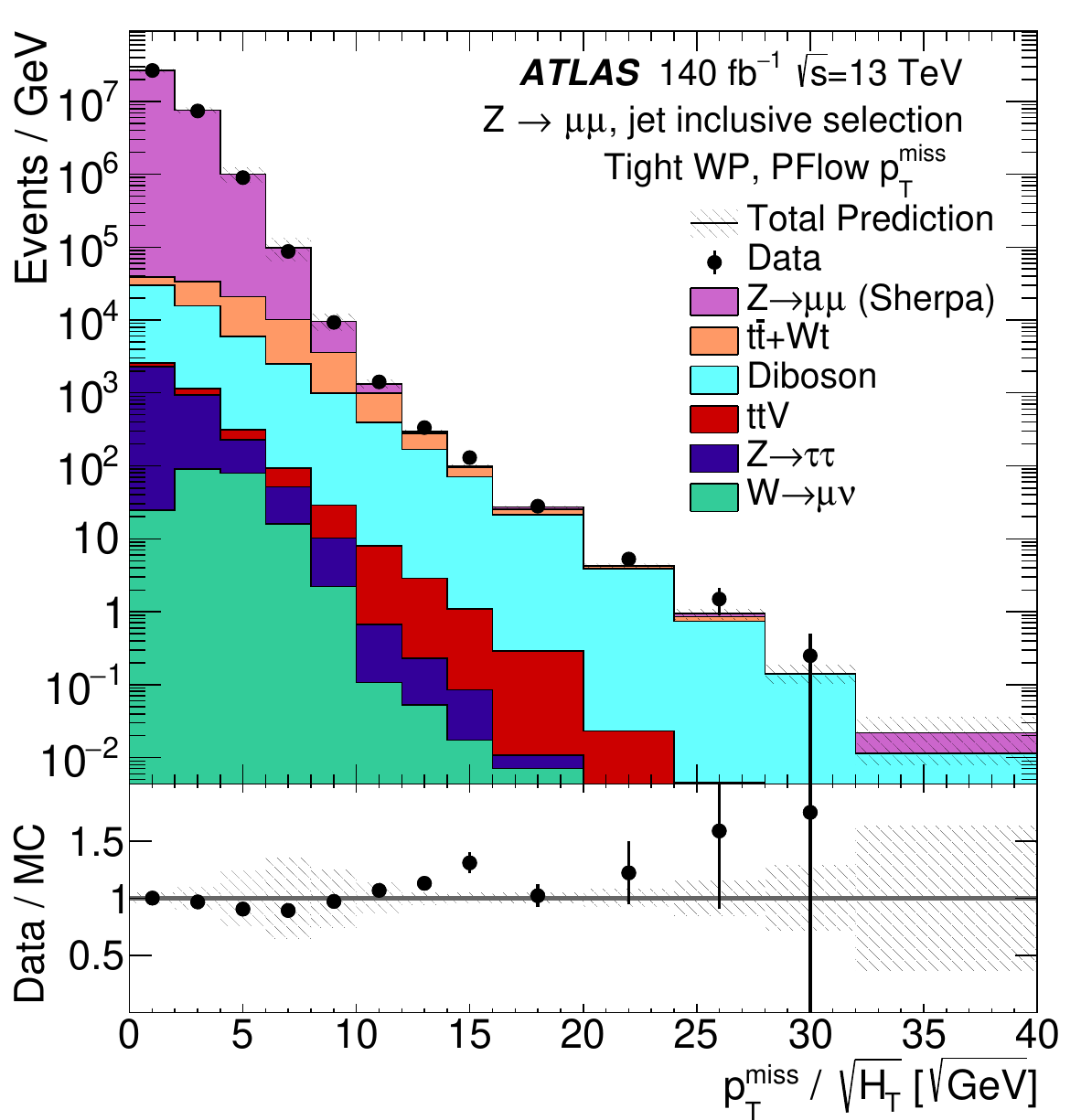}}\\
\subfloat[]{\label{fig:newmetsigZmm:obj}\includegraphics[width=0.4\textwidth]{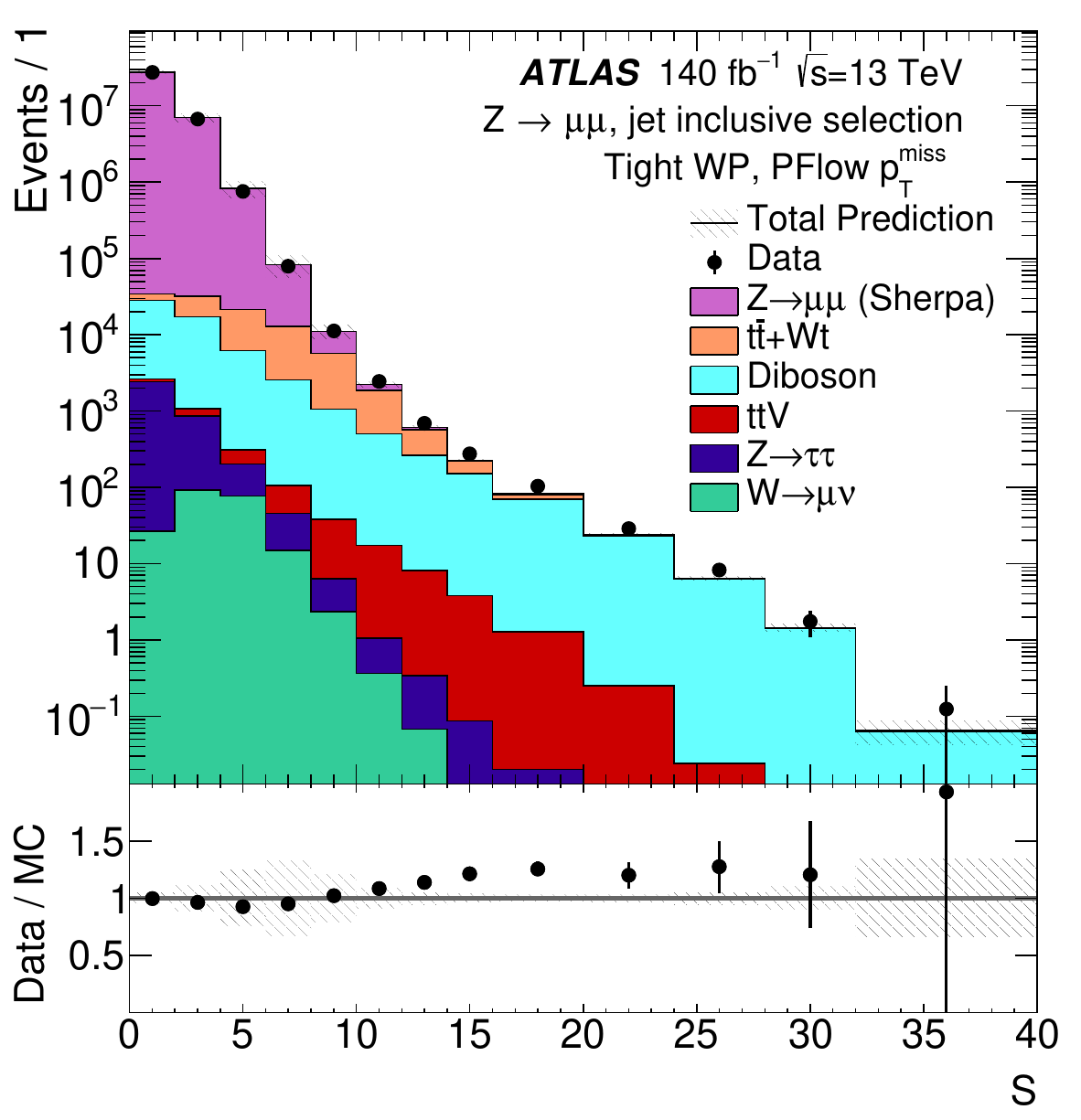}}
\subfloat[]{\label{fig:newmetsigZmm:dir}\includegraphics[width=0.4\textwidth]{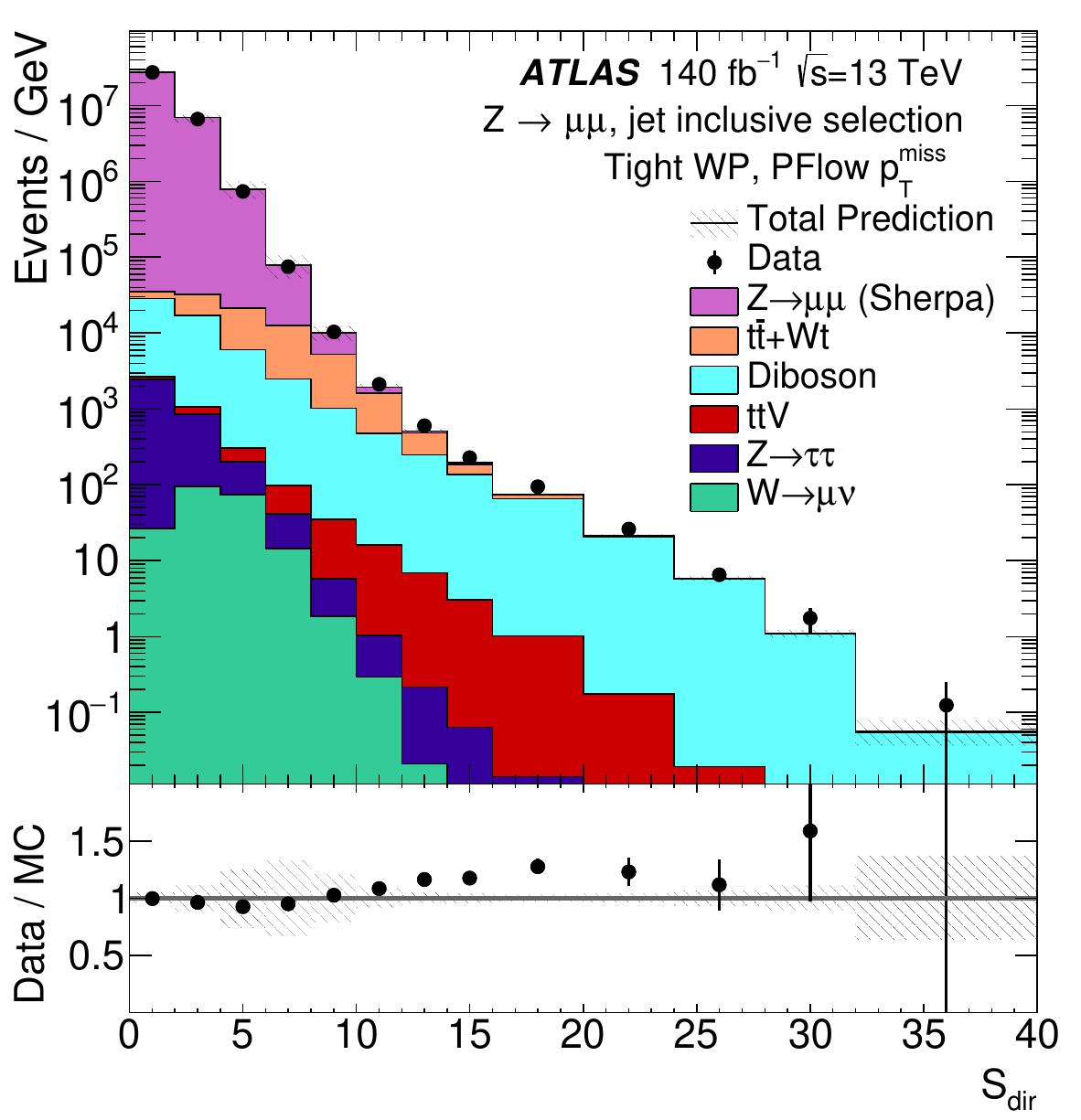}}\\
\caption{Event-based proxies for \ptmiss\ significance ((a) and (b)), Object-based \ptmiss\ significance (c), and its directional form (d), in \Zmm events. \ptmiss\ is built using the \workingPoint{Tight} working point and PFlow jets. \SHERPA\ is used to generate the \Zmm\ events. The error band includes MC statistical, luminosity and detector uncertainties.}
\label{fig:oldmetsigZmm}
\end{figure}

The performance of the various \ptmiss\ significance definitions at discriminating between processes with real and fake \ptmiss\ is assessed next.
This is done by calculating ROC curves for each definition, to determine background rejection against signal efficiency, as shown in Figure~\ref{fig:METSigROC}.
The comparison is made in the \ttbar\ event selection, considering semileptonically decaying \ttbar\ MC simulated events as the real \ptmiss\ signal, and MC simulated \Zmm\ events as the fake \ptmiss\ background which contaminates the event selection.
In an ATLAS analysis, a common use of \ptmiss\ significance would be as a selection requirement on the events entering the analysis region, and ideally the signal efficiency and background rejection should both be maximised through a particular threshold on the significance value.
The ROC curves demonstrate that discrimination power improves with the object-based significance measures in comparison to the event-based definitions.
The directional significance \(\mathcal{S}_{\mathrm{dir}}\) has a very similar definition to the object-based significance and shows a comparable, although marginally worse, performance.

\begin{figure}
\centering
\includegraphics[width=0.45\textwidth]{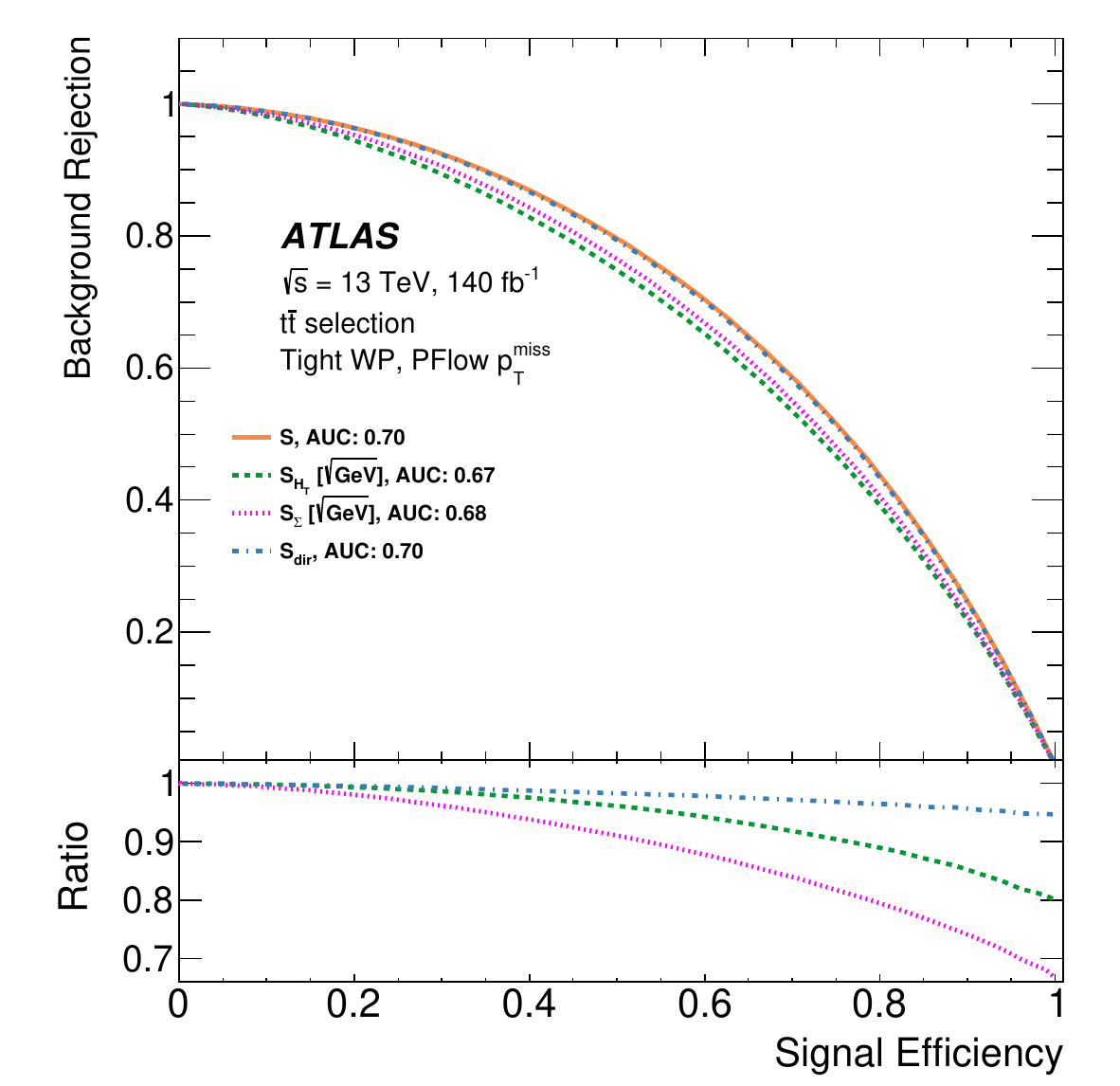}
\caption{Background rejection versus signal efficiency in simulated \Zmm and \ttbar events. Both samples have a \ttbar event selection applied. All events passing the selections are used to calculate background rejection and signal efficiency using the \workingPoint{Tight} and PFlow jets to build \ptmiss. The Area Under the Curve (AUC) value is shown beside each significance definition in the legend.}
\label{fig:METSigROC}
\end{figure}


\FloatBarrier

\section{Conclusion}
\label{sec:conclusion}

This paper presents the performance of missing transverse momentum and its significance in \(140~\ifb\) of proton-proton collisions recorded at a center-of-mass energy of 13\,\TeV, acquired by the ATLAS experiment between 2015 and 2018.
A complete description of \ptmiss\ reconstruction is given, including the update to the particle flow jet collection, and the definitions of four working points to allow more stringent removal of pile-up contamination for analyses that require it.
The state-of-the-art object-based \ptmiss\ significance is derived, in comparison to earlier event-based approximations.
Comparisons of MC simulation and data are shown for various \ptmiss\ quantities, with a \Zll\ selection applied.
There is generally good agreement, particularly in the overall \ptmiss\ distribution for all \ptmiss\ working points, jet definitions and MC generators considered.
The \ptmiss\ significance modelling is also satisfactory, and showed a better separation between topologies with real and fake \ptmiss\ in comparison to the event-based approximations.

Firstly, the dependence of the \ptmiss\ resolution on pile-up is shown by comparing different jet selections --- demonstrating that almost all pile-up dependence originates from jets in the \ptmiss\ calculation, as expected.
Secondly, \ptmiss\ working points are compared, demonstrating success at improving the otherwise degraded resolution at high pile-up by up to 39\% as the working points are tightened from \workingPoint{Loose} to \workingPoint{Tenacious}.
The resolution is considered for several processes to demonstrate that all working points are useful.
The comparison of reconstructed and truth \ptmiss\ is made for each working point, as a function of the truth \ptmiss.
All working points behave similarly here, with reconstructed \ptmiss\ overestimating the truth \ptmiss\ for low values of truth \ptmiss, and estimating it well at higher values.
Finally the \ptmiss\ scale is shown to be similar between data and MC simulation in \Zee\ events, with both showing an underestimation of the hadronic recoil.

Systematic uncertainties in the \ptsoft\ scale and resolution are calculated using \Zee\ events, by considering how well data and MC simulation meet the expectation of a perfect balance between \pthard\ and \ptsoft\ in events with zero real \ptmiss.
The uncertainties are calculated as the maximal disagreement between data and MC simulation in three different \Zll\ generators, in bins of \pthard.
The uncertainty values are reduced throughout the \pthard\ distribution, by up for $76\%$ for scale and up to $51\%$ for resolution.

\runtwo\ \ptmiss\ reconstruction at ATLAS is observed to be resilient against rising pile-up, overall the modelling is good and the disagreement in the \ptsoft\ modelling is evaluated and taken into account with systematic uncertainties.
As an important detector signature for ATLAS, \ptmiss\ will continue to be a robust component of many physics analyses to come.


%

%
\section*{Acknowledgements}
%

%
%

%
%

We thank CERN for the very successful operation of the LHC and its injectors, as well as the support staff at
CERN and at our institutions worldwide without whom ATLAS could not be operated efficiently.

The crucial computing support from all WLCG partners is acknowledged gratefully, in particular from CERN, the ATLAS Tier-1 facilities at TRIUMF/SFU (Canada), NDGF (Denmark, Norway, Sweden), CC-IN2P3 (France), KIT/GridKA (Germany), INFN-CNAF (Italy), NL-T1 (Netherlands), PIC (Spain), RAL (UK) and BNL (USA), the Tier-2 facilities worldwide and large non-WLCG resource providers. Major contributors of computing resources are listed in Ref.~\cite{ATL-SOFT-PUB-2025-001}.

We gratefully acknowledge the support of ANPCyT, Argentina; YerPhI, Armenia; ARC, Australia; BMWFW and FWF, Austria; ANAS, Azerbaijan; CNPq and FAPESP, Brazil; NSERC, NRC and CFI, Canada; CERN; ANID, Chile; CAS, MOST and NSFC, China; Minciencias, Colombia; MEYS CR, Czech Republic; DNRF and DNSRC, Denmark; IN2P3-CNRS and CEA-DRF/IRFU, France; SRNSFG, Georgia; BMBF, HGF and MPG, Germany; GSRI, Greece; RGC and Hong Kong SAR, China; ISF and Benoziyo Center, Israel; INFN, Italy; MEXT and JSPS, Japan; CNRST, Morocco; NWO, Netherlands; RCN, Norway; MEiN, Poland; FCT, Portugal; MNE/IFA, Romania; MESTD, Serbia; MSSR, Slovakia; ARIS and MVZI, Slovenia; DSI/NRF, South Africa; MICINN, Spain; SRC and Wallenberg Foundation, Sweden; SERI, SNSF and Cantons of Bern and Geneva, Switzerland; NSTC, Taipei; TENMAK, T\"urkiye; STFC/UKRI, United Kingdom; DOE and NSF, United States of America.

Individual groups and members have received support from BCKDF, CANARIE, CRC and DRAC, Canada; CERN-CZ, PRIMUS 21/SCI/017 and UNCE SCI/013, Czech Republic; COST, ERC, ERDF, Horizon 2020, ICSC-NextGenerationEU and Marie Sk{\l}odowska-Curie Actions, European Union; Investissements d'Avenir Labex, Investissements d'Avenir Idex and ANR, France; DFG and AvH Foundation, Germany; Herakleitos, Thales and Aristeia programmes co-financed by EU-ESF and the Greek NSRF, Greece; BSF-NSF and MINERVA, Israel; Norwegian Financial Mechanism 2014-2021, Norway; NCN and NAWA, Poland; La Caixa Banking Foundation, CERCA Programme Generalitat de Catalunya and PROMETEO and GenT Programmes Generalitat Valenciana, Spain; G\"{o}ran Gustafssons Stiftelse, Sweden; The Royal Society and Leverhulme Trust, United Kingdom.

In addition, individual members wish to acknowledge support from CERN: European Organization for Nuclear Research (CERN PJAS); Chile: Agencia Nacional de Investigaci\'on y Desarrollo (FONDECYT 1190886, FONDECYT 1210400, FONDECYT 1230812, FONDECYT 1230987); China: National Natural Science Foundation of China (NSFC - 12175119, NSFC 12275265, NSFC-12075060); Czech Republic: PRIMUS Research Programme (PRIMUS/21/SCI/017); EU: H2020 European Research Council (ERC - 101002463); European Union: European Research Council (ERC - 948254, ERC 101089007), Horizon 2020 Framework Programme (MUCCA - CHIST-ERA-19-XAI-00), European Union, Future Artificial Intelligence Research (FAIR-NextGenerationEU PE00000013), Italian Center for High Performance Computing, Big Data and Quantum Computing (ICSC, NextGenerationEU); France: Agence Nationale de la Recherche (ANR-20-CE31-0013, ANR-21-CE31-0013, ANR-21-CE31-0022, ANR-22-EDIR-0002), Investissements d'Avenir Labex (ANR-11-LABX-0012); Germany: Baden-Württemberg Stiftung (BW Stiftung-Postdoc Eliteprogramme), Deutsche Forschungsgemeinschaft (DFG - 469666862, DFG - CR 312/5-1); Italy: Istituto Nazionale di Fisica Nucleare (ICSC, NextGenerationEU), Ministero dell'Università e della Ricerca (PRIN - 20223N7F8K - PNRR M4.C2.1.1); Japan: Japan Society for the Promotion of Science (JSPS KAKENHI JP21H05085, JSPS KAKENHI JP22H01227, JSPS KAKENHI JP22H04944, JSPS KAKENHI JP22KK0227); Netherlands: Netherlands Organisation for Scientific Research (NWO Veni 2020 - VI.Veni.202.179); Norway: Research Council of Norway (RCN-314472); Poland: Polish National Agency for Academic Exchange (PPN/PPO/2020/1/00002/U/00001), Polish National Science Centre (NCN 2021/42/E/ST2/00350, NCN OPUS nr 2022/47/B/ST2/03059, NCN UMO-2019/34/E/ST2/00393, NCN \& H2020 MSCA 945339, UMO-2020/37/B/ST2/01043, UMO-2021/40/C/ST2/00187, UMO-2022/47/O/ST2/00148); Slovenia: Slovenian Research Agency (ARIS grant J1-3010); Spain: BBVA Foundation (LEO22-1-603), Generalitat Valenciana (Artemisa, FEDER, IDIFEDER/2018/048), La Caixa Banking Foundation (LCF/BQ/PI20/11760025), Ministry of Science and Innovation (MCIN \& NextGenEU PCI2022-135018-2, MICIN \& FEDER PID2021-125273NB, RYC2019-028510-I, RYC2020-030254-I, RYC2021-031273-I, RYC2022-038164-I), PROMETEO and GenT Programmes Generalitat Valenciana (CIDEGENT/2019/023, CIDEGENT/2019/027); Sweden: Swedish Research Council (VR 2018-00482, VR 2022-03845, VR 2022-04683, VR grant 2021-03651), Knut and Alice Wallenberg Foundation (KAW 2017.0100, KAW 2018.0157, KAW 2018.0458, KAW 2019.0447); Switzerland: Swiss National Science Foundation (SNSF - PCEFP2\_194658); United Kingdom: Leverhulme Trust (Leverhulme Trust RPG-2020-004); United States of America: U.S. Department of Energy (ECA DE-AC02-76SF00515), Neubauer Family Foundation.

%
%


%
%
%
%
%
%
%

%
\clearpage
\appendix
\part*{Appendix}
\addcontentsline{toc}{part}{Appendix}
\section{\ptmiss\ significance}
\label{sec:app_metsig}

Section~\ref{sec:reco} introduced the concept of an object-based approach to \ptmissvec, described in Eq.~(\ref{eq:reco:underbraceptmiss}).
An analogous object-based approach can be used to define an improved, object-based, \ptmiss\ significance.
This significance encodes the resolutions of all reconstructed objects while also accounting for the correlations between each object in an event.

The relative resolution of each hard object as a function of their \pT\ motivates the use of an object-based approach to the significance in Figure \ref{fig:RelObjRes}.
The relative resolutions can vary by a large amount across the \pT\ range and even in what \abseta\ region the candidate object is in.

\begin{figure}[H]
\centering
\includegraphics[width=0.5\textwidth]{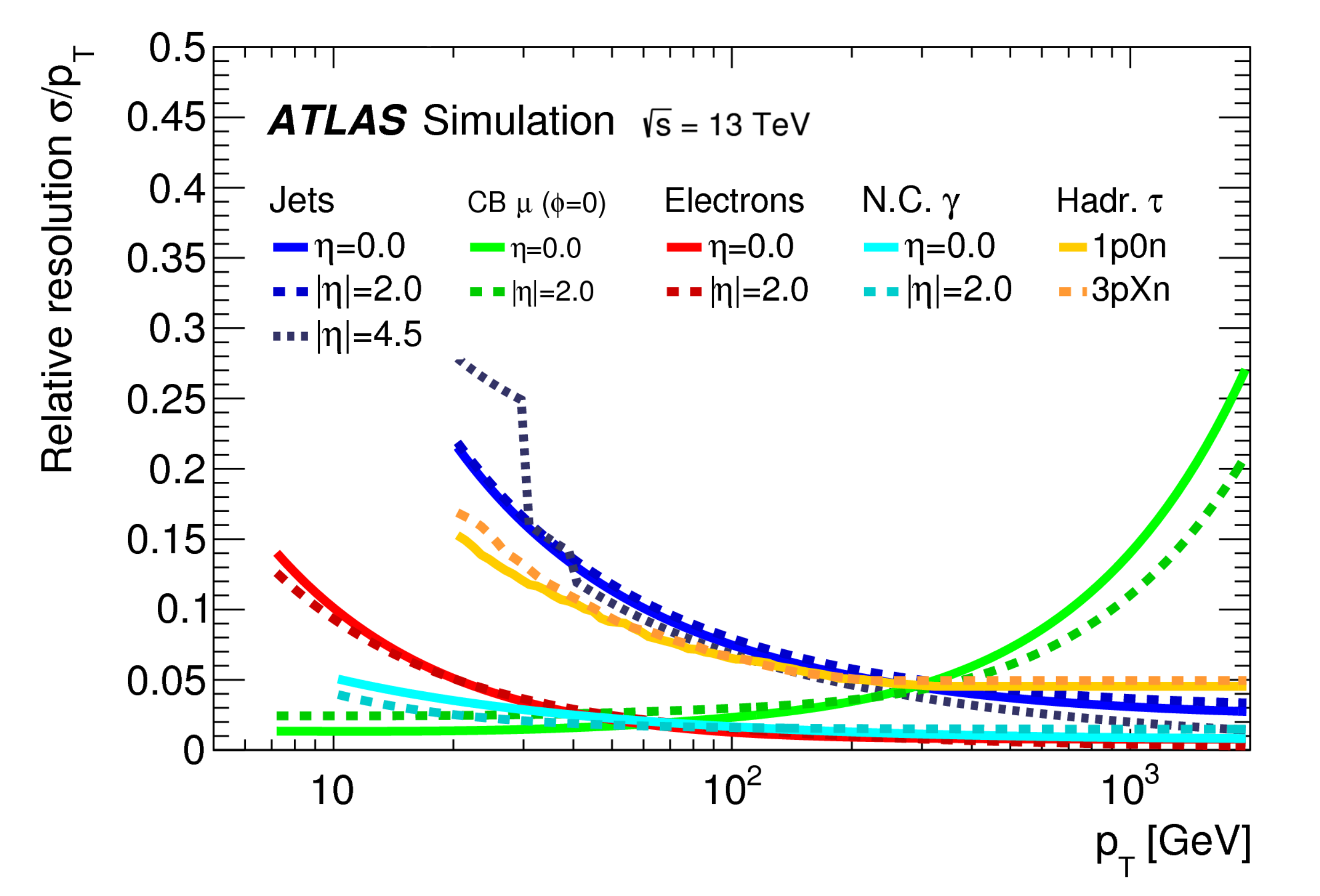}
\caption{Each of the relative resolutions (\(\sigma / \pT\)) for the objects entering the \ptmiss, defined in Section~\ref{sec:reco}. The lines are split by \abseta\ conditions and run with the \pT\ of the object in question. The muons are said to be combined (CB) meaning that they come from combined inner detector tracks and muon spectrometer hits. The photons are those which have not converted into an \(e^{+}e^{-}\) pair. The jet curves include the contribution from pile-up, which is binned in \pt, giving the sharp shape for \(|\eta|=4.5\). More detail on object definitions is found in Section~\ref{sec:objects}.}
\label{fig:RelObjRes}
\end{figure}

With the objects and their respective resolutions used in Eq.~(\ref{eq:reco:underbraceptmiss}) in mind, one can formulate a true significance.
To determine if the observed missing transverse momentum, \ptmissvec, is due to a real invisible particle, or instead caused by resolution effects and mismeasurement of detector objects, a hypothesis test between there being no momentum carried by invisible particles (\(\ptmisstruevec = 0\)) against there being genuine \pT\ carried by invisible particles (\(\ptmisstruevec \neq 0\)) is defined.
This test forms the missing transverse momentum significance (\ptmisssig) definition,

\begin{equation*}\label{auxeq:METSigHypoTest}
\mathcal{S}^2 = 2 \mathrm{ln} \left(\frac{\mathrm{max}_{\ptmisstruevec \neq 0} \mathcal{L}\left( \ptmissvec | \ptmisstruevec \right)}{\mathrm{max}_{\ptmisstruevec = 0} \mathcal{L}\left( \ptmissvec| \ptmisstruevec \right)} \right).
\end{equation*}

This log likelihood ratio, based on the Neyman-Pearson lemma \cite{Neyman:1933wgr}, assumes that each of the likelihoods depends on all the objects measured in an event, their multiplicities, types and kinematic properties.
In other words \(\mathcal{S}\) is an event-by-event evaluation of the \(p\)-value that the observed \ptmissvec, is consistent with the null hypothesis that there is no truth \ptmissvec, \(\pTobjtruevec = 0\),
\begin{equation}\label{auxeq:METSigHypoTestMaxed}
\mathcal{S}^2 = 2 \mathrm{ln} \left(\frac{\mathcal{L}\left( \ptmissvec | \ptmisstruevec \right)}{\mathcal{L}\left( \ptmissvec| 0 \right)} \right).
\end{equation}

In addition to this, the functional form of \(\mathcal{L}\left( \ptmissvec | \ptmisstruevec \right)\) is required to calculate \ptmisssig. This can be found following a few assumptions.
Firstly, the \pT\ measurement for each object, \pTobjvec, is assumed to be independent of all others (where \(\mathrm{Obj \in \{ e, \, \gamma, \, \tau, \, \mu, \, \mathrm{jet} \}}\)).
Each of the objects measuring \pTobjvec\ (given a true value of \pTobjtruevec) is taken to follow a particular probability distribution of the form \(f \left( \pTobjvec - \pTobjtruevec \right)\).
The probability distribution for each object is assumed to be Gaussian and has a covariance matrix labelled \metsigcovobj.
This is the sum of covariances quantifying the resolutions of each object, in \pT\ and \(\phi\), entering the \ptmissvec\ calculation.
Finally, conservation of momentum in the transverse plane means that if the true momentum of each measured particle were to be summed this would balance with the negative signed invisible particle momentum: \(\sum_{\mathrm{Objects}}\pTobjvec = - \ptmisstruevec.\)

With these assumptions made, the form of the likelihood is
\begin{equation*}\label{auxeq:Likelihood}
\mathcal{L}\left( \ptmissvec | \ptmisstruevec \right) \propto \exp\left[ - \frac{1}{2} \left( \ptmissvec - \ptmisstruevec \right)^{\mathrm{T}} \left( \sum_{\mathrm{Objects}} \metsigcovobj \right)^{-1}  \left( \ptmissvec - \ptmisstruevec \right) \right],
\end{equation*}
which is a two dimensional Gaussian distribution.
Entering this into the maximised log likelihood ratio, Eq.~(\ref{auxeq:METSigHypoTestMaxed}), results in the cancellation of any preceding coefficients and leaves:
\begin{equation}\label{auxeq:METSigChiSquare}
\mathcal{S}^2 = \left( \ptmissvec \right)^{\mathrm{T}} \left( \sum_{\mathrm{Objects}} \metsigcovobj \right)^{-1} \left( \ptmissvec \right).
\end{equation}

This is now a sum of independent standard normal variables in two dimensions, or more simply a \chiSq\ hypothesis test in two dimensions.
Equation (\ref{auxeq:METSigChiSquare}) links \ptmissvec\ to all the object resolutions which are encoded in the covariance matrix summation.

In this format the results of the \chiSq\ test are easily interpreted with a single value that indicates how likely it is that the null hypothesis (\(\ptmisstruevec = 0\)) holds.
Low values of \(\mathcal{S}^2\) indicate that the \ptmissvec\ comes from fake sources like mismeasurement or resolution effects while high values show that it is likely the \ptmissvec\ comes from a real invisible particle leaving the detector without interaction.

The covariance matrix for each object is defined with an axis along the measured transverse momentum vector of the object under consideration, \pTobjvec.
This allows each object's covariance matrix to be simply written in terms of the resolution of the magnitude of \pTobjvec\ and the resolution in the azimuthal angle,
\begin{equation}\label{auxeq:CovMatrixObject}
\metsigcovobj =
\begin{pmatrix}
\sigma^{2}_{\pTobj} & 0 \\
0 & {\pTobj}^{2}\sigma^{2}_{\phiobj}
\end{pmatrix},
\end{equation}
under the condition that \pTobj\ and \phiobj\ are independent measurements.

So far only the well defined hard objects have been considered but as was seen there is a soft term in Eq.~(\ref{eq:reco:underbraceptmiss}) with its own resolution.
The covariance matrix for the soft term is defined in a similar fashion to the objects in Eq.~(\ref{auxeq:CovMatrixObject}) and allows the complete covariance matrix to be written as:

\begin{equation*}\label{auxeq:CovMatrixTot}
\metsigcov = \sum_{\mathrm{Objects}} \metsigcovobj + \metsigcovsoft.
\end{equation*}

The soft term is included in the \(\mathrm{Obj}\) set with the other hard objects.
The total covariance matrix can be rotated using the two dimensional rotation matrix \covrot\ in the azimuthal plane,
\begin{equation*}\label{auxeq:CovMatrixRotxy}
\metsigcovxy = \sum_{\mathrm{Objects}} \covrotinv \, \metsigcovobj \, \covrot =
\begin{pmatrix}
\sigma_{x}^{2} & \sigma_{xy}^{2} \\
\sigma_{xy}^{2} & \sigma_{y}^{2} \\
\end{pmatrix}.
\end{equation*}
Here the \(\sigma\) terms are now the combined resolutions of \ptmiss\ in \(x\) and \(y\).
To simplify the situation even further it is prudent to again rotate the system to the frame of \ptmissvec.
In this frame there are two components to the total \ptmissvec\ resolutions; one longitudinal (or parallel) ``L'' and another transverse (or perpendicular) to \ptmissvec\ ``T''.
To do this another two dimensional rotation matrix is applied, \covrotLT, to end up with:
\begin{equation*}\label{auxeq:CovMatrixRotLT}
\metsigcovlt = \covrotLT \, \metsigcovxy \, \covrotLTinv \ =
\begin{pmatrix}
\sigLsq & \rhoLT \sigL \sigT \\
\rhoLT \sigL \sigT & \sigTsq \\
\end{pmatrix}.
\end{equation*}

The longitudinal variance is \sigL, the transverse variance is \sigT\ and \rhoLT\ represents the covariance between measurements in the longitudinal and transverse directions.
Equation~(\ref{auxeq:METSigChiSquare}) takes the inverse of \metsigcov, this can be retrieved using the following relation for a two-by-two matrix,
\begin{equation*}\label{auxeq:MatrixInversion}
{\metsigcov}^{-1} = \frac{1}{\det \metsigcov} \left[ (\mathrm{tr} \metsigcov) \textbf{I} -\metsigcov \right].
\end{equation*}
Which gives,
\begin{equation}\label{auxeq:CovMatrixLTInv}
\metsigcovltinv = \frac{1}{\sigLsq\sigTsq - \rhoLTsq\sigLsq\sigTsq}
\begin{pmatrix}
\sigTsq & - \rhoLT \sigL \sigT \\
- \rhoLT \sigL \sigT & \sigLsq \\
\end{pmatrix}.
\end{equation}
This can finally be substituted into a slightly more expanded version (for clarity) of Eq.~(\ref{auxeq:METSigChiSquare}) with the total covariance matrix in the ``LT'' frame, as defined above,
\begin{equation*}\label{auxeq:METSigChiSquareComps}
\mathcal{S}^2 = \left( \ptmiss, 0 \right) \metsigcovltinv
\begin{pmatrix}
\ptmiss \\
0\\
\end{pmatrix}.
\end{equation*}

Finally entering Eq.~(\ref{auxeq:CovMatrixLTInv}) and multiplying out the matrix one ends up with the much simpler definition of \ptmisssig:
\begin{equation*}\label{auxeq:METSigDef}
\mathcal{S}^2 = \frac{|p_{\mathrm{T}}^{\mathrm{miss}}|^{2}}{\sigma_{\mathrm{L}}^{2} \left( 1 - \rho_{\mathrm{LT}}^2\right)}
\end{equation*}
or
\begin{equation}\label{auxeq:METSigDefroot}
\ptmisssig = \frac{\ptmiss}{\sqrt{\sigma_{\mathrm{L}}^{2} \left( 1 - \rho_{\mathrm{LT}}^2\right)}}.
\end{equation}

Equation~(\ref{auxeq:METSigDefroot}) is the final object-based missing transverse momentum significance and is a true significance.
This variable contains the measured quantity in the numerator along with information on the variance of its measurement in the denominator in a dimensionless way.%


%

%
\section{\ptmiss\ with EMTopo jets}
\label{sec:app_EMTopo}

EMTopo jets are reconstructed from topo-clusters, using the anti-\(k_t\) algorithm with \(R=0.4\).
The topo-clusters are calibrated at the EM energy scale, and fully calibrated~\cite{JETM-2018-05}.
Requirements of \(\pt>\SI{20}{\GeV}\) and \(|\eta|<4.5\) are made on the calibrated EMTopo jets.
Tracks are matched to jets using ghost-association~\cite{PERF-2015-04}.
This consists of repeating the jet clustering process with the addition of `ghost' versions of tracks with the same direction but infinitesimal \pt.
A track is ghost-associated if it is contained within the re-clustered jet.
After full calibration, EMTopo jets are subject to JVT requirements that are the same as those for PFlow jets, except that \(\text{JVT}>0.59\) is used to achieve the same efficiency.

The reconstruction of \ptmiss\ when EMTopo jets are used follows the procedure defined in Section~\ref{sec:reco}.
Similar to PFlow-based \ptmiss\ (illustrated in Table~\ref{tab:MET_WPs}), four working points are supported, and shown in Table~\ref{tab:MET_WPs_EMTopo}.

\begin{table}[h]
\centering
\small
\caption{Selections for the \ptmiss\ working points supported for EMTopo jets.}
\begin{tabular}{ l| l l l l }
\hline
& \multicolumn{4}{|c}{Selections} \\
& \multicolumn{2}{|c}{\(\pT\ [\GeV]\) for } & & fJVT for jets with \\
& \multicolumn{2}{|c}{jets with:} & JVT for jets with & \(2.5<|\eta|<4.5\) \& \\
Working point	& \(|\eta|<2.4\)	& \(2.4 < |\eta| < 4.5\)	& \(|\eta|<2.4\)	& \(\pt<120~\)\GeV \\
\hline
\workingPoint{Loose}		& \(>20\) & \(>20\)	& \(>0.59\text{ for } \pt<\SI{60}{\GeV}\)	& - \\
\workingPoint{Tight}		& \(>20\) & \(>30\)	& \(>0.59\text{ for } \pt<\SI{60}{\GeV}\)	& \(<0.4\)\\
\workingPoint{Tighter}		& \(>20\) & \(>35\)	& \(>0.59\text{ for } \pt<\SI{60}{\GeV}\)	& -\\
\workingPoint{Tenacious}	& \(>20\) & \(>35\)	& \(>0.91\text{ for } 20<\pt<\SI{40}{\GeV}\) 	& \(<0.5\) \\
& & & \(>0.59\text{ for } 40<\pT<\SI{60}{\GeV}\) & \\
& & & \(>0.11\text{ for } 60<\pT<\SI{120}{\GeV}\) & \\
\hline
\end{tabular}
\label{tab:MET_WPs_EMTopo}
\end{table}

\begin{figure}[htb!]
\centering
\subfloat[]{\label{fig:termsZmmEMTopo:ptmiss}\includegraphics[width=0.35\textwidth]{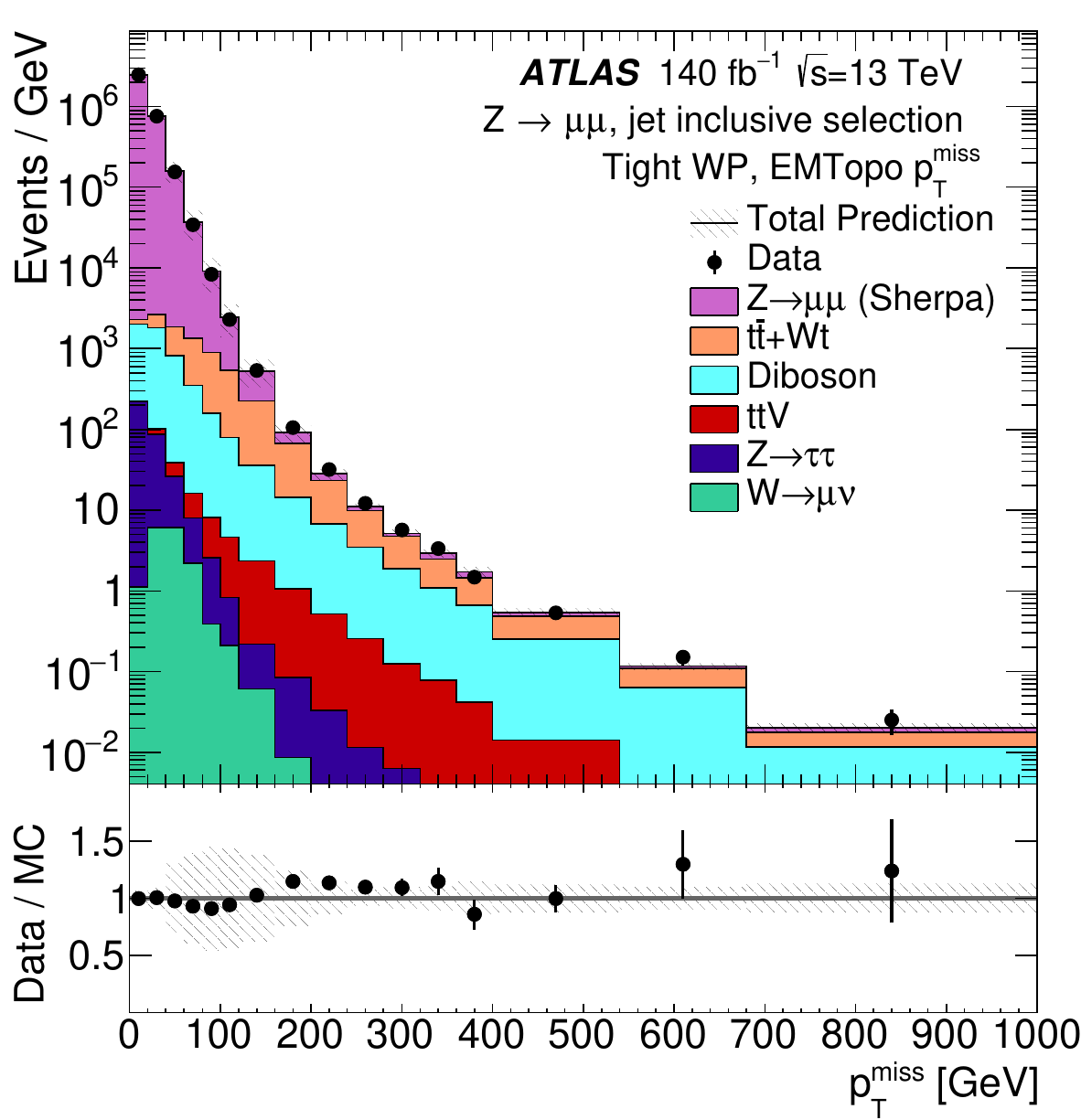}}\qquad
\subfloat[]{\label{fig:termsZmmEMTopo:soft}\includegraphics[width=0.35\textwidth]{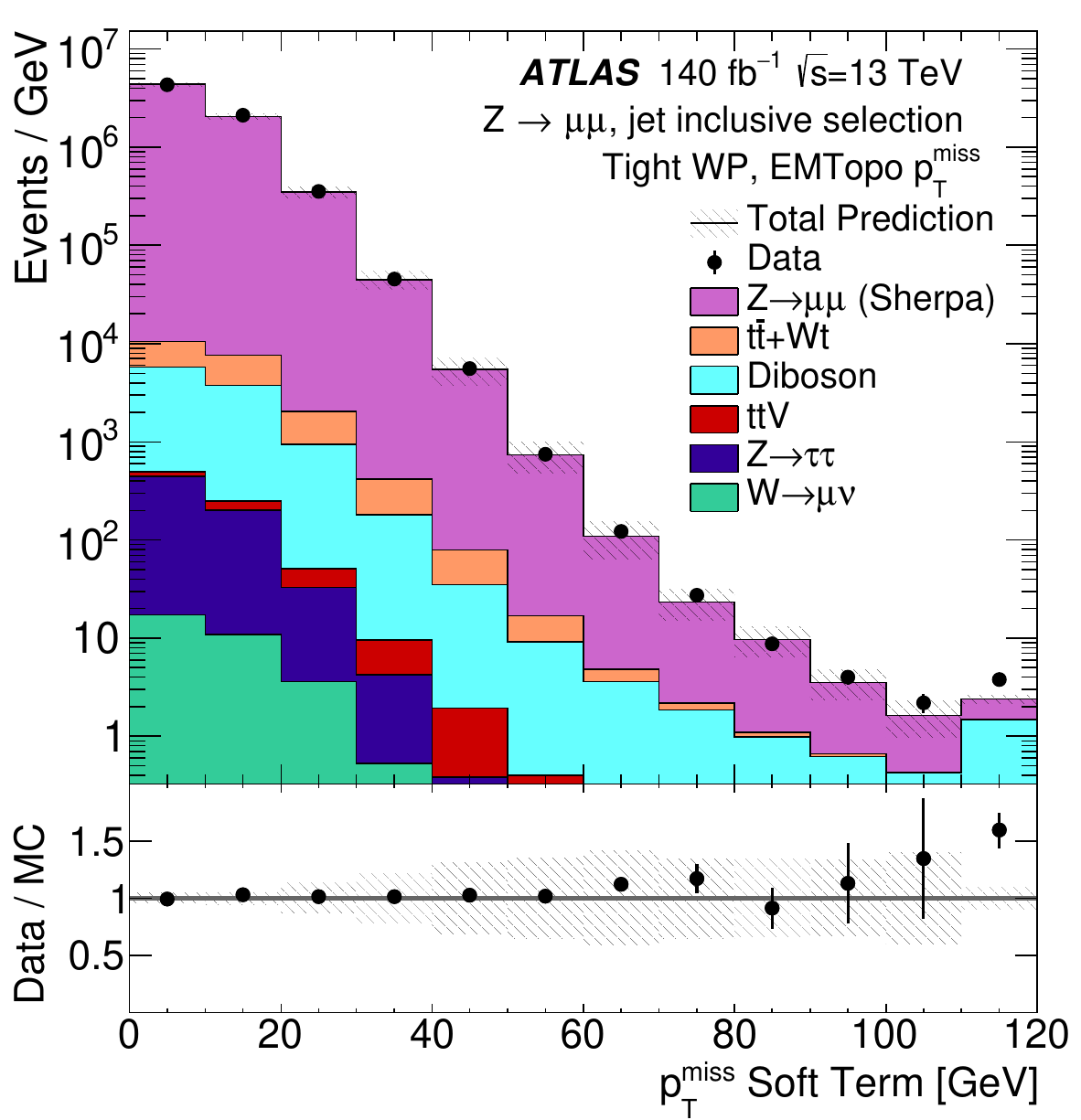}}
\caption{Distributions of \ptmiss\ (a) and its constituent soft (b) terms in MC and data. Events satisfy a \Zmm\ selection. EMTopo jets are used with a jet inclusive selection and the \workingPoint{Tight} \ptmiss\ working point. \SHERPA\ is used to generate the \Zmm\ events. The error band includes MC statistical, luminosity and detector uncertainties.}
\label{fig:termsZmmEMTopo}
\end{figure}

Figure~\ref{fig:termsZmmEMTopo} shows the \ptmiss\ and \ptsoft\ distributions, for \ptmiss\ built from EMTopo jets satisfying a \Zmm\ selection.
These show a similar level of agreement between data and Monte-Carlo simulation in comparison to the PFlow-based distributions shown in Figure~\ref{fig:termsZmmZee}.
The soft term has a smaller tail when PFlow jets are used to build \ptmiss\ compared to using EMTopo jets, attributed to the particle flow algorithm's improved ability to reject pile-up.

\begin{figure}
\centering
\subfloat[]{\label{fig:resZmmjets:muEMTopo}\includegraphics[width=0.35\textwidth]{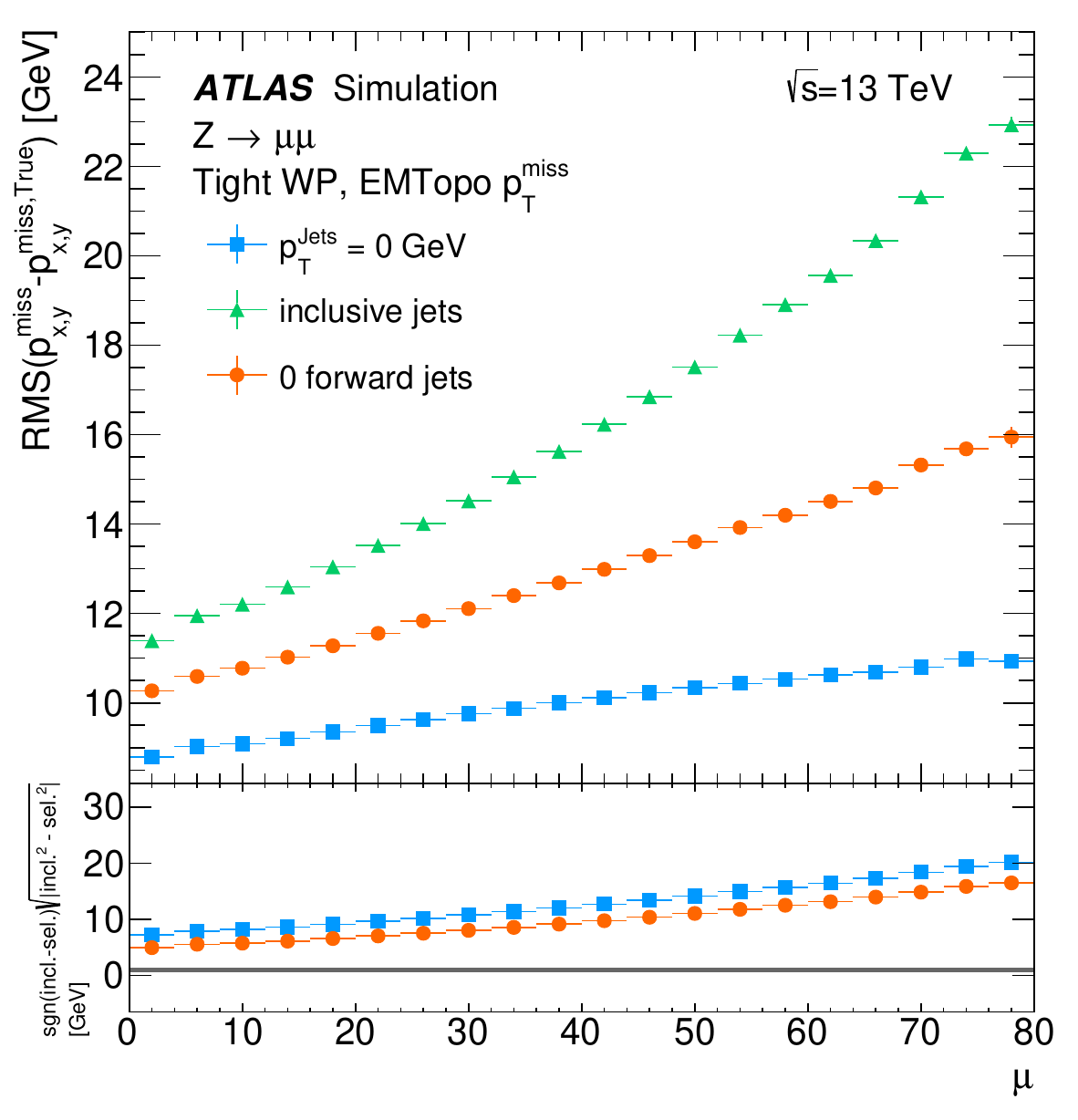}}\qquad
\subfloat[]{\label{fig:resZmmwps:muEMTopo}\includegraphics[width=0.35\textwidth]{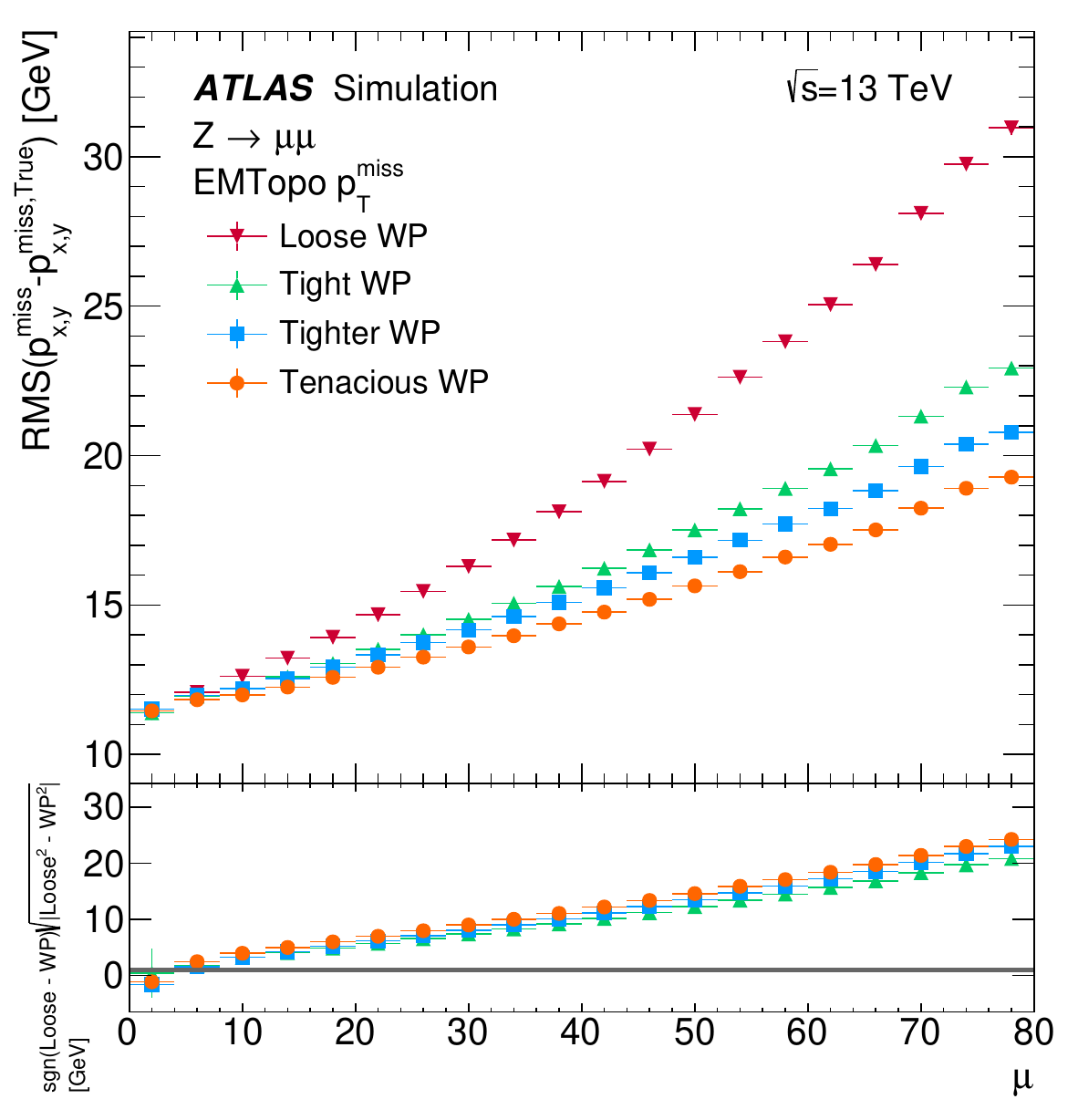}}\\
\caption{The \pxmiss\ and \pymiss\ resolution for different jet selections (sel.) (a) and different \ptmiss\ working points (b) as a function of \(\mu\). The \workingPoint{Tight} \ptmiss\ working point is used and \SHERPA\ \Zmm\ MC simulated events are used. EMTopo jets are used. The error bars include the MC statistical uncertainty. In the y-axis label of the lower panels, `incl.' refers to the inclusive jet selection, `sel.' to the alternate jet selection under consideration and `WP' to the working point under consideration.}
\label{fig:resZmm_EMTopo}
\end{figure}

The values of the soft term uncertainties calculated for EMTopo, are given Figure~\ref{fig:systs:summaryEMTopo}.
The EMTopo uncertainties are generally larger than for PFlow (shown in Figure~\ref{fig:systs:summary}), attributed to PFlow's better rejection of poorly modelled pile-up.

\begin{figure}[!htb]
\centering
\includegraphics[width=0.4\textwidth]{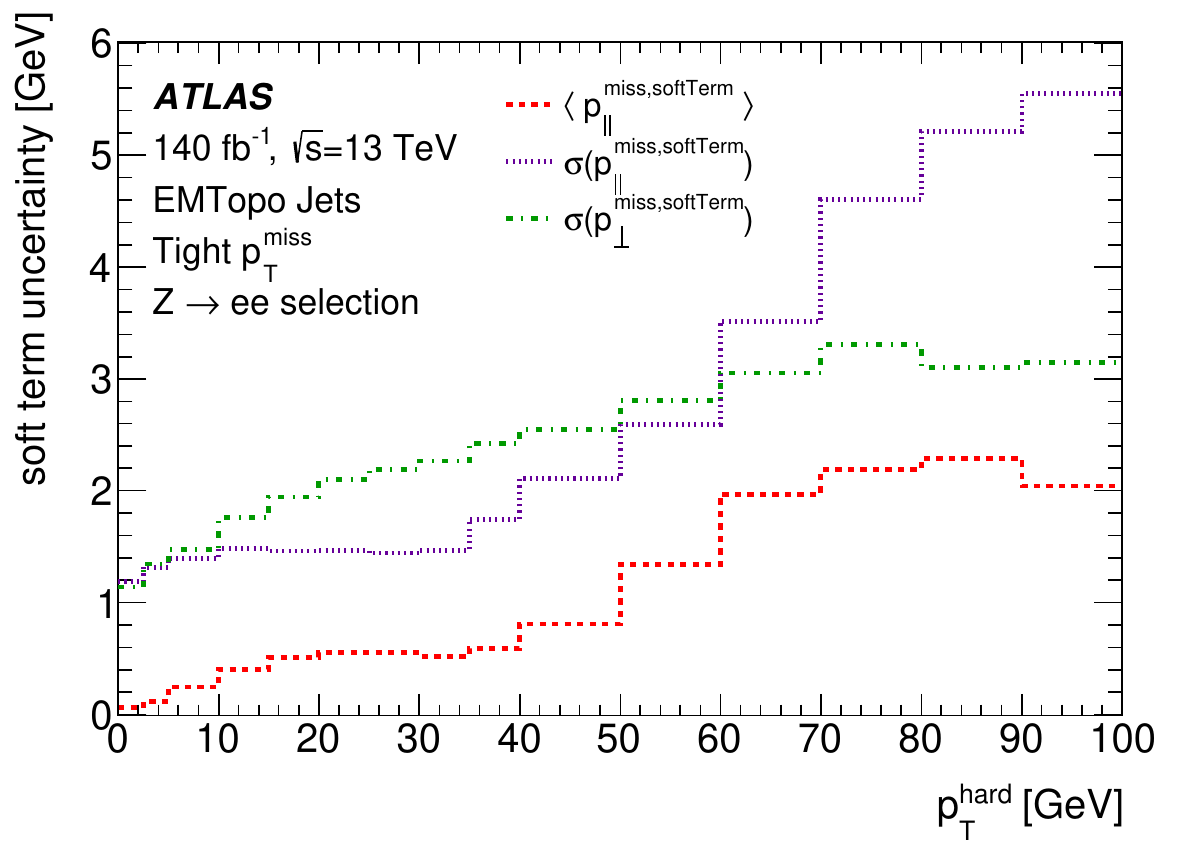}
\caption{A summary of the \ptsoft\ systematic uncertainties for \ptmiss\ built with EMTopo jets. The parallel scale (\scalepara), parallel resolution (\resopara) and transverse resolution (\resoperp) of the \ptsoft\ projection onto the \pthard, binned in \pthard. Full \runtwo\ data and Monte Carlo samples are shown with a \Zee\ event selection applied.}
\label{fig:systs:summaryEMTopo}
\end{figure}


%

%
%
%
\clearpage
\printbibliography
\clearpage
 
\begin{flushleft}
\hypersetup{urlcolor=black}
{\Large The ATLAS Collaboration}

\bigskip

\AtlasOrcid[0000-0002-6665-4934]{G.~Aad}$^\textrm{\scriptsize 102}$,
\AtlasOrcid[0000-0001-7616-1554]{E.~Aakvaag}$^\textrm{\scriptsize 16}$,
\AtlasOrcid[0000-0002-5888-2734]{B.~Abbott}$^\textrm{\scriptsize 120}$,
\AtlasOrcid[0000-0002-1002-1652]{K.~Abeling}$^\textrm{\scriptsize 55}$,
\AtlasOrcid[0000-0001-5763-2760]{N.J.~Abicht}$^\textrm{\scriptsize 49}$,
\AtlasOrcid[0000-0002-8496-9294]{S.H.~Abidi}$^\textrm{\scriptsize 29}$,
\AtlasOrcid[0000-0002-9987-2292]{A.~Aboulhorma}$^\textrm{\scriptsize 35e}$,
\AtlasOrcid[0000-0001-5329-6640]{H.~Abramowicz}$^\textrm{\scriptsize 152}$,
\AtlasOrcid[0000-0002-1599-2896]{H.~Abreu}$^\textrm{\scriptsize 151}$,
\AtlasOrcid[0000-0003-0403-3697]{Y.~Abulaiti}$^\textrm{\scriptsize 117}$
% ...
\bigskip
\\

% ...
$^{16}$Department for Physics and Technology, University of Bergen, Bergen; Norway.\\
% ...
$^{29}$Physics Department, Brookhaven National Laboratory, Upton NY; United States of America.\\
% ...
$^{35}$$^{(a)}$Facult\'e des Sciences Ain Chock, Universit\'e Hassan II de Casablanca;$^{(b)}$Facult\'{e} des Sciences, Universit\'{e} Ibn-Tofail, K\'{e}nitra;$^{(c)}$Facult\'e des Sciences Semlalia, Universit\'e Cadi Ayyad, LPHEA-Marrakech;$^{(d)}$LPMR, Facult\'e des Sciences, Universit\'e Mohamed Premier, Oujda;$^{(e)}$Facult\'e des sciences, Universit\'e Mohammed V, Rabat;$^{(f)}$Institute of Applied Physics, Mohammed VI Polytechnic University, Ben Guerir; Morocco.\\
% ...
$^{49}$Fakult\"{a}t Physik , Technische Universit{\"a}t Dortmund, Dortmund; Germany.\\
% ...
$^{55}$II. Physikalisches Institut, Georg-August-Universit\"{a}t G\"ottingen, G\"ottingen; Germany.\\
% ...
$^{102}$CPPM, Aix-Marseille Universit\'e, CNRS/IN2P3, Marseille; France.\\
% ...
$^{117}$Department of Physics, New York University, New York NY; United States of America.\\
% ...
$^{120}$Homer L. Dodge Department of Physics and Astronomy, University of Oklahoma, Norman OK; United States of America.\\
% ...
$^{151}$Department of Physics, Technion, Israel Institute of Technology, Haifa; Israel.\\
$^{152}$Raymond and Beverly Sackler School of Physics and Astronomy, Tel Aviv University, Tel Aviv; Israel.\\
% ...
$^{a}$ Also Affiliated with an institute covered by a cooperation agreement with CERN.\\
$^{b}$ Also at An-Najah National University, Nablus; Palestine.\\
$^{c}$ Also at Borough of Manhattan Community College, City University of New York, New York NY; United States of America.\\
$^{d}$ Also at Center for High Energy Physics, Peking University; China.\\
$^{e}$ Also at Center for Interdisciplinary Research and Innovation (CIRI-AUTH), Thessaloniki; Greece.\\
$^{f}$ Also at Centro Studi e Ricerche Enrico Fermi; Italy.\\
% ...

\end{flushleft}


%

%
%

%

%
%
%
%

\end{document}